\documentclass[structabstract]{aa} 
\usepackage{graphicx}
\usepackage{txfonts}
\usepackage{natbib}
\usepackage{rotating} 
\usepackage{epstopdf}
\bibpunct{(}{)}{;}{a}{}{,} 

\begin{document}
\title{Herschel-PACS observations of discs in the $\rm \eta$ Chamaeleontis association.\thanks{{\it Herschel} is an ESA space observatory with
    science instruments provided by European-led Principal
    Investigator consortia and with important participation from
    NASA.}}
   \author{P. Riviere-Marichalar\inst{1}, P. Elliott\inst{2,3}, I. Rebollido\inst{1}, A. Bayo\inst{4,5}, A. Ribas\inst{1}, B. Mer\'in\inst{1},  I. Kamp\inst{6}, W. R. F. Dent\inst{7}, B. Montesinos\inst{8}
}
\institute{European Space Astronomy Centre (ESA), P.O. Box 78, 28691 Villanueva de la Ca\~nada, Madrid, Spain
\\
             \email{priviere@sciops.esa.int}
\and European Southern Observatory, Alonso de Cordova 3107, Vitacura Casilla 19001, Santiago 19, Chile      
\and School of Physics, University of Exeter, Stocker Road, Exeter, EX4 4QL 
\and Instituto de F\'isica y Astronom\'ia, Facultad de Ciencias, Universidad de Valpara\'iso, Av. Gran Breta\~na 1111, 5030 Casilla, Valpara\'iso, Chile 
\and ICM nucleus on protoplanetary discs, Universidad de Valpara\'iso, Av. Gran Breta\~na 1111, Valpara\'iso, Chile 
\and Kapteyn Astronomical Institute, University of Groningen, P.O. Box 800, 9700 AV Groningen, The Netherlands 
 \and ALMA, Avda Apoquindo 3846, Piso 19, Edificio Alsacia, Las Condes, Santiago, Chile 
\and Depto. Astrof\'isica , Centro de Astrobiolog\'{\i}a (CAB, INTA--CSIC), P.O. Box 78, ESAC Campus, 28691 Villanueva de la Ca\~nada, Madrid, Spain 
}
   \authorrunning{The team}
   \date{}

 \abstract
{Protoplanetary discs are the birthplace for planets. Studying protoplanetary discs is the key to constraining theories of planet formation. By observing dust and gas in associations at different ages we can study the evolution of these discs, their clearing timescales, and their physical and geometrical properties. The stellar association $\rm \eta$ Cha is peculiar; some members still retain detectable amounts of gas in their discs at the late age of $\rm \sim$ 7 Myr, making it one of the most interesting young stellar associations in the solar neighbourhood.} 
{We characterise the properties of dust and gas in protoplanetary and transitional discs in the $\rm \eta$ Cha young cluster, with special emphasis on explaining the peculiarities that lead to the observed high disc detection fraction and prominent IR excesses at an age of $\rm \sim7$ Myr.} 
{We observed 17 members of the $\rm \eta$ Cha association with \textit{Herschel}-PACS in photometric mode and line spectroscopic mode. A subset of members were also observed in range spectroscopic mode. The observations trace [OI] and $\rm H_{2}O$ emissions at 63.18 and 63.32 $\rm \mu m$, respectively, as well as CO, OH, $\rm CH^{+}$ and [CII] at different wavelengths for those systems observed in range mode. The photometric observations were used to build complete spectral energy distributions (SEDs) from the optical to the far-IR.  High-resolution multi-epoch optical spectra with high signal-to-noise ratios were also analysed  to study the multiplicity of the sources and look for further gas (accreting) and outflow indicators.}
{We detect four out of fifteen sources observed at 70 $\rm \mu m$, four out of six at 100 $\rm \mu m$, and six out of sixteen at 160 $\rm \mu m$. Only one system shows [OI] emission at 63 $\rm \mu m$, namely RECX 15 or J0843.3-7905. None of them shows far-IR line emission at any other wavelength. The [OI] emission toward RECX 15 points to the presence of an outflow; however, the emission is not extended. We study $\rm H_{\alpha}$ emission among $\rm \eta$ Cha members and conclude that RECX 4, 5, 9, 11, and 15 are actively accreting in at least  one epoch. }   

   \maketitle
\section{Introduction} \label{Sec:Introduction}
Planets are born in protoplanetary discs that surround young stars.  The study of both gas and dust evolution in protoplanetary discs is therefore a major step to constrain planet formation theories. In the last three decades, we have acquired extensive knowledge on the evolution of the dust phase \citep[see][]{Williams2011}. One of the most prominent discoveries related to the dust evolution is that protoplanetary discs develop inner disc opacity holes, leading to the formation of the so-called transitional discs \citep{Strom1989}. These transitional discs seem to be a transient stage before the fast dissipation \citep[$\rm <0.5$ Myr,][]{Skrutskie1990} of the protoplanetary disc in timescales $\rm <$ 10 Myr \citep[see][for a review]{Williams2011}.

Due to both observational and computational difficulties, we still lack a good description of the evolution of the gas phase, even if gas dominates the mass budget of protoplanetary discs. The most abundant gas species, $\rm H_{2}$, lacks a permanent electric dipole, making IR rotational transitions weak. After $\rm H_{2}$, CO is the most abundant molecule and it has been traditionally  used to study the gas mass. However, CO lines are usually optically thick. Furthermore, CO can be frozen in the mid plane of the disc, and photodissociated in the disc atmosphere \citep{Aikawa1996,Gorti2008}, making it a poor tracer of the total disc gas mass in some cases. The gas-to-dust ratio has a dramatic influence in shaping the disc and in the subsequent planet formation \citep{Pinte2014}.  Therefore, determining the gas-to-dust ratio in different disc conditions is key to understanding planet formation.  Typically, a gas-to-dust ratio of 100 has been assumed, based on the ISM value. However, recent studies show that the actual ratio might be smaller \cite[see, e. g. ][]{Thi2010}. \cite{Williams2014} show that the combination of $\rm ^{13}CO$ and $\rm C^{18}O$ provides a good estimate of the gas mass. To further constrain the geometry of the disc, we need observations of other atomic and molecular species, typically in the near- to far-IR, such as O and $\rm H_{2}O$. 

\begin{table}[!t]
\caption{Stellar parameters for $\eta$ Cha members observed with \textit{Herschel}-PACS}             
\label{tableStar}      
\centering          
\begin{tabular}{lllc}     
\hline\hline       
Source & Sp. Type & T$\rm _{eff}$ &  L$_*$ \\ 
\hline
 --	& -- &(K) & L$_{\sun}$  \\ 
\hline                    
\object{RECX~1}   & K4 & 4100  & 0.96$\rm \pm$0.52\\
\object{RECX~3}   & M3 & 3200 & 0.09$\rm \pm$0.05\\
\object{RECX~4}   & K7 & 3600 & 0.23$\rm \pm $0.13 \\
\object{RECX~5}   & M5 & 3000 & 0.06$\rm \pm$0.03\\
\object{RECX~6}   & M2 & 3300 & 0.11$\rm \pm 0.06$  \\
\object{RECX~7}   & K3 & 4100 & 0.70$\rm \pm 0.38$  \\
\object{RECX~8}   & A7 & 6600 & 23 $\rm \pm $ 2 \\
\object{RECX~9}   & M4 & 3300 & 0.09$\rm \pm$0.05 \\
\object{RECX~10} & M0 & 3800 & 0.22$\rm \pm 0.12$  \\
\object{RECX~11} & K5 & 4100 & 0.55$\rm \pm$0.30  \\
\object{RECX~12} & M3 & 3300 & 0.24$\rm \pm$0.13 \\
\object{RECX~13}  (HD 75505) & A1 & 8000 & 7.2 $\rm \pm $ 4.0  \\
\object{RECX~14}  (J0841.5-7853) & M4 & 2900 & 0.02$\rm \pm$0.01 \\
\object{RECX~15}  (J0843.3-7905) & M2 & 3300 & 0.08$\rm \pm$0.04 \\
\object{RECX~16}  (J0844.2-7833) & M5.5 & 2900 & 0.02$\rm \pm$0.01  \\
\object{RECX~17}  (J0838.9-7916) & M5 & 3100 & 0.04$\rm \pm$0.02  \\
\object{RECX~18}  (J0836.2-7908) & M5.5 & 2900 & 0.02$\rm \pm$0.01 \\
\hline                  
\end{tabular}
\tablefoot{Spectral types from \cite{ZuckermanSong2004} .}
\end{table}

An additional caveat in studying the evolution of gas in protoplanetary discs is that the age of individual systems is hard to constrain \citep{Soderblom2014}. Therefore, we must use stars belonging to young clusters and stellar associations, because the mean age of their members is well known. We can compare the disc frequency and properties of discs in stellar associations at different ages to perform evolutionary studies. 

The $\rm \eta$ Cha cluster is a young stellar association with an age of 5-9 Myr \citep{Mamajek1999,Lawson2001,Luhman2004}, located at a distance of 97 pc to the Sun.  \cite{Luhman2004} propose a list of 18 cluster members within a radius of 0.5 pc from the cluster centre, with spectral types in the range B8-M5.5, with most members (15) belonging to types K and M.

The cluster shows a deficit of low-mass stars \citep[$\rm M < 25 M_{Jup}$][]{Lyo2006}. However, it is not clear whether the census of members is complete. By comparing the IMF of $\eta$ Cha with that of the Trapezium cluster, \cite{Lyo2004} predict 20-29 undiscovered low-mass stars  and brown dwarfs ($\rm 0.025<M_{*}/M_{\odot}<0.15$), meaning that the number of undetected members could be similar to the number of actual members. Different surveys \citep{Luhman2004,Song2004,Lyo2006} looked for new cluster members up to 2.6 pc from the centre, with no success. More recently, \cite{Murphy2010} proposed three new candidate members at distances between 2.6 and 10 pc from the cluster centre in the mass range $\rm 0.08 < M_{*}/M_{\odot} <0.3$. The presence of mass segregation was shown by \cite{Lyo2004}, who claimed that 50 \% of the cluster mass is located inside the inner 0.17 pc. Low-mass members could be located at longer distances (making them harder to be detected), in a low-mass cluster halo, either due to dynamical interactions or to mass segregation during cluster relaxation. 
 
 \cite{Gautier2008} computed a disc fraction of 56\% at 24 $\rm \mu m$, and a lower limit of 31\% at 70 $\rm \mu m$, considered high for a $\rm \sim$7 Myr old cluster. This high disc detection fraction was later confirmed by \cite{Sicilia-Aguilar2009} making use of \textit{Spitzer}-IRS spectroscopic observations of the cluster. However, when discussing the detection fraction one has to remember that the census of $\rm \eta$ Cha members might be incomplete (see previous paragraph). If the low-mass members escaping detection are mostly disc-less stars the disc fraction could dramatically fall. \cite{Fang2013} showed that evolution in loose environments, such as in $\rm \eta$ Cha, proceeds slowly compared to  more crowded associations, a fact that may explain the high number of disc detections in $\rm \eta$ Cha. Furthermore, the presence of a disc seems to be linked to single stars, as only one $\rm \eta$ Cha disc is detected surrounding a binary system \citep{Bouwman2006}. The authors showed that the system is a wide binary, and the disc is most likely circumprimary, and  concluded that binarity results in shorter disc clearing timescales. 
 
In this paper we present photometric and spectroscopic \textit{Herschel}-PACS observations of the $\rm \eta$ Cha cluster, as well VLT/UVES spectroscopic observations. We discuss the implications of our observations for the peculiarities of the $\rm \eta$ Cha cluster discussed in the previous paragraph. In Sec. \ref{Sec:Sample} we describe the sample properties and observations performed, as well as the data reduction process. In Sec.  \ref{Sec:Results} we present the main results from our photometric and spectroscopic survey. In Sec. \ref{Sec:Discussion} we discuss the main implications of our results and in Sec. \ref{Sec:Summary} we overview the contents of this study.

\begin{figure*}[!t]
\begin{center}
\includegraphics[scale=0.35,trim=0mm 0mm 0mm 0mm,clip]{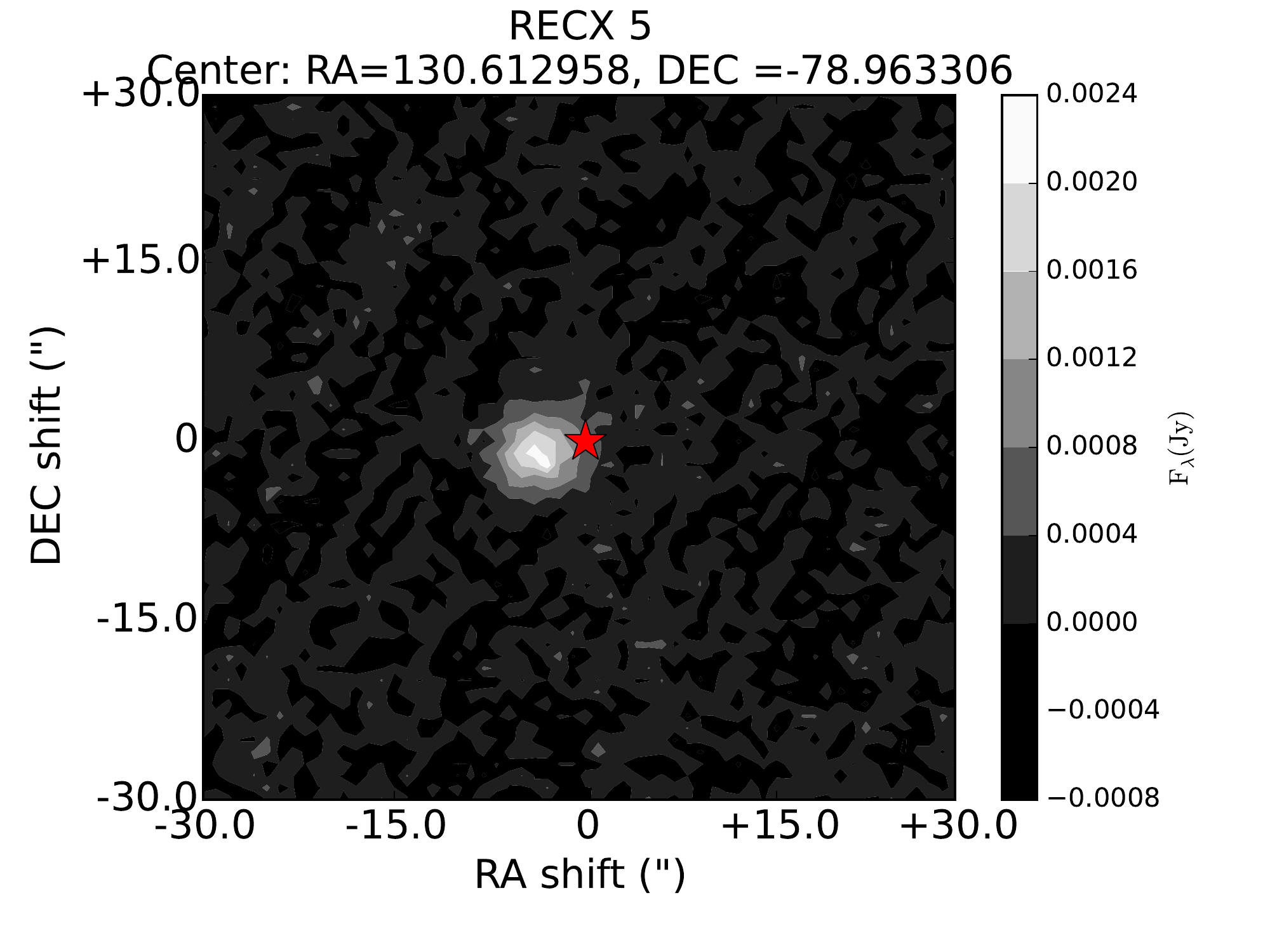}\includegraphics[scale=0.35,trim=0mm 0mm 0mm 0mm,clip]{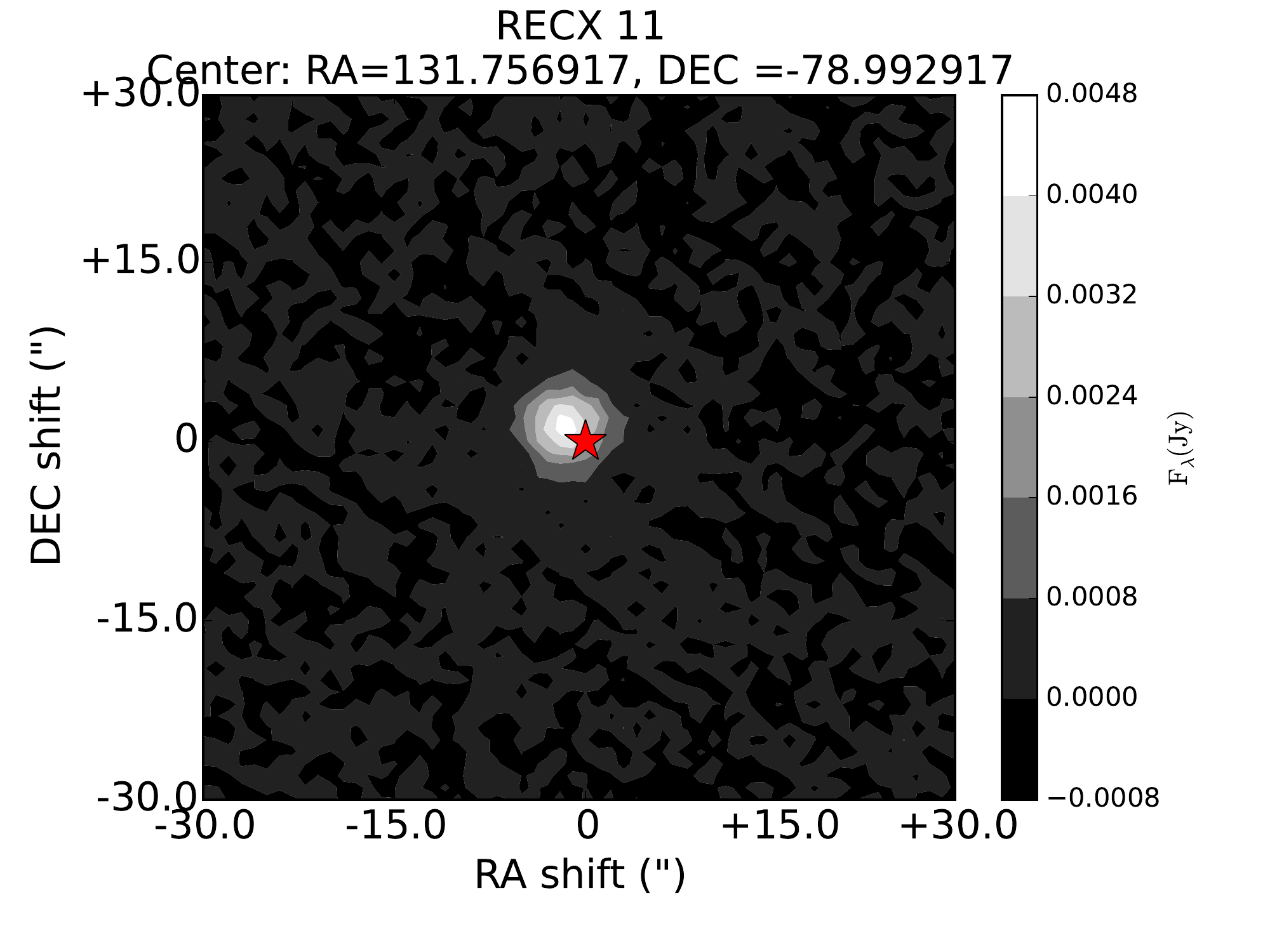}\\
\includegraphics[scale=0.35,trim=0mm 0mm 0mm 0mm,clip]{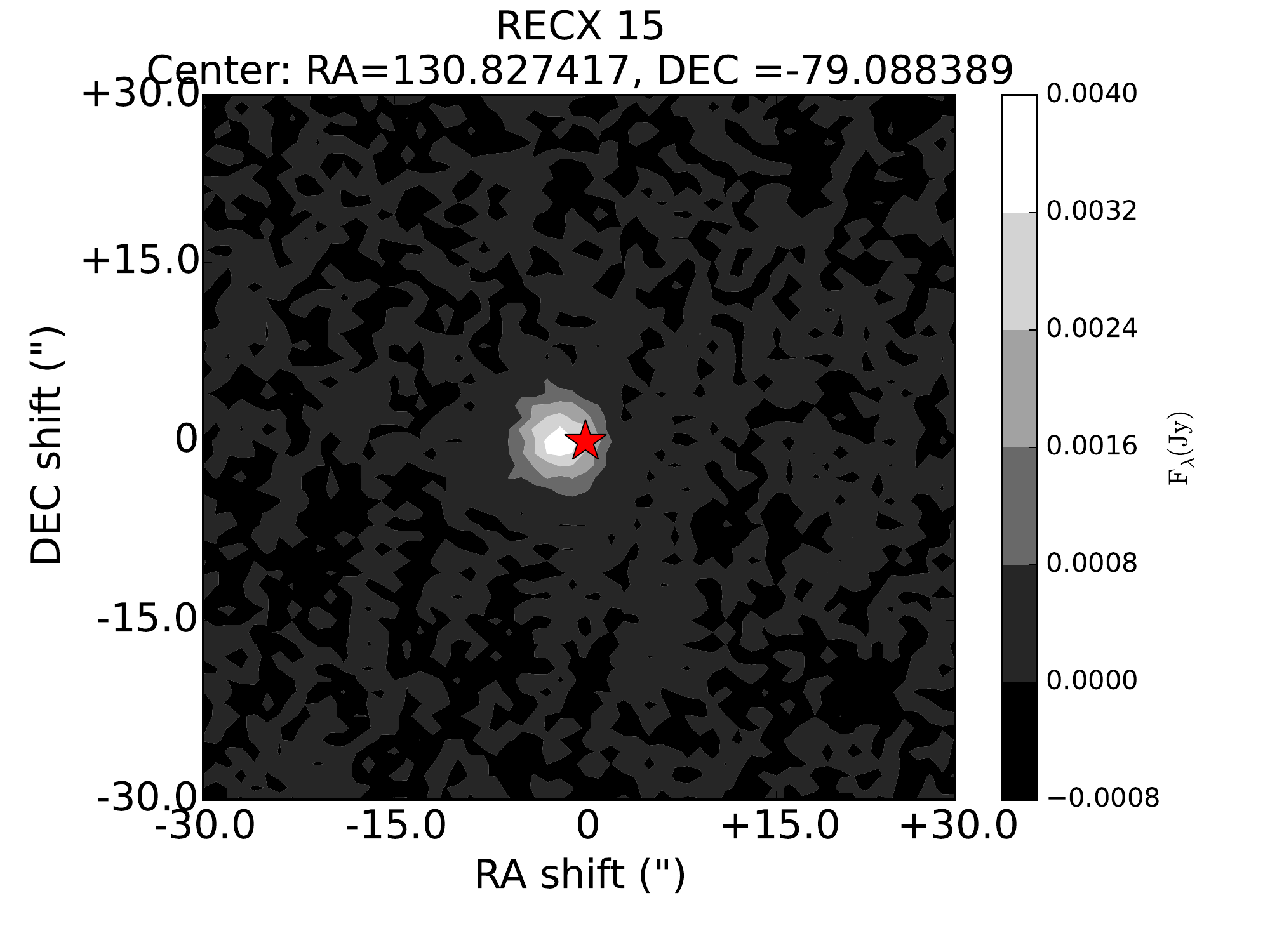}\includegraphics[scale=0.35,trim=0mm 0mm 0mm 0mm,clip]{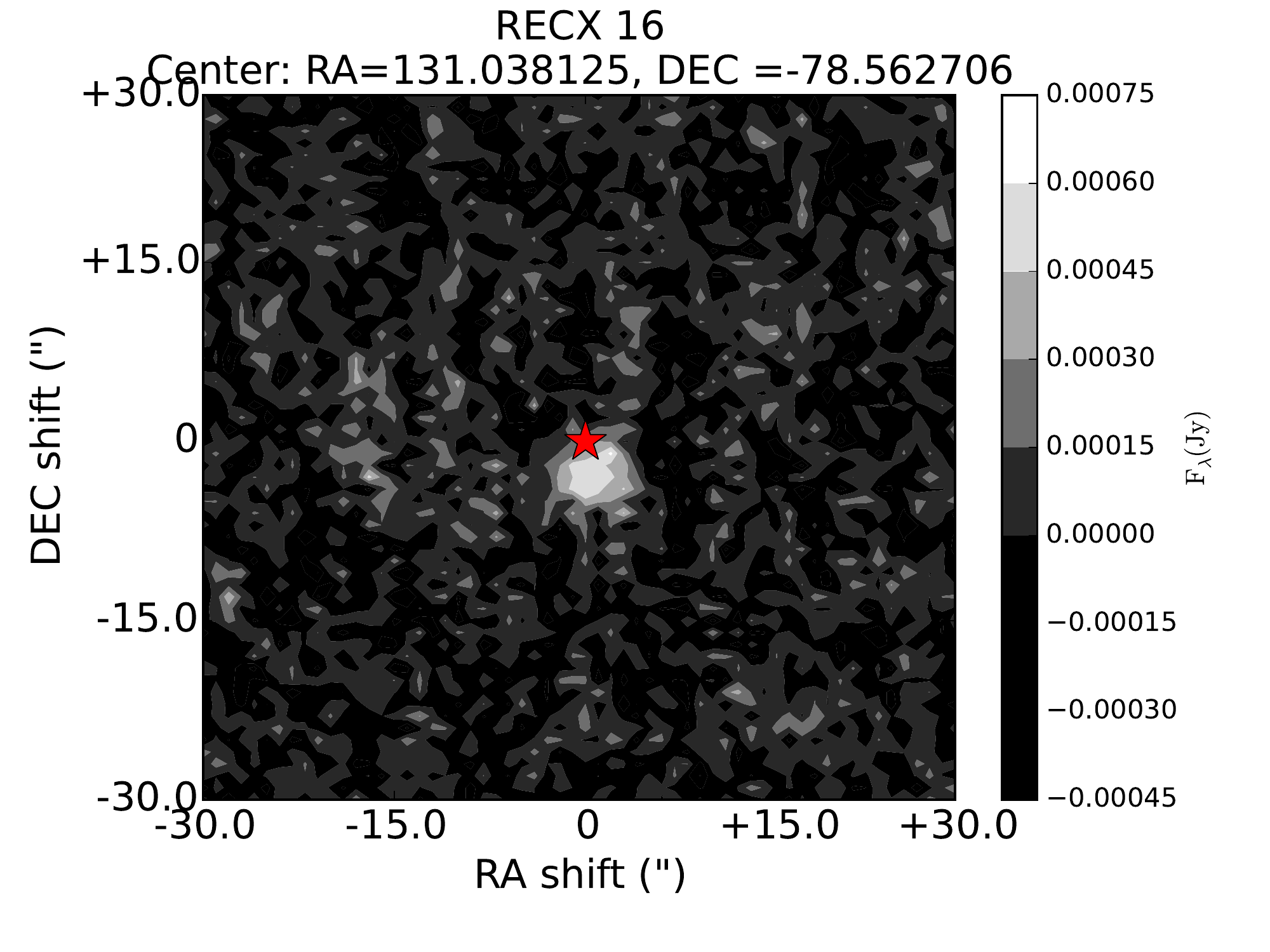}\\
\caption{Contour plots for PACS images at 70 $\rm \mu m$. The red star marks the nominal position of the star.}
   \label{Fig:EtaCha_cont_70}
\end{center}
\end{figure*}

\begin{figure*}[!t]
\begin{center}
\includegraphics[scale=0.35,trim=0mm 0mm 0mm 0mm,clip]{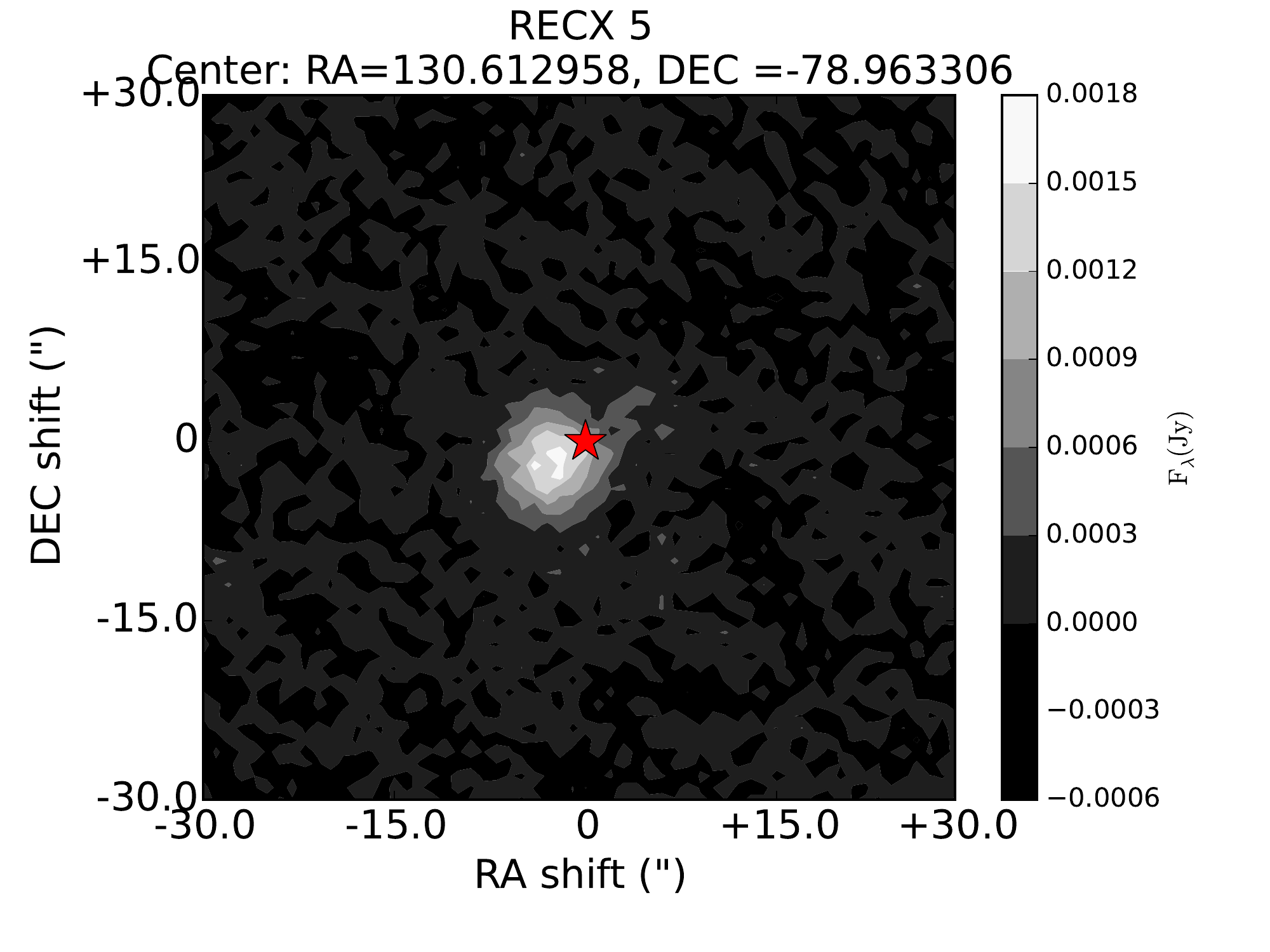}\includegraphics[scale=0.35,trim=0mm 0mm 0mm 0mm,clip]{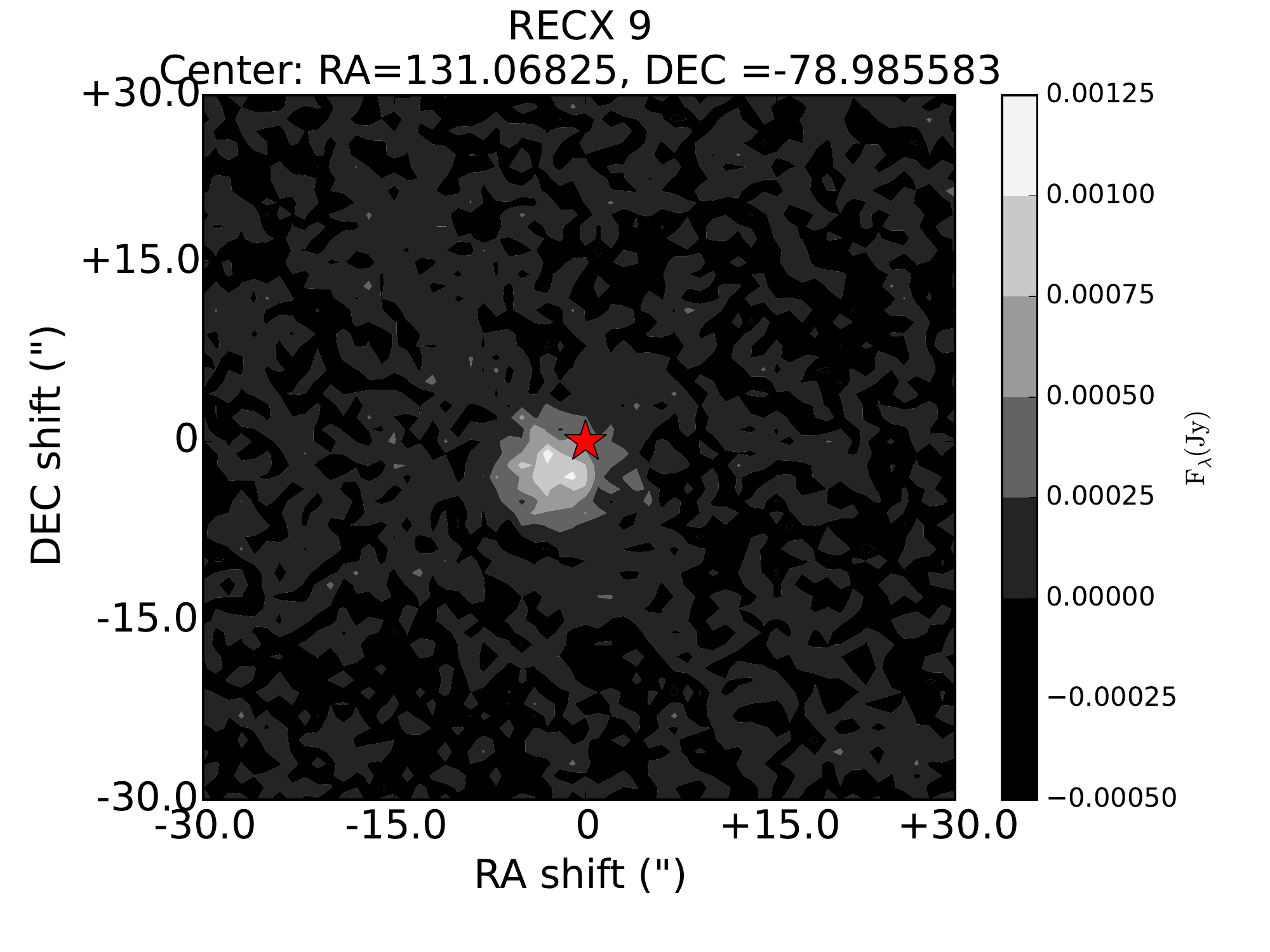}\\
\includegraphics[scale=0.35,trim=0mm 0mm 0mm 0mm,clip]{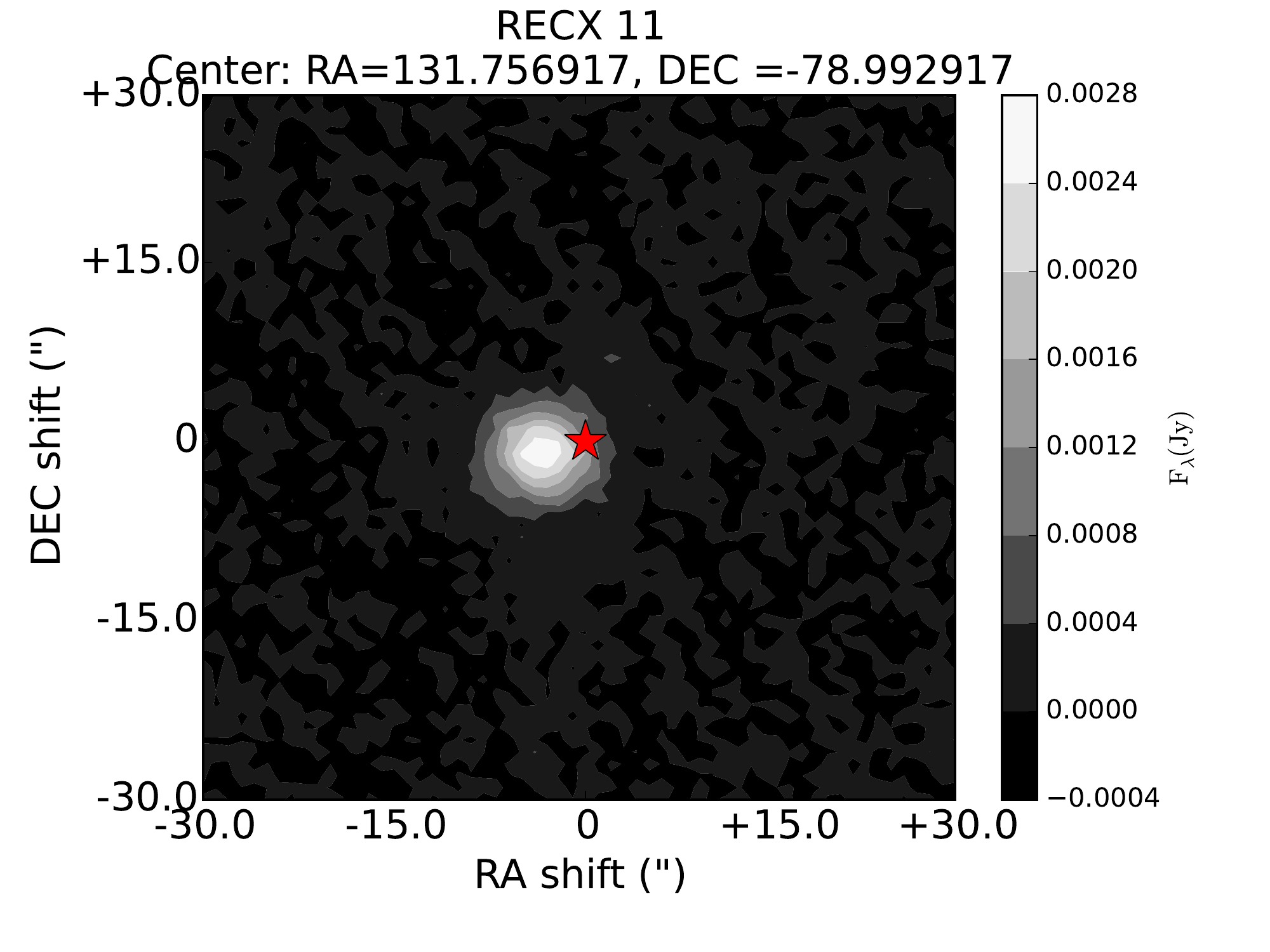}\includegraphics[scale=0.35,trim=0mm 0mm 0mm 0mm,clip]{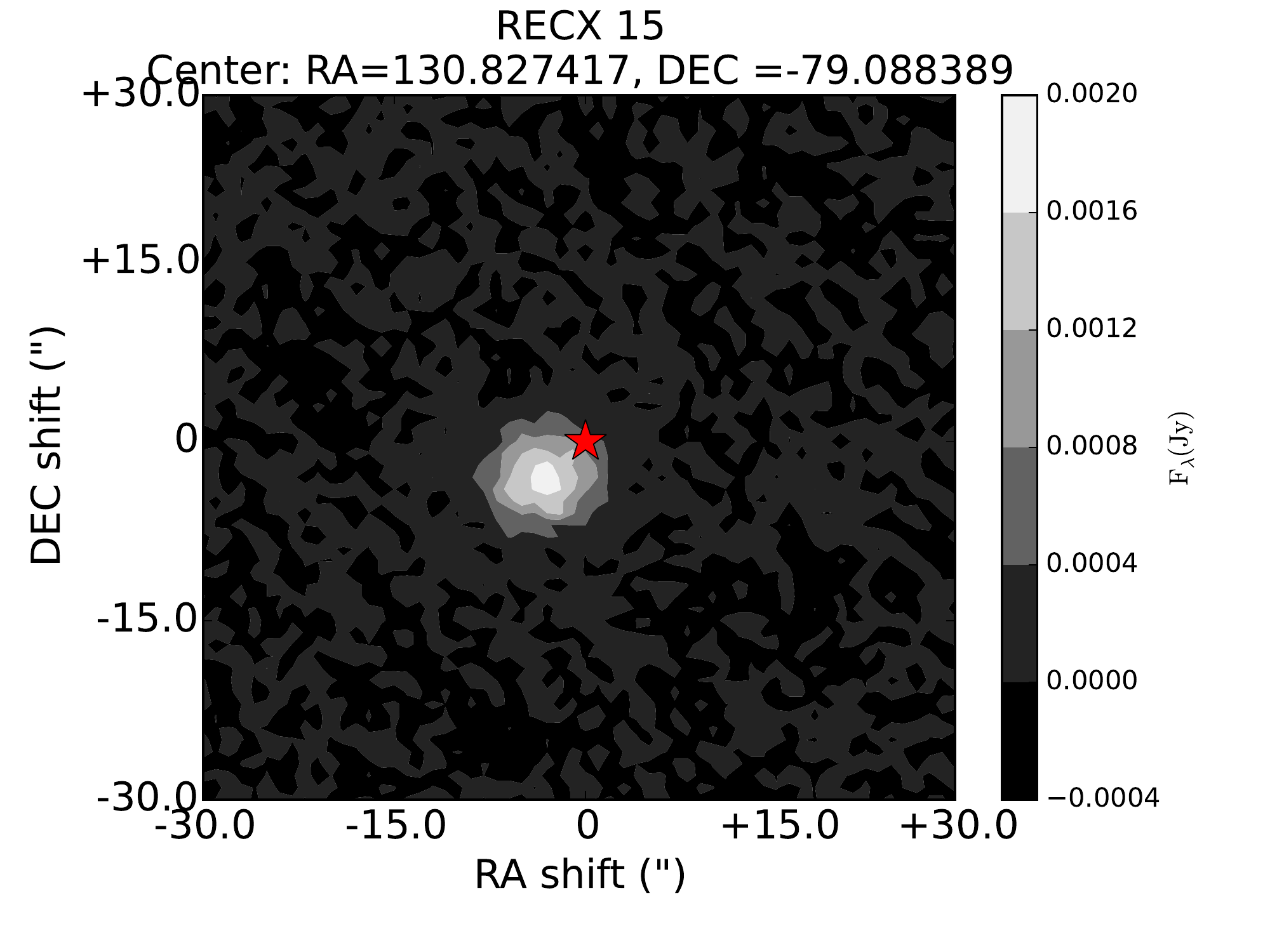}\\
\caption{Contour plots for PACS images at 100 $\rm \mu m$. The red star marks the nominal position of the star.}
   \label{Fig:EtaCha_cont_100}
\end{center}
\end{figure*}

\begin{figure*}[!t]
\begin{center}
\includegraphics[scale=0.35,trim=0mm 0mm 0mm 0mm,clip]{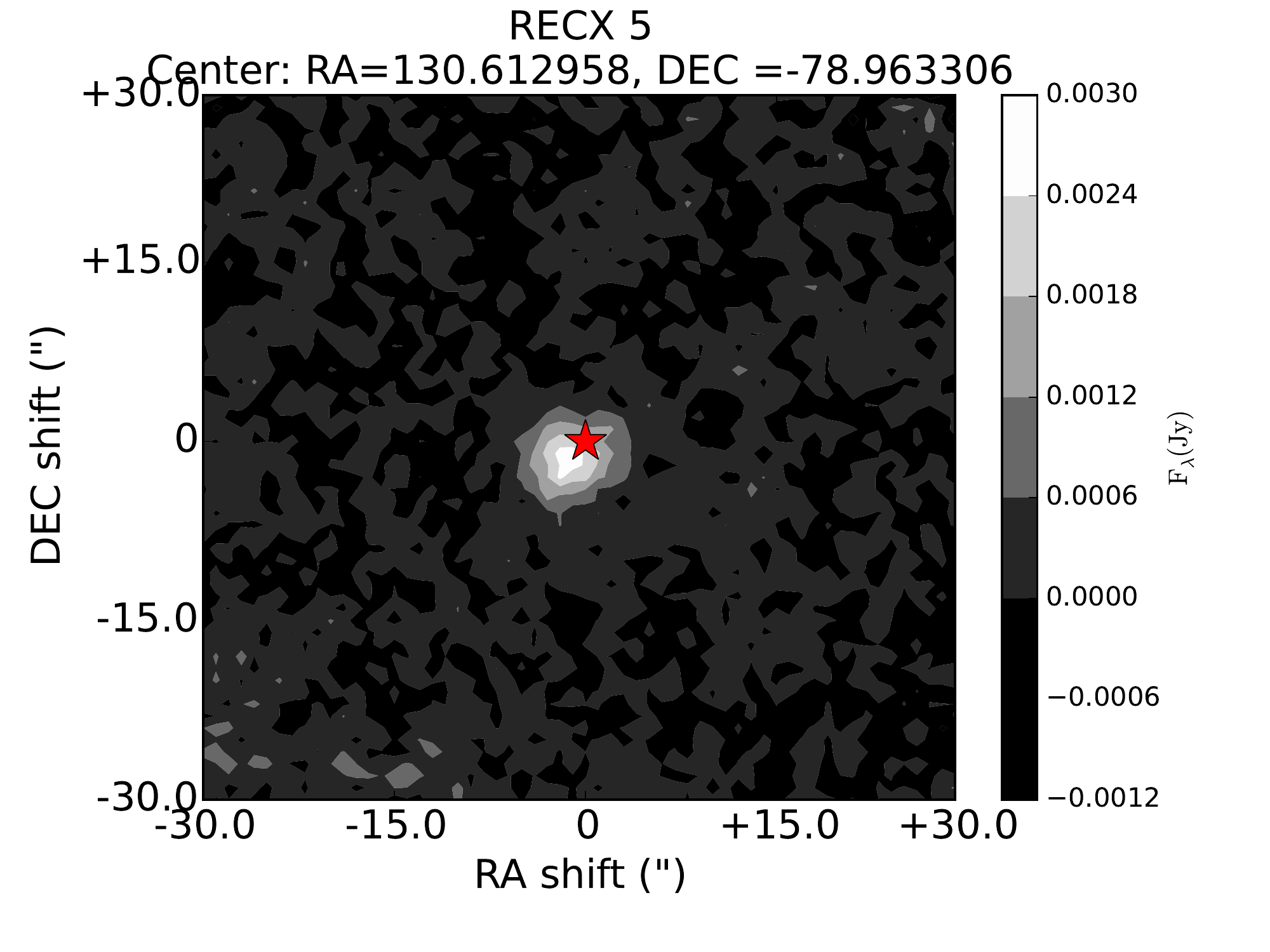}\includegraphics[scale=0.35,trim=0mm 0mm 0mm 0mm,clip]{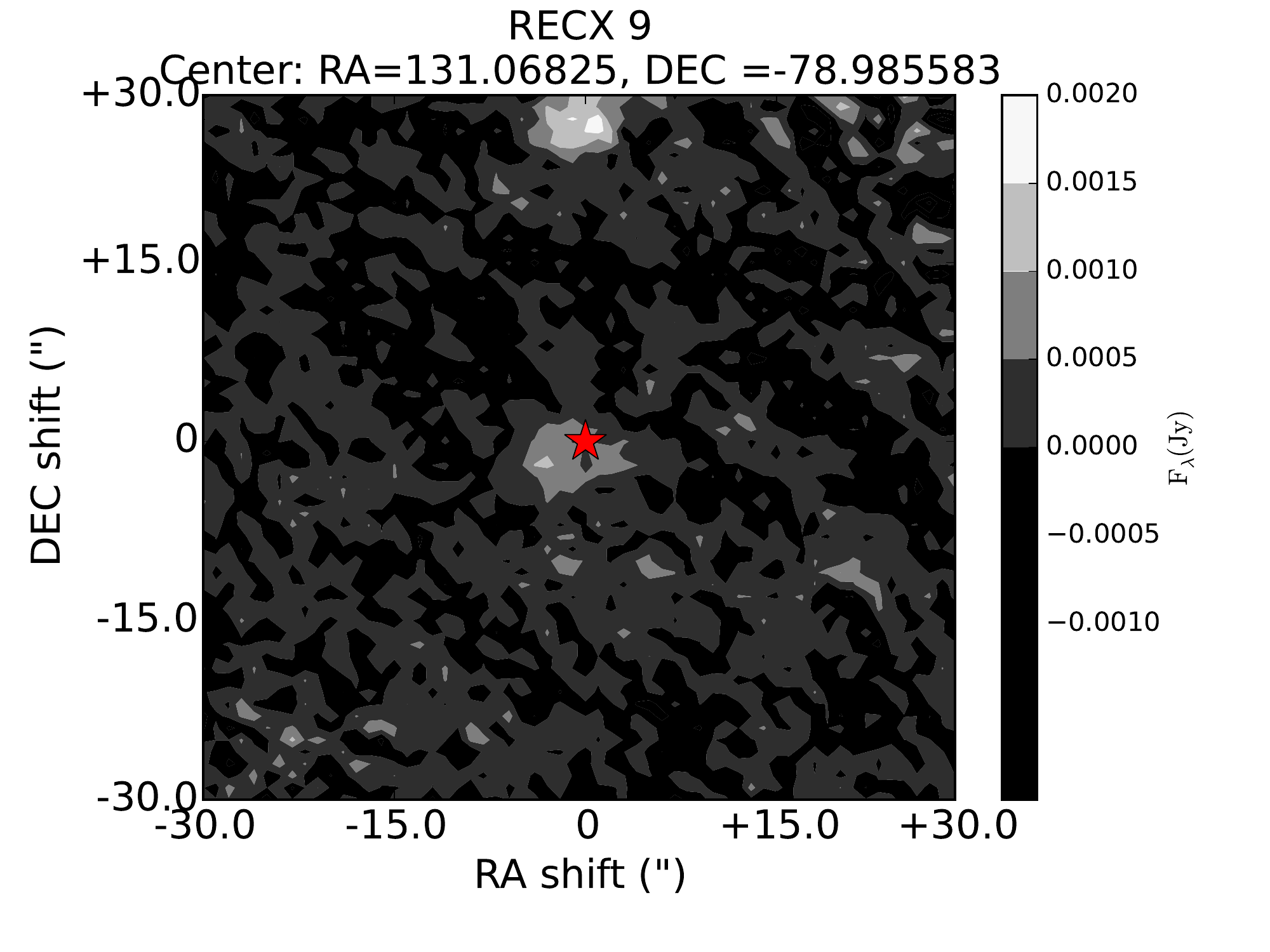}\\
\includegraphics[scale=0.35,trim=0mm 0mm 0mm 0mm,clip]{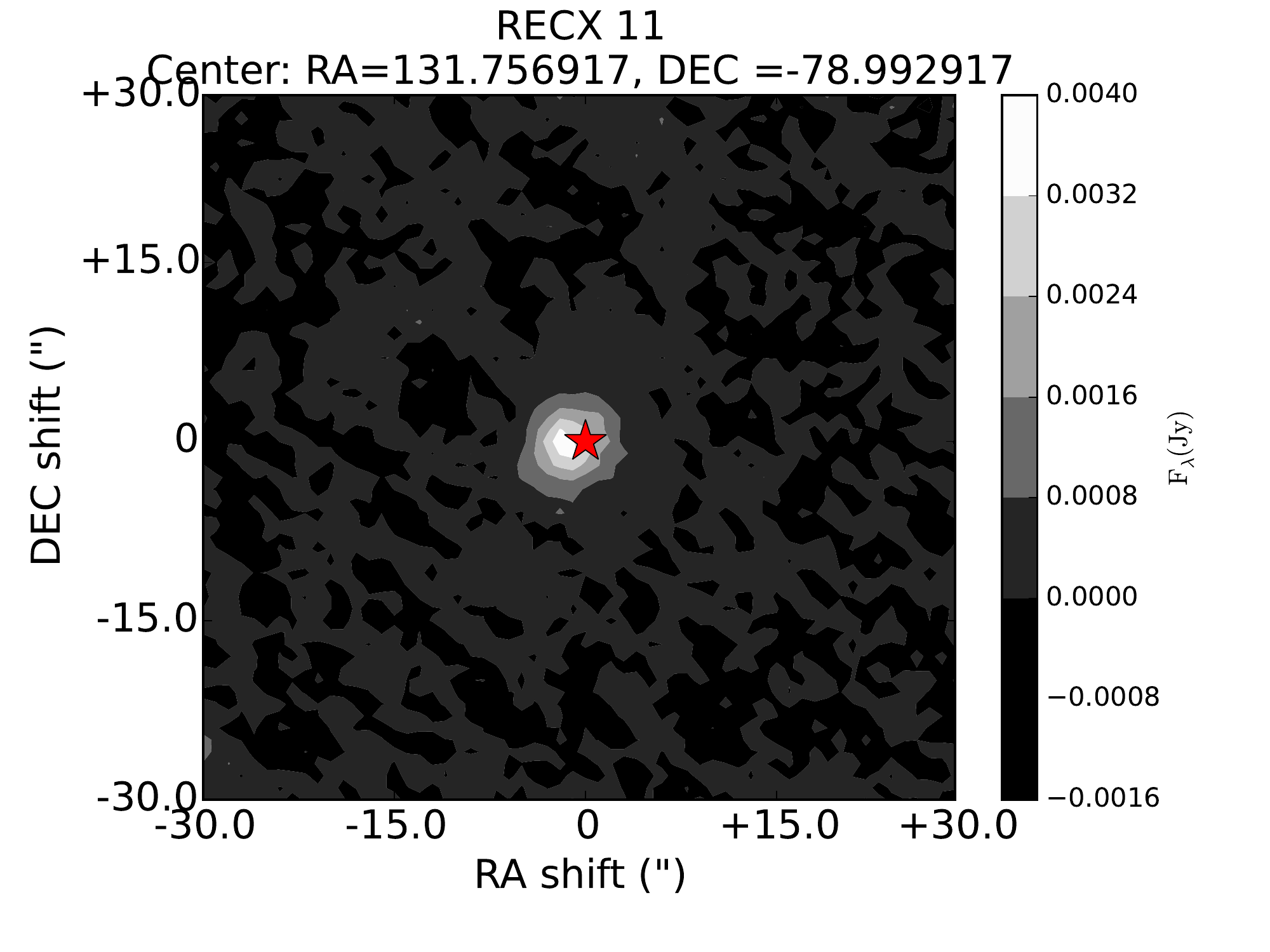}\includegraphics[scale=0.35,trim=0mm 0mm 0mm 0mm,clip]{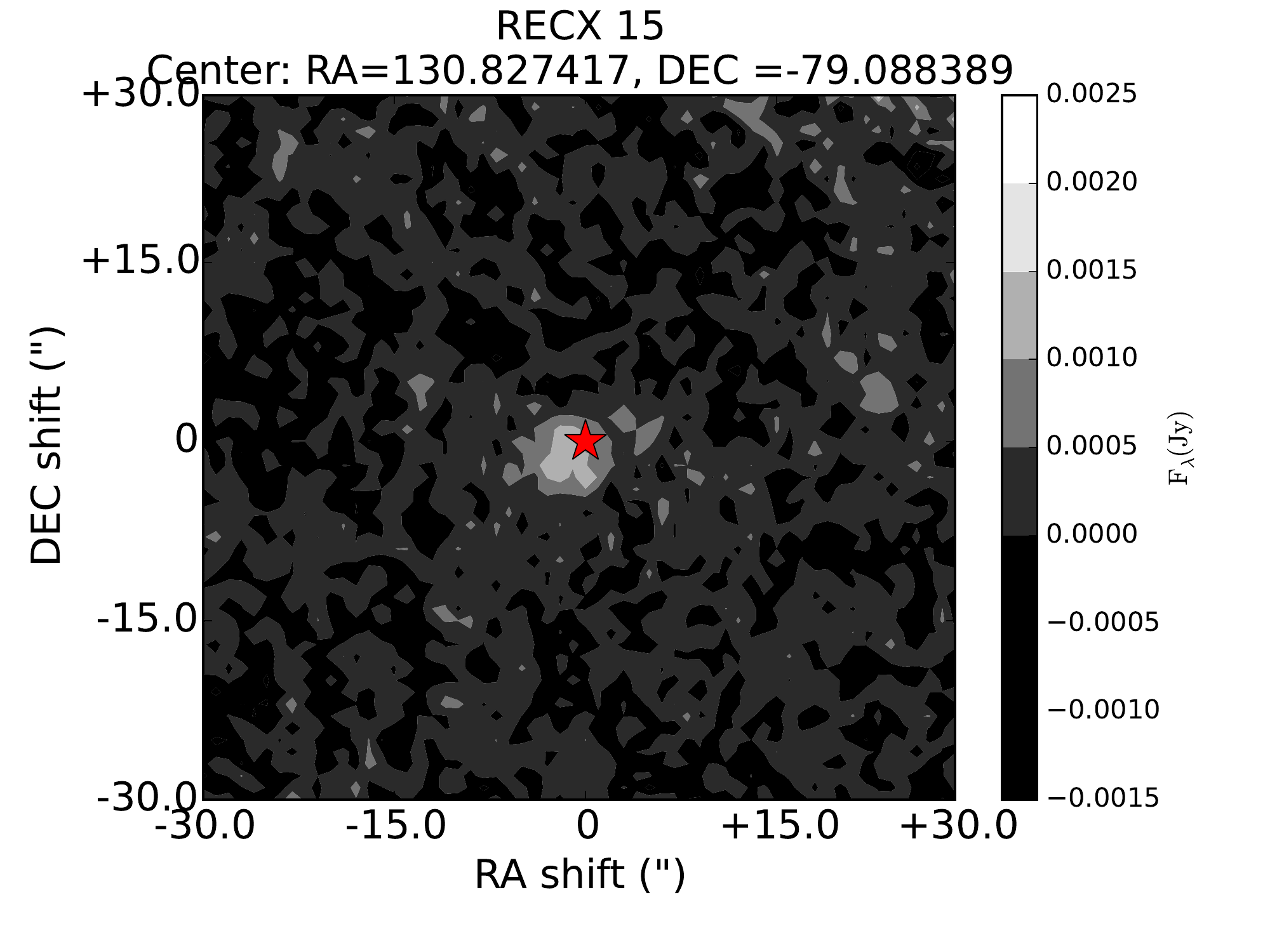}\\
\includegraphics[scale=0.35,trim=0mm 0mm 0mm 0mm,clip]{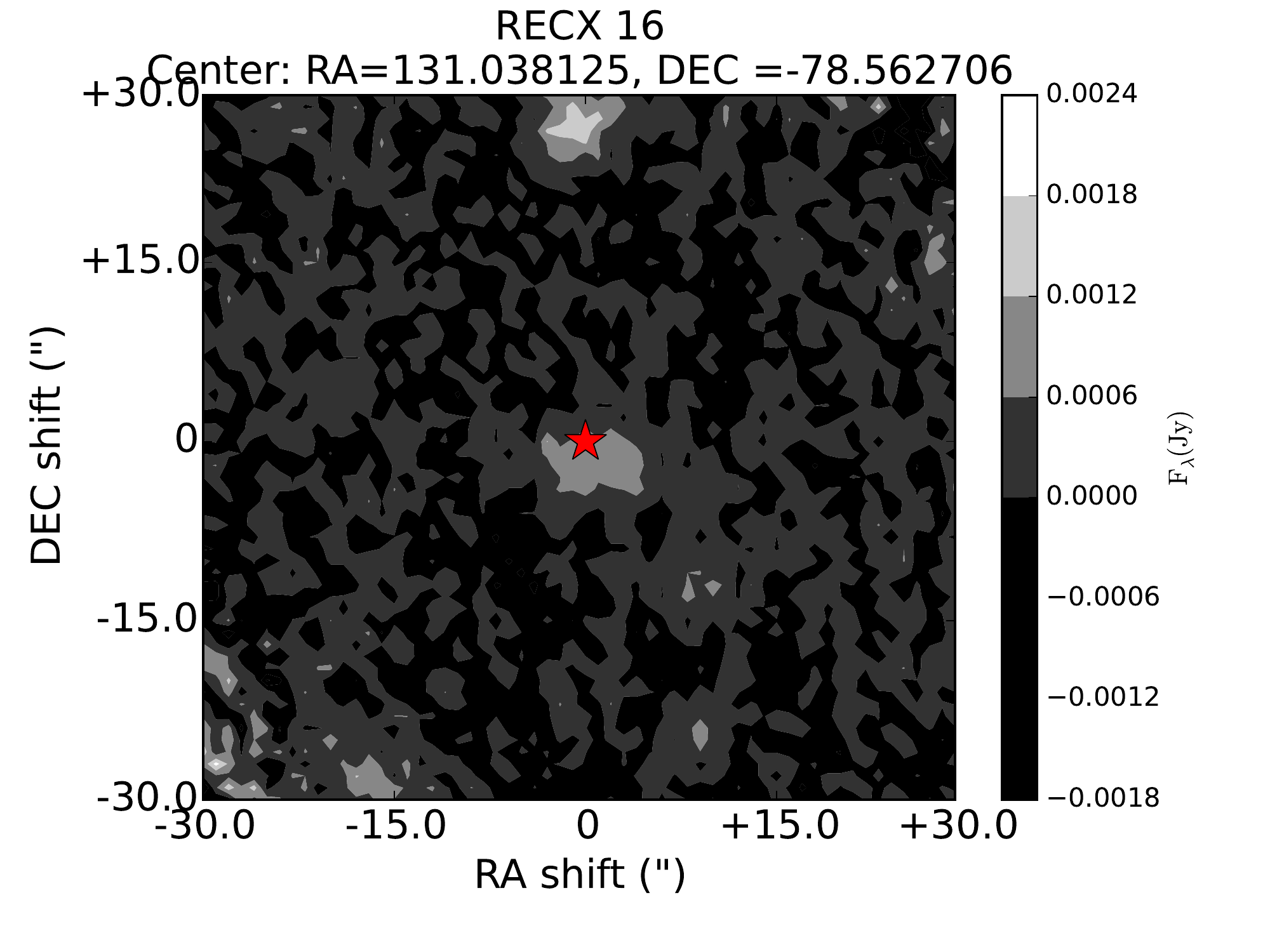}

\caption{Contour plots for PACS images at 160 $\rm \mu m$. The red star marks the nominal position of the star.}
   \label{Fig:EtaCha_cont_160}
\end{center}
\end{figure*}

\section{Sample and  observations}\label{Sec:Sample}
The GASPS program \citep[GAs Survey of Protoplanetary Systems,][]{Dent2013} observed more than 250 stars in seven young stellar associations, with ages in the range 1-- 30 Myr, including 17 young stellar objects belonging to the $\rm \eta$ Cha stellar association. We included all the $\eta$ Cha members with spectral types in the range A1-M5.5. We compiled archival and literature photometry for each $\rm \eta ~Cha$ member in the sample, including 2MASS \citep{Skrutskie2006}, WISE \citep{Wright2010}, AKARI-IRC \citep{Ishihara2010}, optical photometry \citep{Bessel1990,Lawson2001,Lyo2004} and Spitzer photometry \citep{Gautier2008,Megeath2005}.

We obtained PACS photometric observations of 17 sources: 16 objects were observed at 70 $\rm \mu m$,  7 were also observed at 100 $\rm \mu m$ and one source was only observed at 100 $\rm \mu m$ (the PACS observational IDs for each source are given in Table \ref{ObsLog}). Since PACS simultaneously observes at either 70 or 100 $\rm \mu m$ and 160 $\rm \mu m$, we always have observations at 160 $\rm \mu m$. The photometric observations were made in ScanMap mode. For most objects, at least two images were obtained in the blue channel (70 or 100 $\rm \mu m$) with complementary angles (70$\rm ^{\circ}$ and 110$\rm ^{\circ}$) to increase the SNR once combined. Total integration times, after co-adding, range from 133 s to 2464 s at 70 $\rm \mu m$, 552 s to 1122 s at 100 $\rm \mu m$ and 552 s to 2464 s at 160 $\rm \mu m$. The GASPS photometry, together with archival and literature photometry, was used to compile spectral energy distributions (SEDs) for $\rm \eta$ Cha members from optical to far-IR wavelengths. 

We spectroscopically observed all the $\eta$ Cha members known to show IR-excess (seven sources) plus a subsample of six members without a known IR-excess, for a total of thirteen members observed. Spectroscopic observations aimed to detect emission from the [OI] $\rm ^{3}P_{1} \rightarrow$$\rm^{3}P_{2}$ transition at 63.185 $\rm \mu m$ (spectroscopic observational IDs are given in Table \ref{ObsLog}). We also obtained PACS range spectroscopic observations of the three sources with largest disc fractional luminosity, namely RECX 5, RECX 11 and RECX 15, to detect CO, OH, $\rm H_{2}O$ and [CII] emission. They were observed at 72 and 78 $\rm \mu m$ in the blue bands and at 145 and 158 $\rm \mu m$ in the red bands. RECX 15 was also observed at 89 and 190 $\rm \mu m$. 

Aiming to better characterise gas emission in $\rm \eta$ Cha and to look for close, spectroscopic binaries (SB) we queried the ESO archive for high-resolution optical spectroscopic observations, finding publicly available data obtained with VLT/UVES.  Ten out of seventeen objects had previously been observed. We have three epochs of observations for most sources, exception made of RECX 1, which only has two epochs, and RECX 3 and 5, which have four epochs. The time coverage was not homogeneous, covering timelines from one day to one month.

We used the Virtual Observatory SED Analyzer \citep[VOSA][]{Bayo2008,Bayo2014} to enlarge the wavelength range of the SED of our targets. VOSA queries a number of VO-compliant services to look not only for photometry but also other relevant parameters as distances, extinction measurements in the line of sight, etc. Our most complete SEDs cover from  Str\"omgren photometry to far infrared Spitzer/MIPS bands. VOSA offers the possibility to automatically look for infrared excess in the compiled SEDs. The original methodology was to estimate sequentially the infrared slope $\alpha$, with a starting point at 2 $\mu$m, adding redder photometric points, one at a time. The comparison of this iterative calculated $\alpha$ with the classical Lada \& Lada \citep{Lada1987} parametrization allowed to detect the start of the departure from the purely photospheric behaviour. In the newest version of VOSA (Bayo et al. 2014, 2015, subm.), the method for detecting outliers (data-points with reliable quality flags but for some reason a flux value that seem to deviate from the remaining SED) has been refined, imposing that excess in two consecutive bands is detected before flagging those as non-purely photospheric, and the estimation from the deviation from purely photospheric flux is also provided.

We also took advantage of the two statistical approaches VOSA offers to determine stellar parameters. On the one hand it runs through the grids of models available in the VO and provides the one that minimises the residuals (to be used to build an object by object based panchromatic bolometric correction), and on the other hand it proceeds in a Bayesian framework providing the individual Probability Density Functions (PDFs) for each parameter and object. Both approaches agree within the uncertainties/limitations characterising each one, the grid step and the standard deviation of the PDFs, respectively. Complete SEDs were used to compute the stellar parameters, assuming $\rm A_{V}=0$ \citep[see extinctions for RECX 11 and RECX 15 in ][]{McJunkin2014}. The derived parameters  are shown in Table \ref{tableStar}. Spectral types are taken from \cite{ZuckermanSong2004}. Our stellar parameters for K and M stars agree well with those in \cite{Luhman2004b}, with a mean difference in $\rm T_{eff}$ of 120 K. However, early type stars show a more pronounced difference: 1000 K for RECX 8 and 1230 K for RECX 13. The difference might be due to wrong spectral classification or to a poor coverage of the SED peak emission, which is of main importance to fit models to SEDs. However, the effective temperature of the sources does not affect the results from this paper, and a proper discussion of the differences is out of its scope.

\begin{table}[!t]
\caption{Observation log}             
\label{ObsLog}      
\centering          
\begin{tabular}{lll}     
\hline\hline       
Source &  Phot Obs. ID & Spec Obs. ID  \\ 
 (RECX)           &  (1342000000+)         & (1342000000+) \\
\hline
\hline                    
1 & 209480, 209481 & 210391 \\
3 & 88884, 11967, 211968-211970 & 210389 \\
4 & 211963-211966 & 199241 \\
5 & 195468, 211977, 211978 & 210392 \\
6 & 211981, 211982 & 223114 \\
7 & 189366, 211961, 211962, 187338 & -- \\
8 & 189366, 211961, 211962, 187338 & 223113 \\
9 & 211975, 211976 & 223112 \\
10 & 221134, 221135  & 223111\\
11 & 189387, 211973, 211974 & 223115\\
12 & 211979, 211980 & 223110\\
13 & 189367, 211971, 211972 & -- \\
14 & 189368, 1342221278, 1342221278 & 210390\\
15 & 189366, 211961, 211962, 187338 & 210388\\
16 & 221276, 221277 & 210387 \\
17 & 195469, 209484, 209485 & -- \\
18 & 189365, 209482, 209483 & -- \\ 
\hline                  
\end{tabular}
\tablefoot{Values separated by - indicate ranges of Obs. IDs.}
\end{table}

\begin{figure}[!t]
\begin{center}
\includegraphics[scale=0.45,trim=0mm 0mm 0mm 0mm,clip]{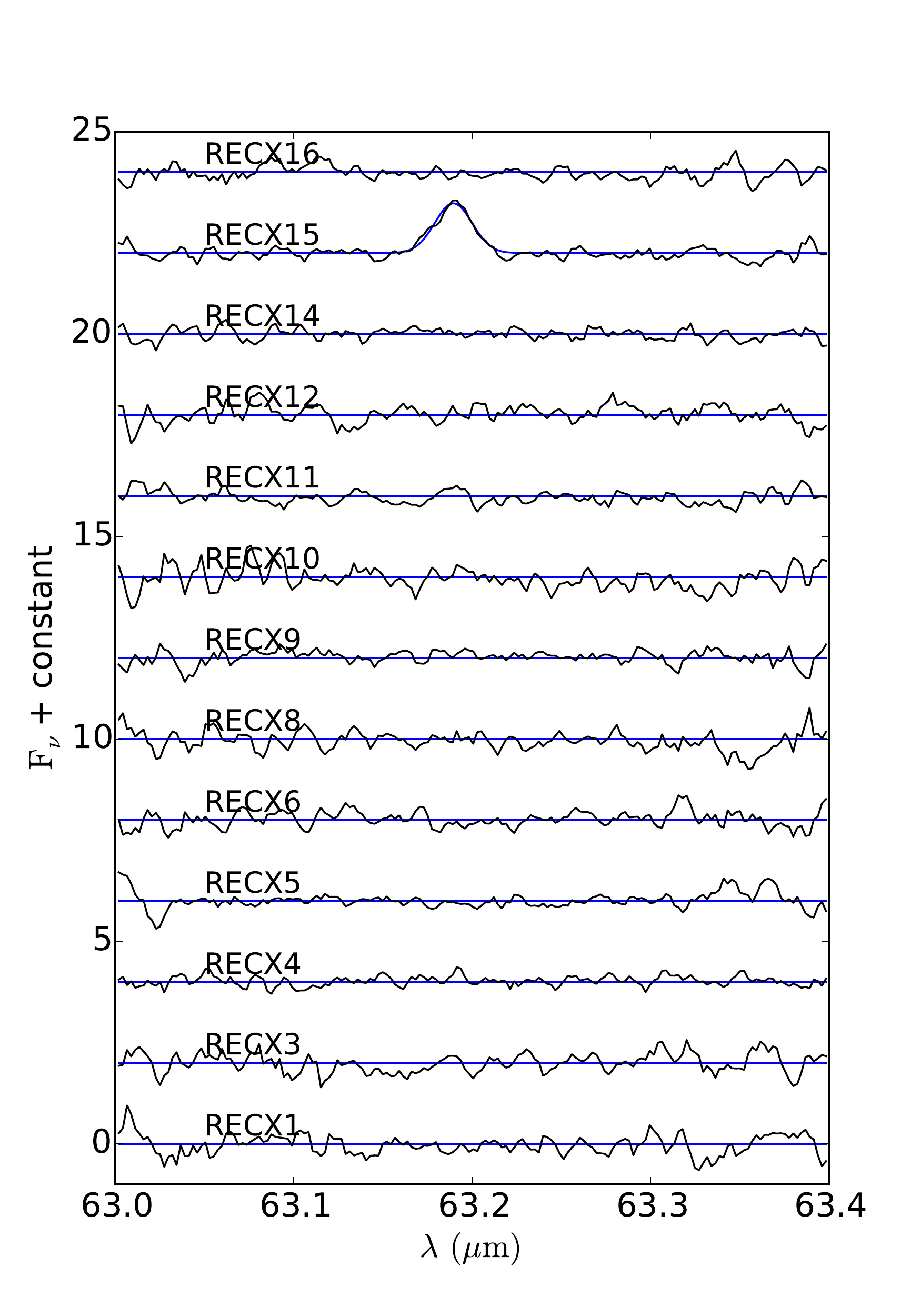}\\
\caption{Continuum subtracted PACS spectra at 63 $\rm \mu m$. The blue line shows a fit to the continuum plus line emission. }
   \label{EtaCha_OI_spec}
\end{center}
\end{figure}

\begin{figure}[!t]
\begin{center}
\includegraphics[scale=0.4,trim=0mm 0mm 0mm 0mm,clip]{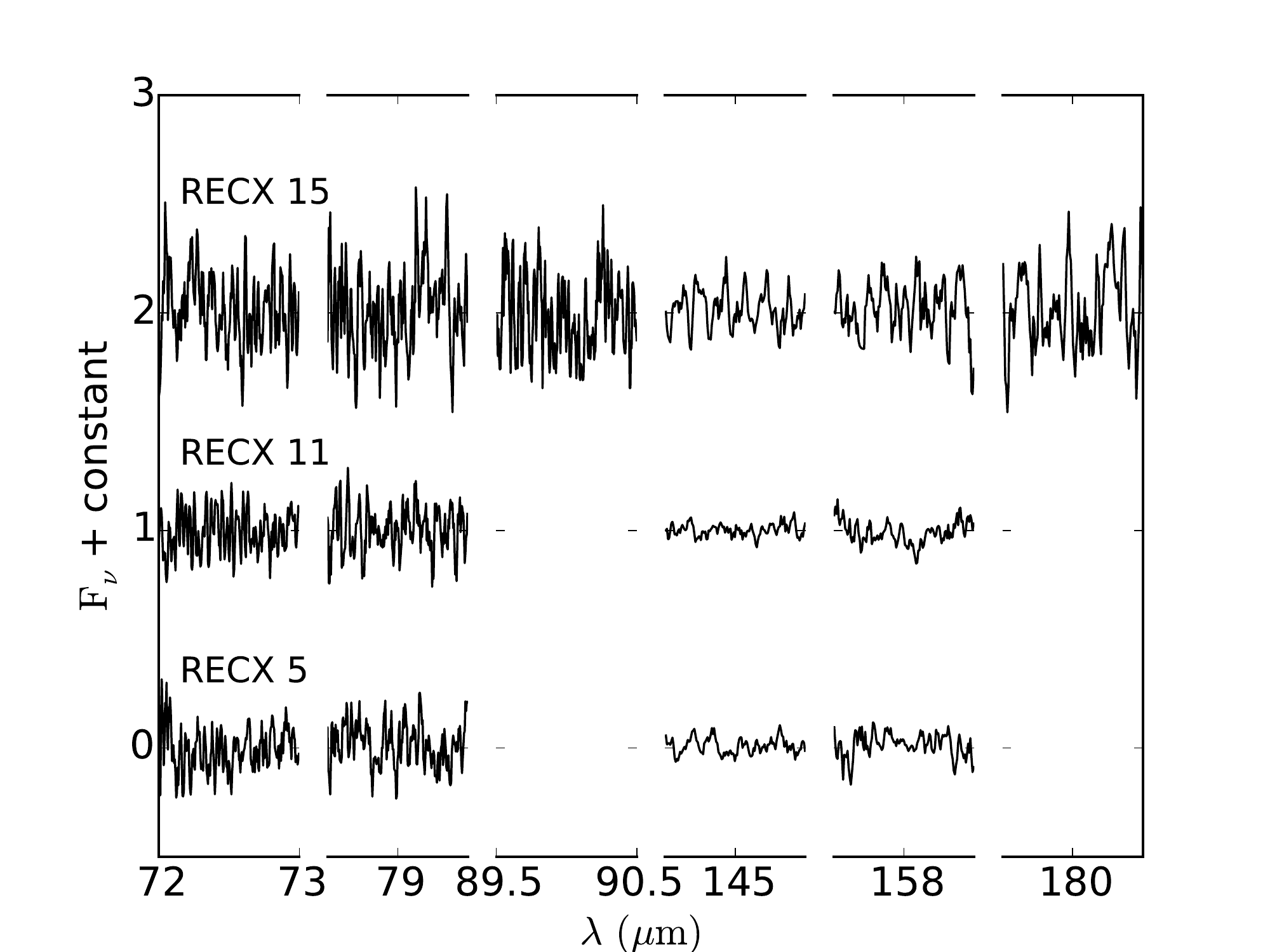}
\caption{From top to bottom: continuum subtracted PACS range spectra for RECX 15, RECX 11 and RECX 5.}
   \label{EtaCha_range_spec}
\end{center}
\end{figure}

\begin{table*}[!t]
\caption{\textit{Herschel}-PACS photometry for $\rm \eta$ Cha members. Upper limits are 3$\sigma$.}             
\label{PACS_phot}      
\centering          
\begin{tabular}{llllc }     
\hline \hline      
Source & $F_{70 \mu m}$  & $F_{100 \mu m}$ &  $F_{160 \mu m}$ & $\rm L_{IR}/L_{*}$ \\ 
\hline
 (RECX) 	& (mJy)  & (mJy)  & (mJy) & -- \\ 
\hline 
1 & $<$ 6 & -- & $<$ 16 & $\rm <  1.8 \times 10^{-4}$  ($\rm <  1.9 \times 10^{-4}$) \\                   
3$\rm ^{*}$ & $<$ 7 ($\rm 6 \pm 1$) & $<$ 10 & $<$ 13 &  $\rm 7.4 \times 10^{-4}$ ($\rm 7.4 \times 10^{-4}$) \\ 
4$\rm ^{*}$ & $<$ 9 ($\rm 9 \pm 1$)  & $<$ 9 & $<$ 14 & $\rm 7.5 \times 10^{-4}$  ($\rm 7.5 \times 10^{-4}$)\\
5 & 113 $\rm \pm$ 24 & 203 $\rm \pm$ 13 & 138 $\rm \pm$ 28 & 0.103 (0.082)\\
6 & $<$ 5 & -- & $<$ 13 & $\rm < 9.7 \times 10^{-4}$  ($\rm < 4.5 \times 10^{-4}$)\\
7 & $<$ 15 & $<$ 11 & $<$ 21 & $\rm < 4.3 \times 10^{-4}$ ($\rm < 3.0 \times 10^{-4}$) \\
8 & $<$ 15 & $<$ 11 & $<$ 21 & $\rm < 2.8 \times 10^{-5}$ ($\rm < 1.2 \times 10^{-5}$)\\
9 & -- &  77 $\rm \pm$ 3 & 53 $\rm \pm$  7 & 0.0158 \\
10 & $<$ 7 & -- & $<$ 14 & $\rm < 7.3 \times 10^{-4}$ ($\rm < 3.4 \times 10^{-4}$) \\
11 & 212 $\rm \pm$ 21 & 205 $\rm \pm$ 13 & 174 $\rm \pm$ 26 & 0.222 (0.269)\\
12 & $<$ 5 & -- & $<$ 13& $\rm < 3.3 \times 10^{-4}$  ($\rm < 1.5 \times 10^{-4}$) \\
13 & $<$ 7 & -- & $<$ 19 & $\rm < 2.6 \times 10^{-5}$ ($\rm < 1.5 \times 10^{-5}$)  \\
14 & $<$ 15 & $<$ 12 & $<$ 25 & $\rm 0.035$ (0.037)\\
15 & 204 $\rm \pm$ 6 & 142 $\rm \pm$ 5 & 71 $\rm \pm$ 7 & 0.399 (0.578) \\
16 & 29 $\rm \pm$ 4 & -- & 59 $\rm \pm$ 8 & 0.068  (0.085) \\
17 & $<$ 8 & -- & $<$ 15 & $\rm < 3.1 \times 10^{-3}$  ($\rm < 2.3 \times 10^{-3}$)\\
18 & $<$ 10 & -- & $<$ 24 & $\rm < 6.3 \times 10^{-3}$ ($\rm < 3.2 \times 10^{-3}$) \\
\hline             
\end{tabular}
\tablefoot{$\rm ^{*}$: we show 70 $\rm \mu m$ PACS fluxes by \cite{Cieza2013} in brackets. For the IR excess, we show in brackets the value computed using photometry with $\rm \lambda \leq 70 \mu m $.}     
\end{table*}

\section{Data reduction}
\subsection{PACS photometric data reduction}
The reduction of PACS photometric data was carried out using HIPE 12.0\footnote{http://herschel.esac.esa.int/hcss-doc-12.0/} (calibration tree version 65) and the high-pass filtering pipeline for small scan maps. The reduction process includes the following main steps: bad and saturated pixel flagging, flat field correction, deglitching and high pass filtering. When several images were available for the same source at the same wavelength, they were combined to increase the signal-to-noise ratio (S/N). The photometric maps were projected onto the final images with a pixel scale of 2 arcsec/pixel for the 70 and 100 $\rm \mu m$ bands and of 3 arcsec/pixel for the 160 $\rm \mu m$ band. The maps were also projected onto images with the native pixel size (3.2, 3.2 and 6.4 arcsec/pixel for the 70, 100 and 160 $\rm \mu m$ bands, respectively) to minimise the impact of correlated noise in photometric uncertainties. The final maps are shown in Fig. \ref{Fig:EtaCha_cont_70}, \ref{Fig:EtaCha_cont_100} and \ref{Fig:EtaCha_cont_160} with a pixel scale of 1\arcsec/pixel.
 
 Aperture photometry was computed using apertures of 6, 6 and 12\arcsec  for the 70, 100 and 160 $\rm \mu m$ bands, respectively. The sky annulus was placed at 25\arcsec  from the source, with a width to 10\arcsec. To retrieve the final fluxes we applied aperture corrections from the proper calibration files (version 65).  
 
Photometric errors consist of the quadratic sum of the noise errors and the calibration errors. Noise errors were computed as the standard deviation of the sky scaled to the size of the aperture radius. The resulting uncertainties were scaled according to the correlated noise correction factor ($cnc$)

\begin{equation}
cnc = a\times(pxs/pxs_{0})^{b}
\end{equation}
where $pxs$ is the pixel size of the processed image, $pxs_{0}$ is the native pixel size (3.2 for the 70 and 100 $\rm \mu m$ bands and 6.4 for the 160 $\rm \mu m$ band), and $a$ and $b$ are numerical factors that depend on the band selected, being $a$=0.95, $b$=1.68  for the 70 and 100 $\rm \mu m$ bands and $a$=0.88, $b$=1.73 for the 160 $\rm \mu m$ band \citep{Balog2014}. Calibration errors are 2.64\%, 2.75\% and 4.15\% for the 70, 100 and 160 $\rm \mu m$ bands, respectively\footnote{http://herschel.esac.esa.int/twiki/bin/view/Public/PacsCalibrationWeb}. In a similar way, upper limits on flux densities for non-detected sources were computed as the standard deviation of the sky annulus centreed at the nominal position of the star. Upper limits were then scaled to the number of pixels inside the aperture and aperture corrected. Photometric  fluxes and uncertainties, together with 3$\sigma$ upper limits for non-detected sources are shown in Table \ref{PACS_phot}.

\subsection{PACS spectroscopic data reduction}
The reduction of PACS spectroscopic data was performed using HIPE 12.0 following the standard procedure for PACS spectroscopic data reduction. A major step when extracting line fluxes from the PACS IFU is to detect where the source is centreed. To that aim, the flux in the continuum in each spaxel must be evaluated, and the flux distribution compared to a theoretical one from model PSFs. For $\rm \eta$ Cha members, the final spectra were extracted from the central spaxel, as all the observations were properly centreed on the source of interest and aperture corrected to account for flux losses. Due to the high noise in the edge of the spectra, we only analyse data in the range $\rm 63.0 < \lambda/\mu m <63.4 $ for line observations. The limits for range observations were: $\rm 72.0 < \lambda/\mu m <73.0 $, $\rm 78.5 < \lambda/\mu m <79.5 $, $\rm 89.5 < \lambda/\mu m <90.5 $, $\rm 144.0 < \lambda/\mu m < 146.0$, $\rm 157.0 < \lambda/\mu m < 159.0$ and $\rm 179.0 < \lambda/\mu m < 181.0$.

Since the spectral ranges are short, we used first order polynomials to fit the continuum in regions where no line emission is expected and performed Gaussian fits to continuum subtracted spectra. The line fluxes were computed as the integral of a Gaussian with the parameters coming from the fit. The uncertainties in line fluxes were computed as the integral of a Gaussian with a peak equal to the noise level, and width equal to that of the fit. When no line emission is detected ($\rm S/N<3$), we compute upper limits in the same way as we compute errors, but with a FWHM equal to the instrumental value. The spectra for $\rm \eta$ Cha sources observed with PACS at 63 $\rm \mu m$ are shown in Fig. \ref{EtaCha_OI_spec}. The spectra for sources observed in the range 72-180 $\rm \mu m$ are shown in Fig. \ref{EtaCha_range_spec}. Line fluxes and 3$\sigma$ upper limits for line and range spectroscopy are shown in Tables \ref{tabSpec} and \ref{tabSpecRange} respectively.

\subsection{UVES spectroscopic data reduction}\label{Subsec:UVES_reduction}

\begin{figure}[!t]
\begin{center}
\includegraphics[scale=0.4,trim=0mm 0mm 0mm 0mm,clip]{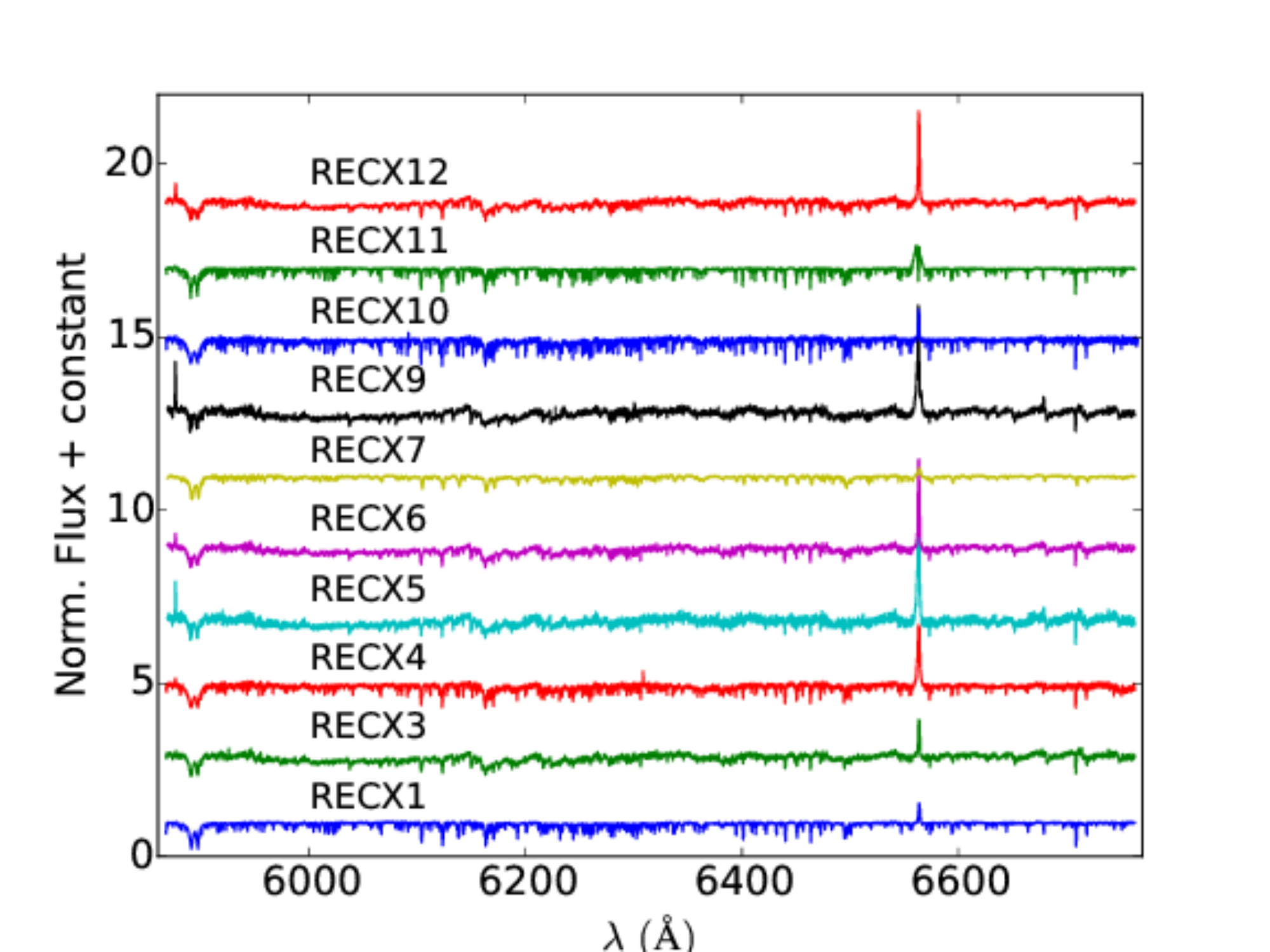}\\
\caption{UVES spectra of $\rm \eta$ Cha members.}
   \label{Fig:EtaCha_UVES}
\end{center}
\end{figure}
UVES performs at a resolution 40,000 with 1\arcsec slit. Observations of ten $\eta$ Cha members were retrieved from the archive. These observations were conducted using the standard setup, a 1\arcsec slit width which covers the wavelength range 3250-6800 $\rm \AA$.  All data were reduced using the UVES pipeline recipe uves\_obs\_redchain with the command-line driven utility esorex (bias corrected, dark current corrected, flat-fielded, wavelength-calibrated and extracted). We show in Fig. \ref{Fig:EtaCha_UVES} the average spectrum from all epochs for each $\rm \eta $ Cha member observed with UVES.

Radial velocity (RV) values were computed by means of computed cross-correlation functions (CCFs) for all reduced spectra. To compute the CCF, the observed spectrum is convolved with a CORAVEL-type numerical mask, as described in \cite{Queloz1995}, K-type or M-type, depending on the spectral type of the star. The shape of the CCF function is approximated by a Gaussian profile.  The RV is the peak of this profile to which the barycentric correction is applied.  The average measured values can be seen in Table \ref{Tab_radVel}. 

\begin{figure}[!t]
\centering
\caption{Top: contours for line (coloured contours) and continuum at 63 $\rm \mu m$ (solid and dashed line contour, where dashed lines represent negative values) emission in RECX 15. Continuum units are in Jy. Bottom: line residual emission for RECX 15 at 63 $\rm \mu m$ (see Sec. \ref{Sec:specResults}) from the tests by \cite{Podio2012}. Continuum contours are also displayed, following the top panel. Red plus signs mark the central positions of the PACS spectrometer spaxels. No significative residual emission is detected.}
\label{RECX15_contours}
\end{figure}

\begin{table}[!h]
\caption{Line fluxes and 3$\sigma$ upper limits of [OI] and o-$\rm H_{2}O$ at 63 $\rm \mu m$}             
\label{tabSpec}      
\centering          
\begin{tabular}{lll}     
\hline \hline      
Source & $\rm F_{[OI]}~(63.18 ~\mu m)$ & $\rm F_{o-H_{2}O}~(63.32 ~\mu m)$ \\ 
\hline
	& $\rm(10^{-18}~W/m^{2})$ & $\rm(10^{-18}~W/m^{2})$ \\ 
\hline 
RECX 1 & $\rm <$11 & $\rm <$11 \\                    
RECX 3 & $\rm <$10 & $\rm <$10 \\ 
RECX 4 & $\rm <$5.0 & $\rm <$5.0 \\ 
RECX 5 &$\rm <$9.2 & $\rm <$9.2 \\ 
RECX 6 & $\rm <$9.0 & $\rm <$9.0 \\ 
RECX 8 & $\rm <$9.8 & $\rm <$9.8 \\
RECX 9 & $\rm <$7.6 & $\rm <$7.6 \\
RECX 10 & $\rm <$13 & $\rm <$13 \\
RECX 11 &  $\rm <$6.6 &  $\rm <$6.6 \\
RECX 12 & $\rm <$11 & $\rm <$11 \\
RECX 14 & $\rm < 6.4$ & $\rm < 6.4$\\
RECX 15 & 24.0$\rm \pm$2.5 & $\rm < 5.6$ \\
RECX 16 & $\rm < 7.3 $  & $\rm < 7.3 $ \\
\hline                  
\end{tabular}
\end{table}

\begin{table}[!t]
\caption{Three-$\sigma$ upper limits from range spectroscopic observations in the 72-180 $\rm \mu m$ range}             
\label{tabSpecRange}      
\centering          
\begin{tabular}{llll}     
\hline \hline      
Species & \multicolumn{3}{c}{Line flux ($\rm W/m^{2}$)} \\
\hline
              & RECX5 & RECX 11 & RECX 15 \\
\hline          
$\rm o-H_{2}O$ \\
71.94 $\rm \mu m$ & $\rm <3.9$ &  $\rm <4.0$  &  $\rm <6.8$ \\ 
78.74 $\rm \mu m$ & $\rm <6.1$ &  $\rm <7.6$  &  $\rm <10.0$ \\ 
179.53 $\rm \mu m$ & -- &  --  &  $\rm <6.7$ \\ 
180.49 $\rm \mu m$ & -- &  --  &  $\rm <6.7$ \\ 
\hline
$\rm p-H_{2}O$ \\
89.99 $\rm \mu m$ & -- &   -- &  $\rm <9.5$ \\ 
144.52 $\rm \mu m$ & $\rm <2.5$ &  $\rm <1.6$  &  $\rm <4.4$ \\  
158.31 $\rm \mu m$ & $\rm <3.3$ &  $\rm <3.6$  &  $\rm <6.9$ \\
\hline
CO \\
72.84 $\rm \mu m$ & $\rm <3.9$ &  $\rm <4.0$  &  $\rm <6.8$ \\ 
79.36 $\rm \mu m$ & $\rm <6.1$ &  $\rm <7.6$  &  $\rm <10.0$ \\ 
90.16 $\rm \mu m$ & -- &   -- &  $\rm <9.5$ \\ 
144.78 $\rm \mu m$ & $\rm <2.5$ &  $\rm <1.6$  &  $\rm <4.4$ \\ 
\hline
OH \\
79.11 $\rm \mu m$ & $\rm <6.1$ &  $\rm <7.6$  &  $\rm <10.0$ \\ 
79.16 $\rm \mu m$ & $\rm <6.1$ &  $\rm <7.6$  &  $\rm <10.0$ \\ 
\hline
$\rm CH^{+}$ \\
72.14 $\rm \mu m$ & $\rm <3.9$ &  $\rm <4.0$  &  $\rm <6.8$ \\ 
90.02 $\rm \mu m$ & -- &   -- &  $\rm <9.5$ \\ 
179.61 $\rm \mu m$ & -- &  --  &  $\rm <6.7$ \\ 
\hline
[OI]
145.52  $\rm \mu m$ & $\rm <2.5$ &  $\rm <1.6$  &  $\rm <4.4$ \\  
\hline
[CII]
157.74 $\rm \mu m$ & $\rm <3.3$ &  $\rm <3.6$  &  $\rm <6.9$ \\
\hline
\end{tabular}
\end{table}

\begin{table}[!t]
\caption{Radial velocities from UVES high resolution spectra}             
\label{Tab_radVel}      
\centering          
\begin{tabular}{llll}     
\hline\hline       
Source &  	$\rm \langle V_{r} \rangle$ & $\rm \sigma_{V_{r}}$ & epochs	\\
          &  (km/s) & (km/s) & \\
\hline
\hline    
RECX 1  &     18.3	   &   1.0	   &         2	 \\ 
RECX 3  &     17.5	   &   0.5	   &         4	 \\ 	
RECX 4  &     17.8	   &   0.5	   &         3	 \\ 	
RECX 5  &     17.2	   &   0.6	   &         4	 \\ 	
RECX 6  &     18.1	   &   1.5	   &         3	 \\ 	
RECX 7  &     43.9	   &   22.6	   &   	   3	 \\ 
RECX 9  &     18.5	   &   0.7	   &         3 	 \\
RECX 10 &     17.5	   &   0.2	   &         3 	 \\	
RECX 11 &     17.5	   &   0.8	   &         3	 \\ 	
RECX 12 &     23.3	   &   0.9	   &         3	 \\ 	
\hline                  
\end{tabular}
\end{table}

\section{Results}\label{Sec:Results}

\begin{figure*}[!t]
\begin{center}
\includegraphics[scale=0.24,trim=8mm 30mm 0mm 0mm,clip]{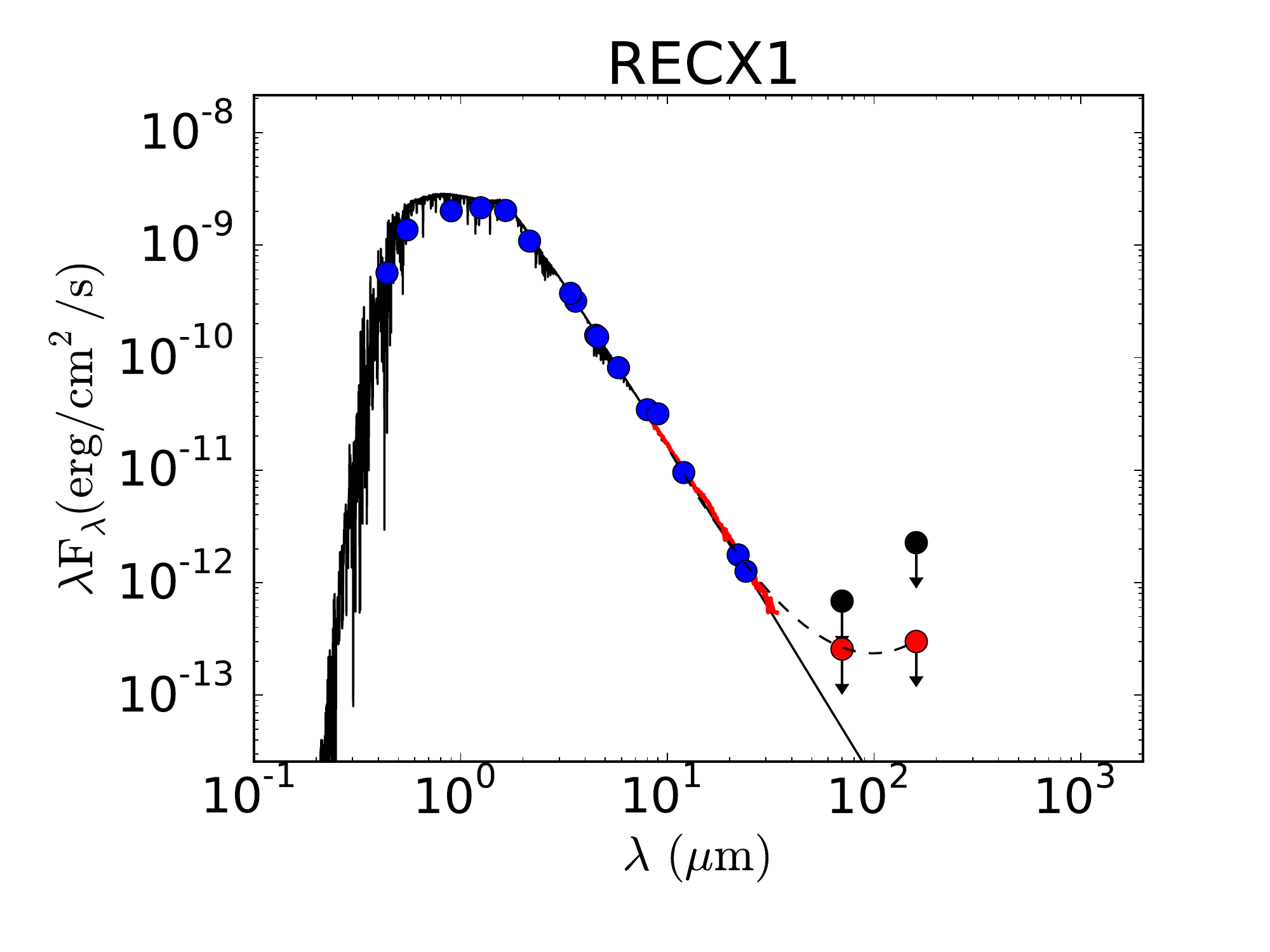}
\includegraphics[scale=0.24,trim=20mm 30mm 0mm 0mm,clip]{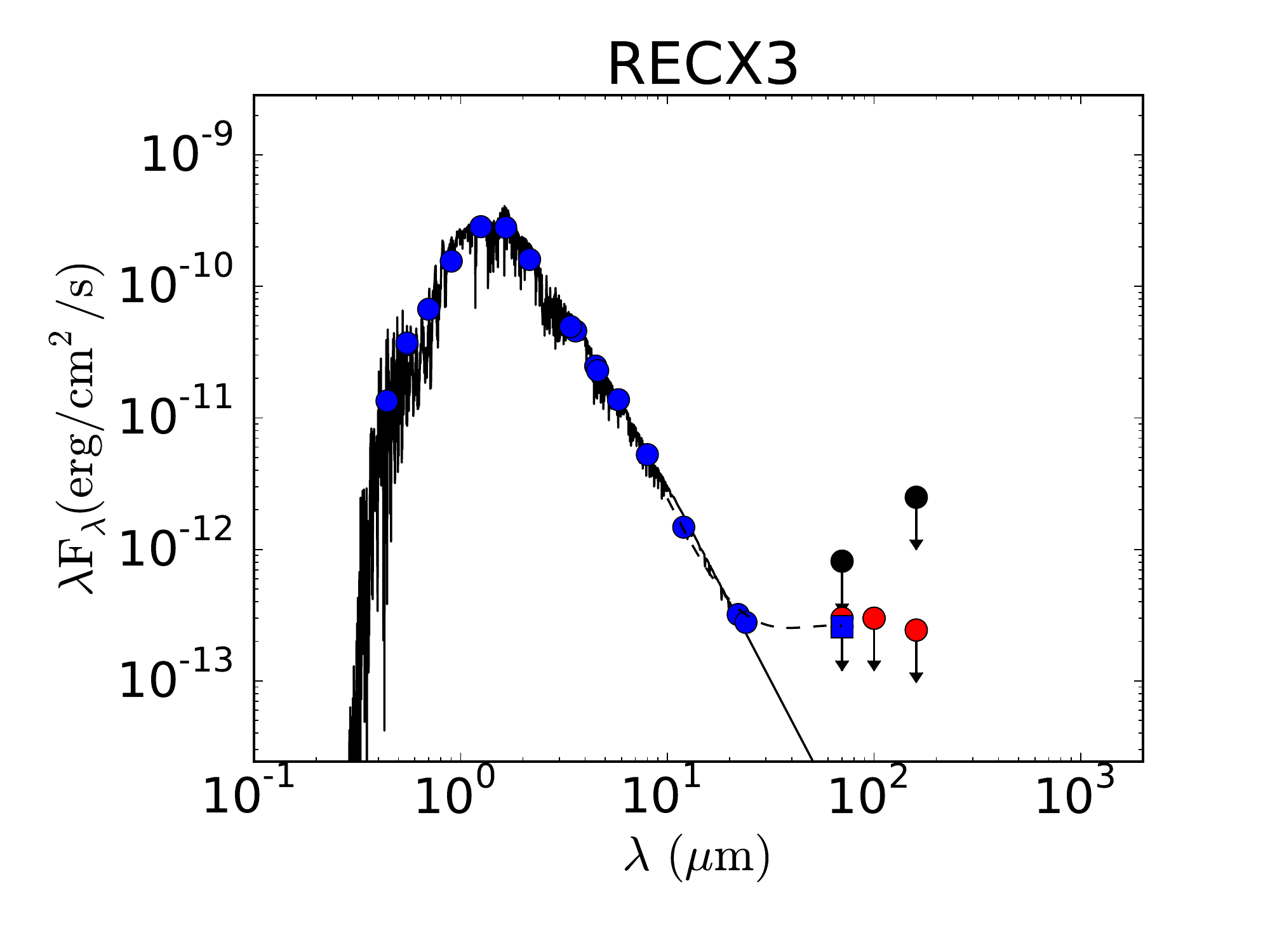}
\includegraphics[scale=0.24,trim=20mm 30mm 0mm 0mm,clip]{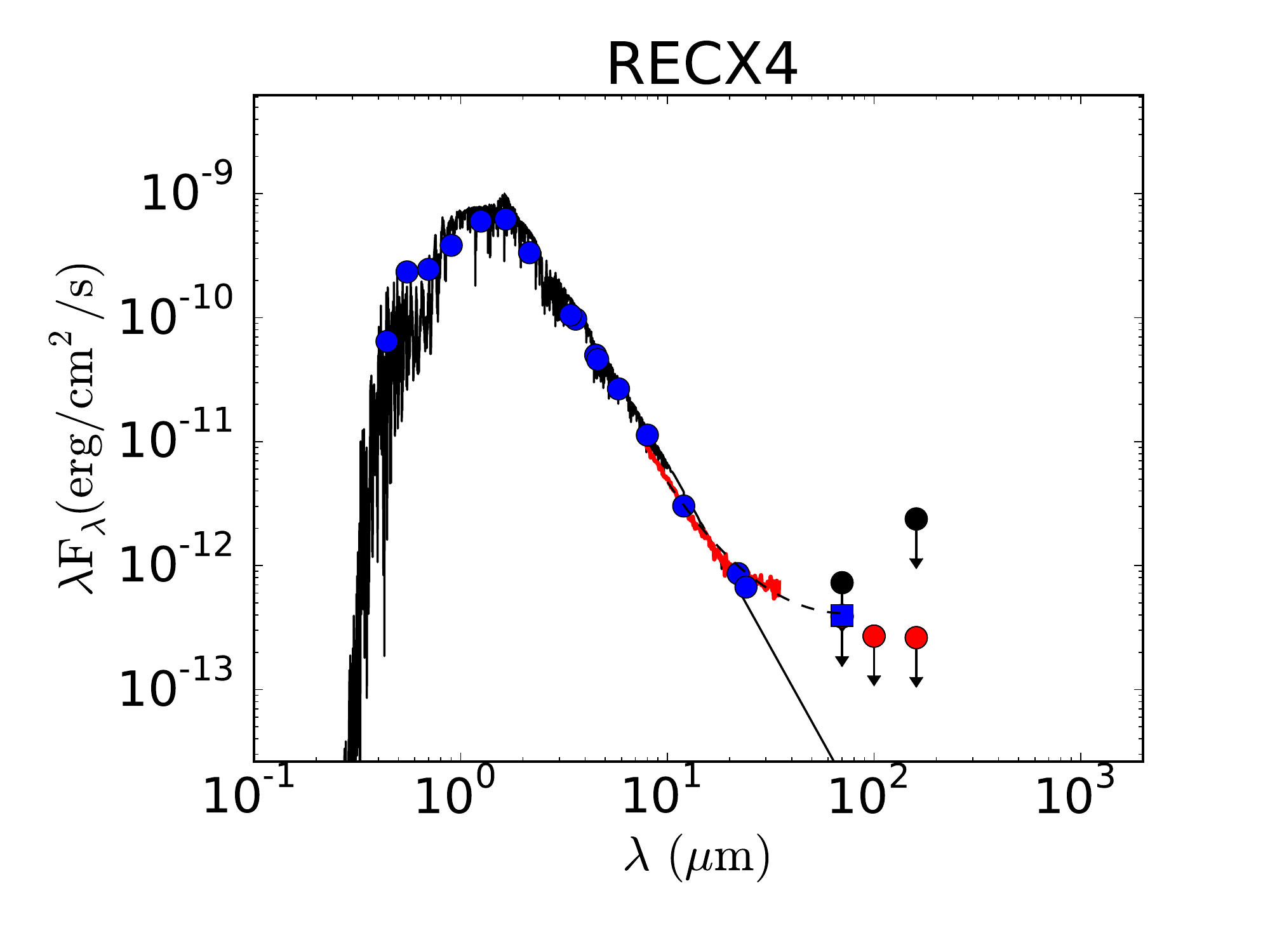} 
\includegraphics[scale=0.24,trim=20mm 30mm 0mm 0mm,clip]{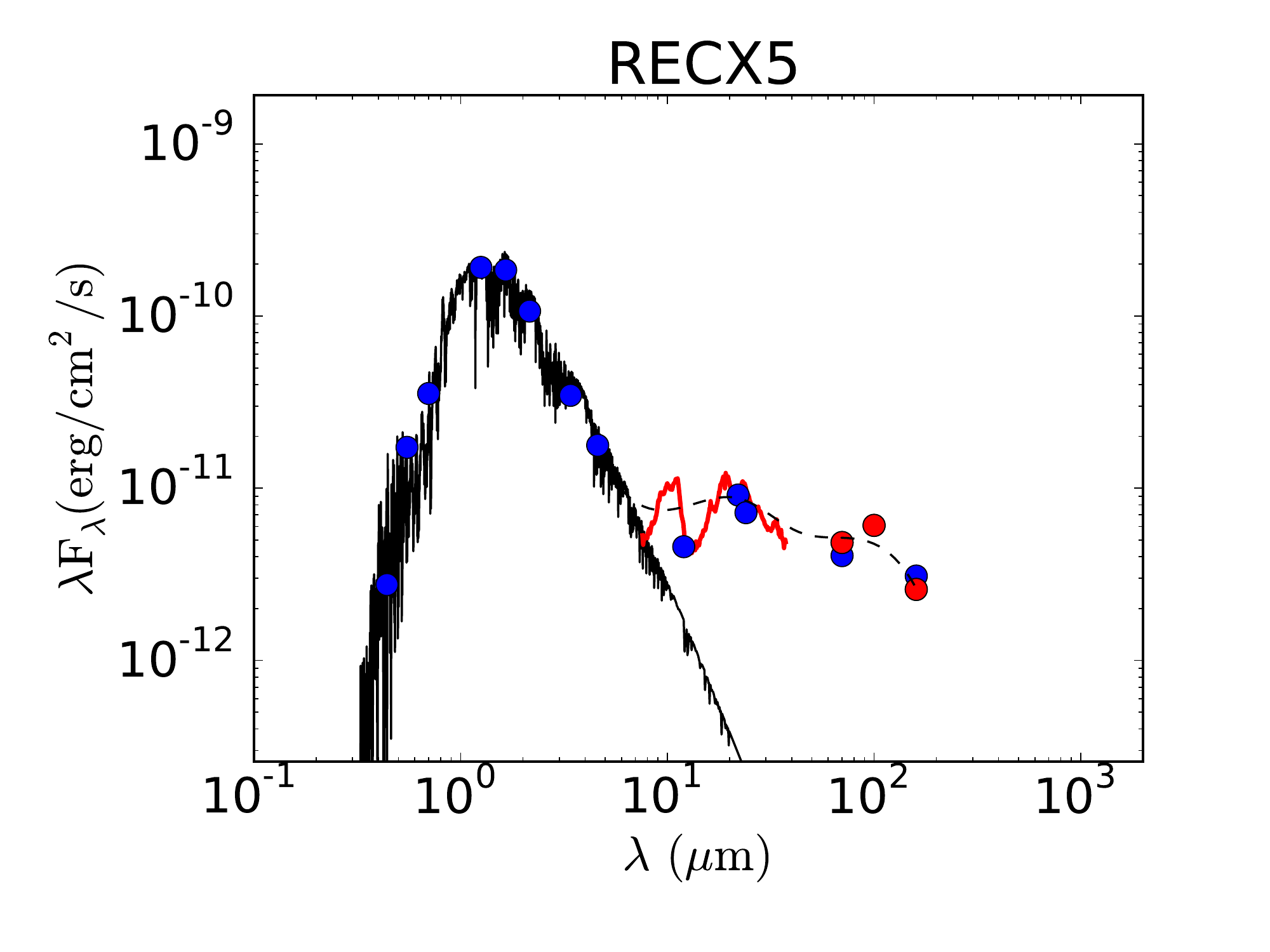}\\
\includegraphics[scale=0.24,trim=8mm 30mm 0mm 0mm,clip]{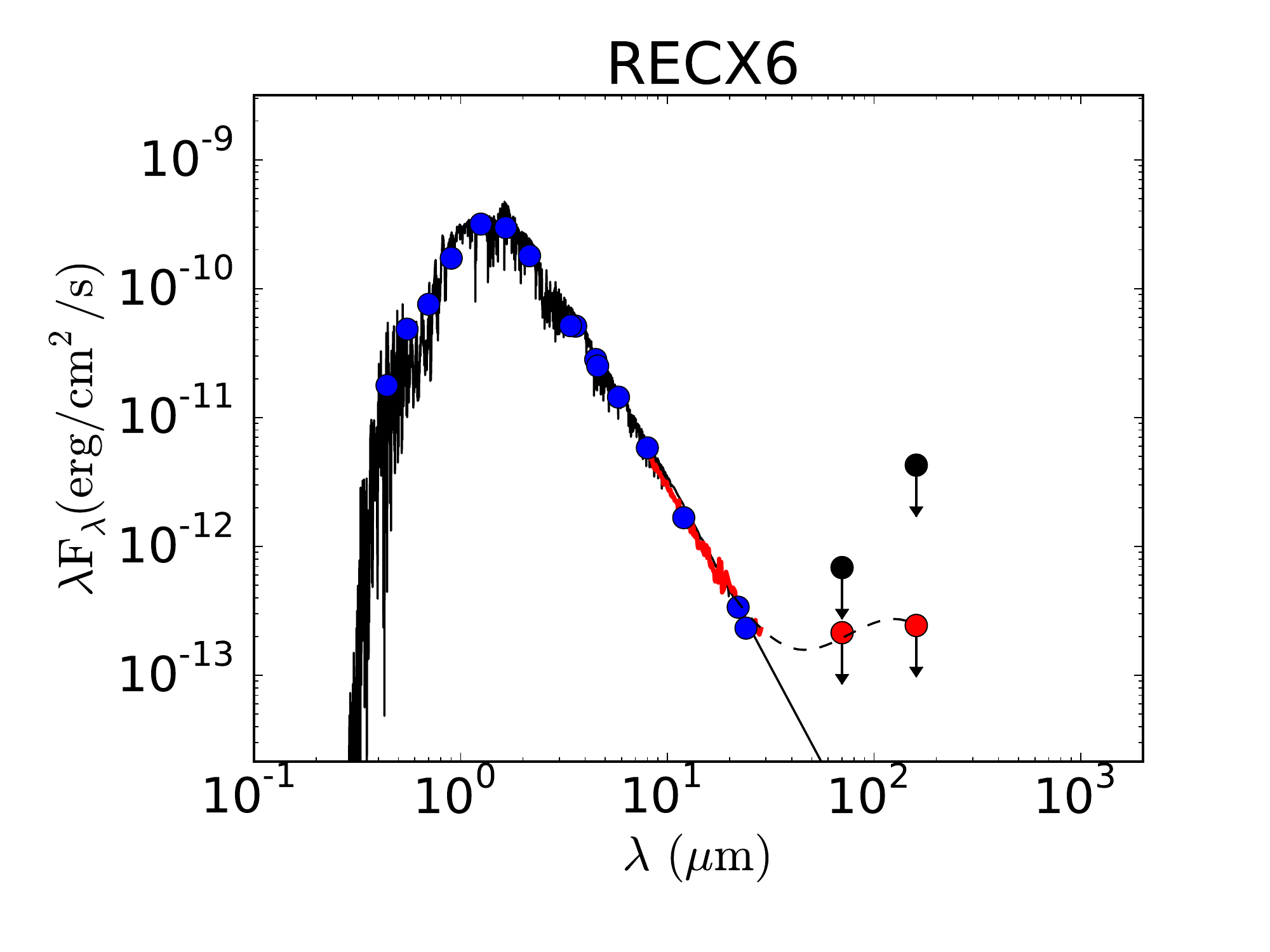}
\includegraphics[scale=0.24,trim=20mm 30mm 0mm 0mm,clip,clip]{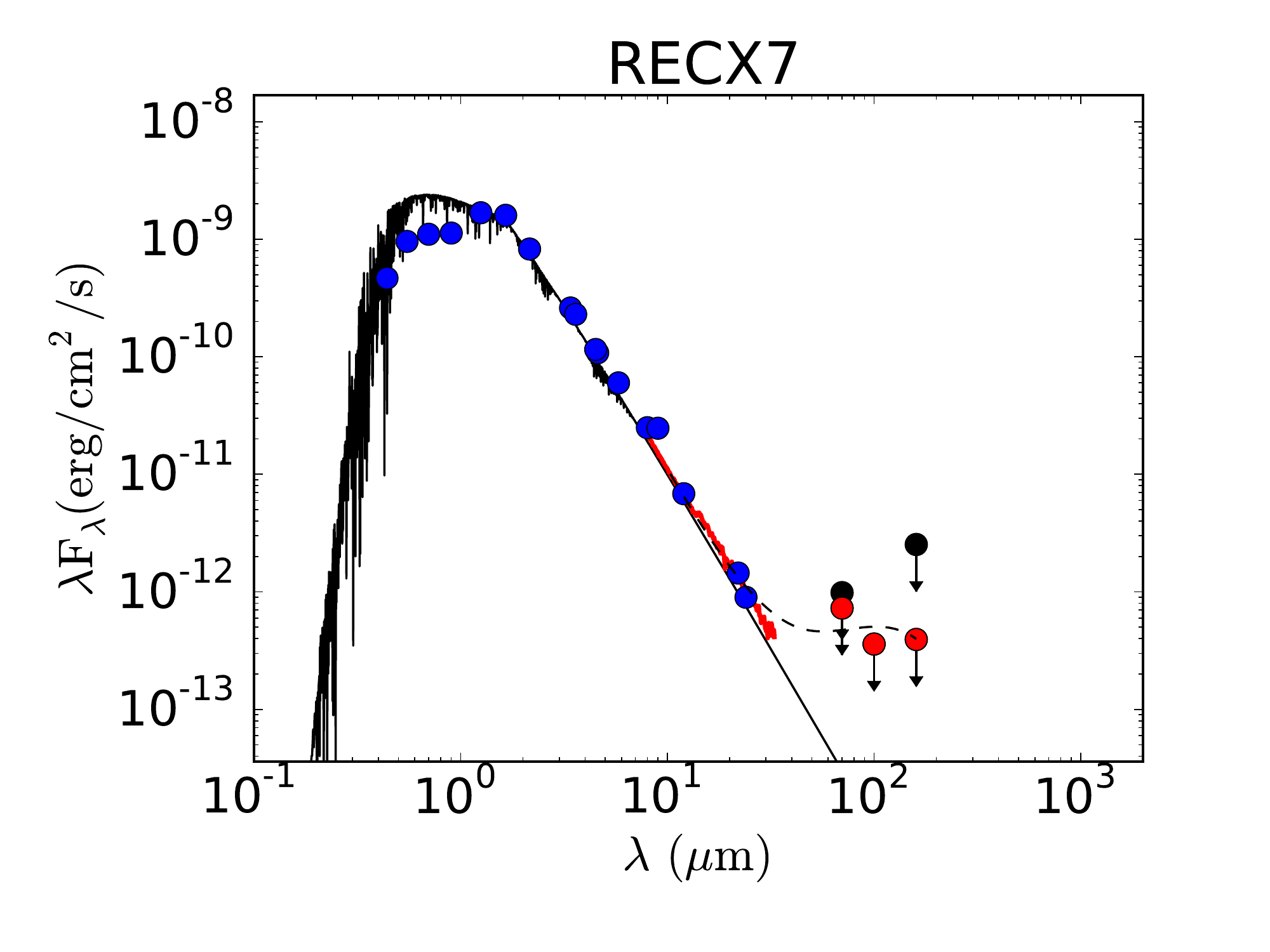}
\includegraphics[scale=0.24,trim=20mm 30mm 0mm 0mm,clip]{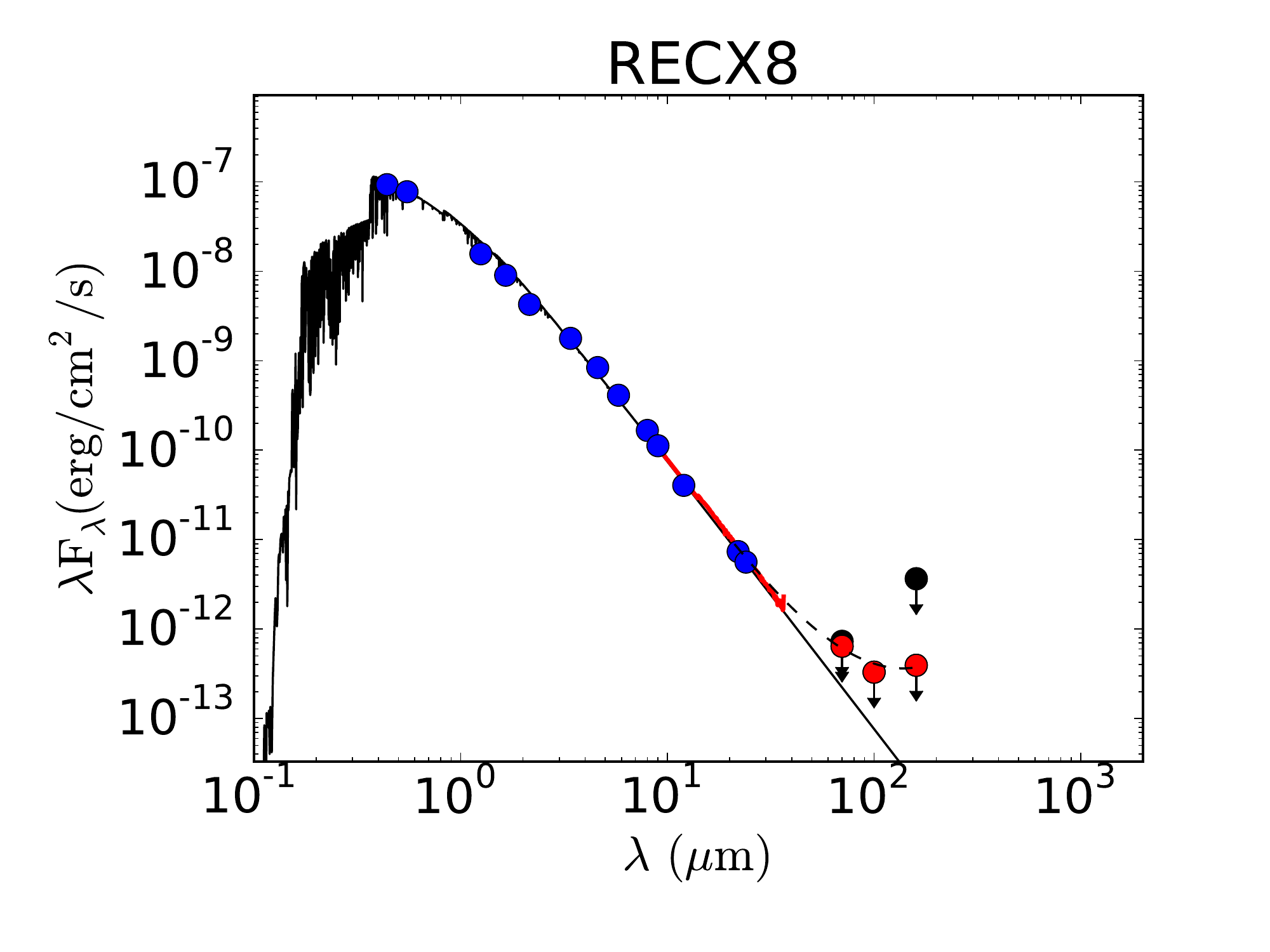}
\includegraphics[scale=0.24,trim=20mm 30mm 0mm 0mm,clip]{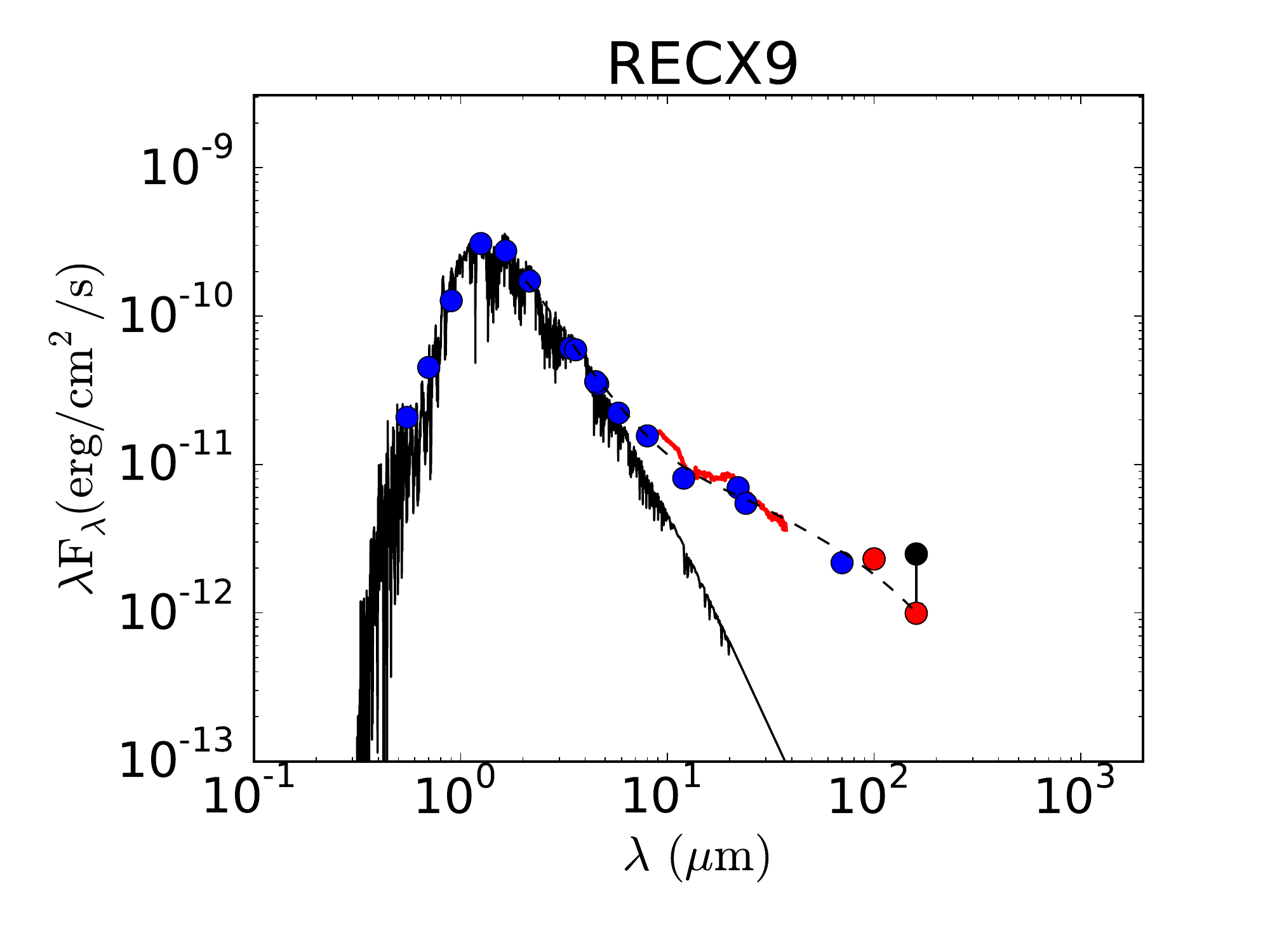}\\
\includegraphics[scale=0.24,trim=8mm 30mm 0mm 0mm,clip]{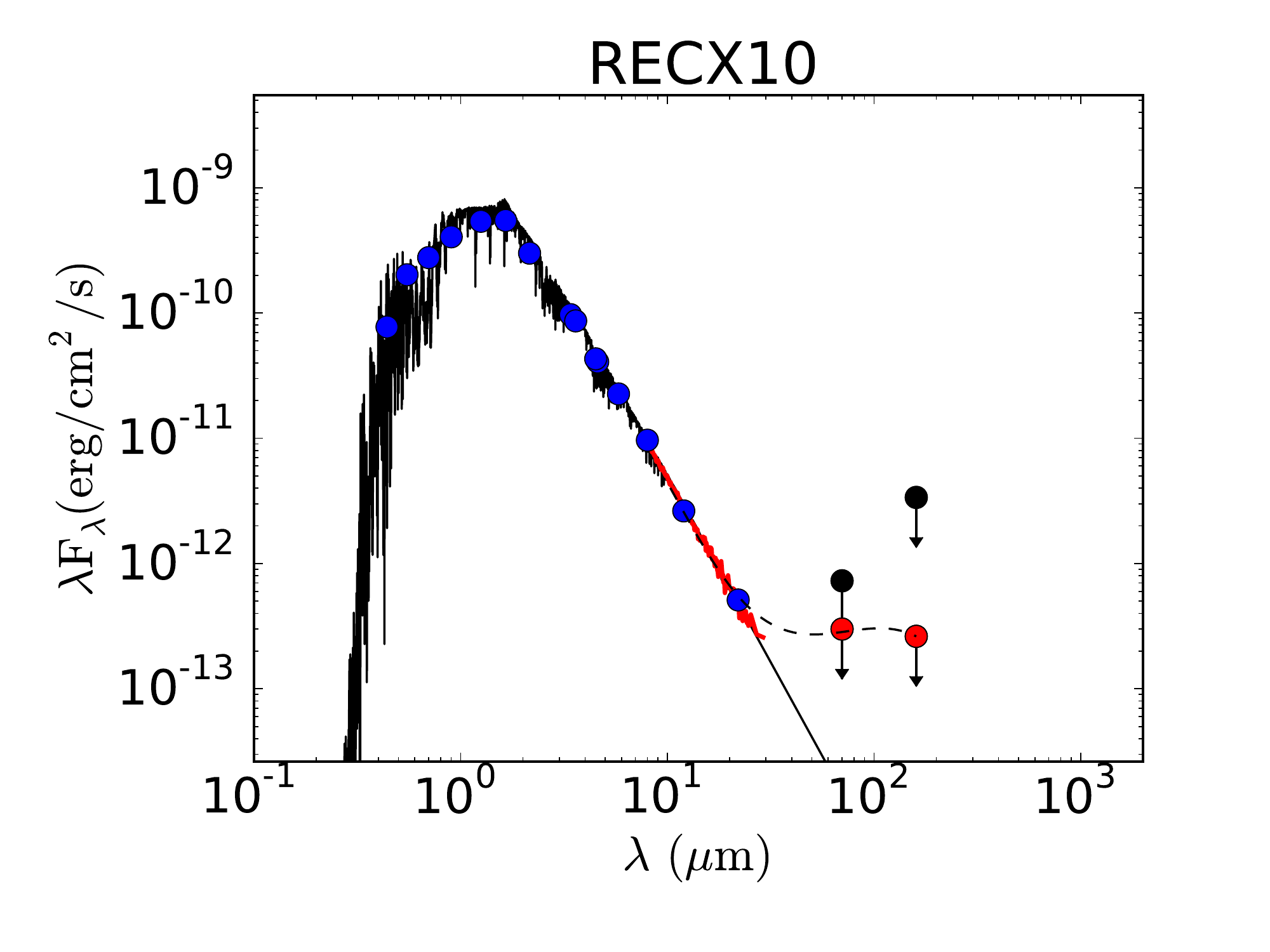}
\includegraphics[scale=0.24,trim=20mm 30mm 0mm 0mm,clip]{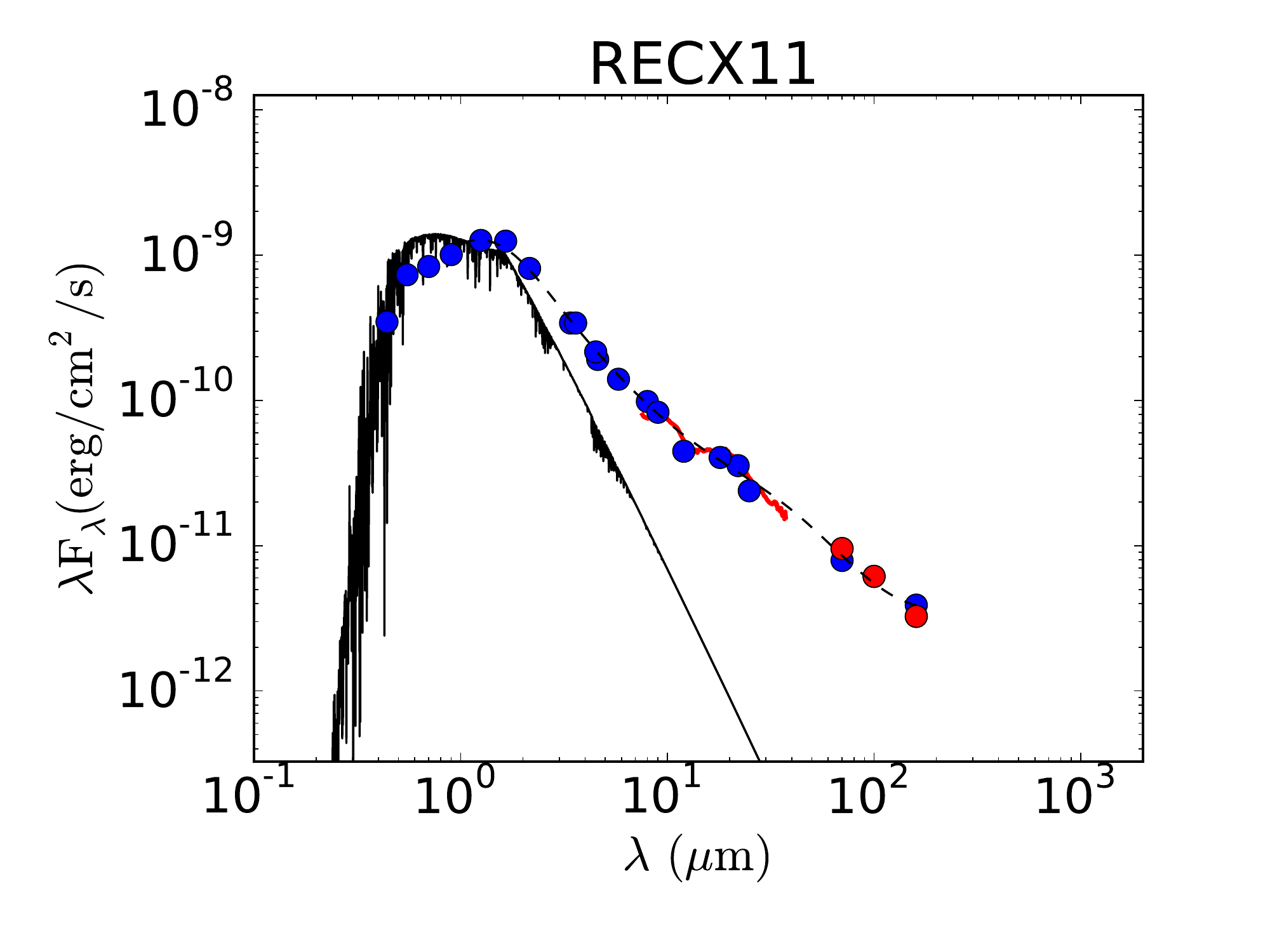}
\includegraphics[scale=0.24,trim=20mm 30mm 0mm 0mm,clip]{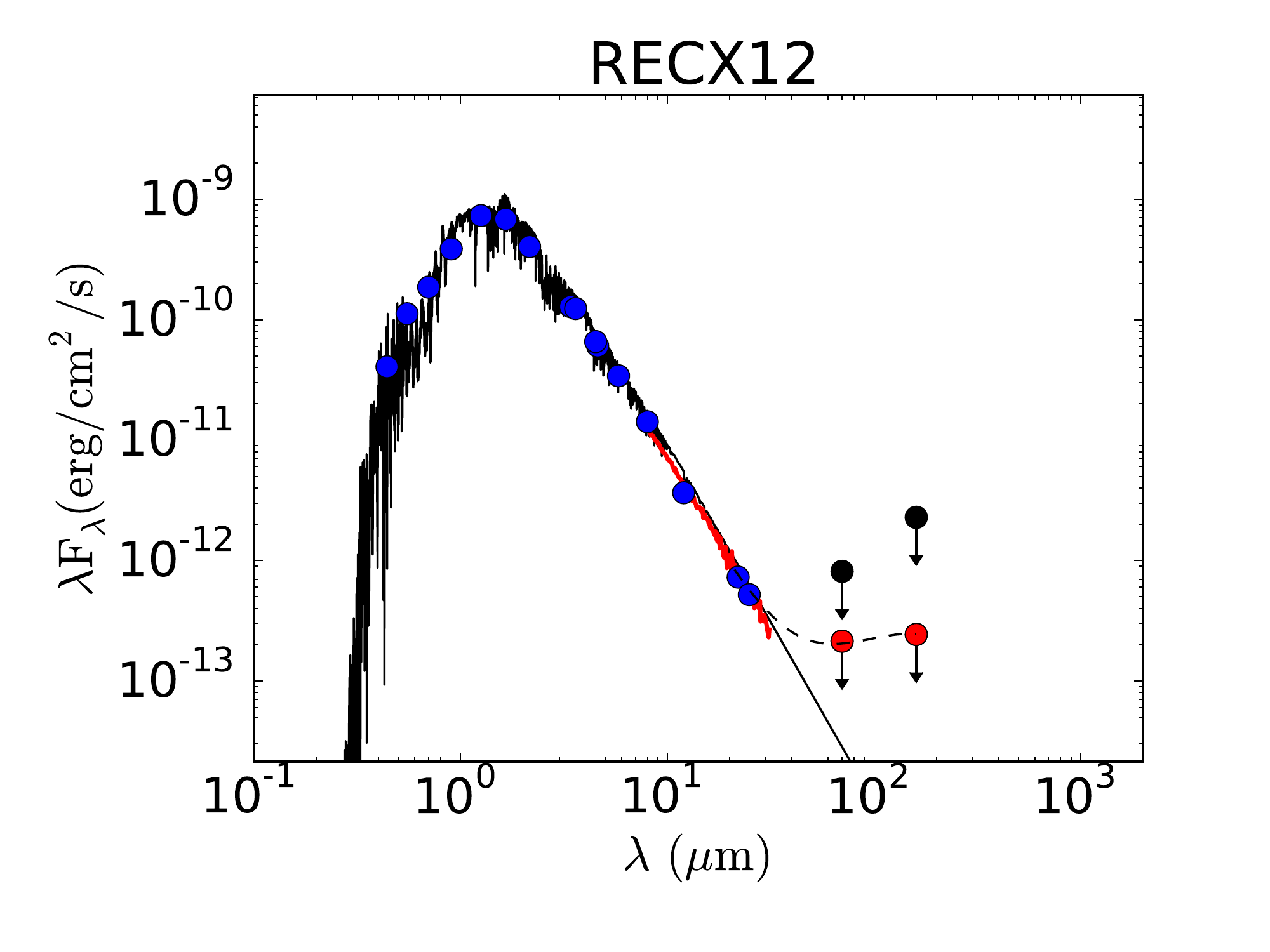}
\includegraphics[scale=0.24,trim=20mm 30mm 0mm 0mm,clip]{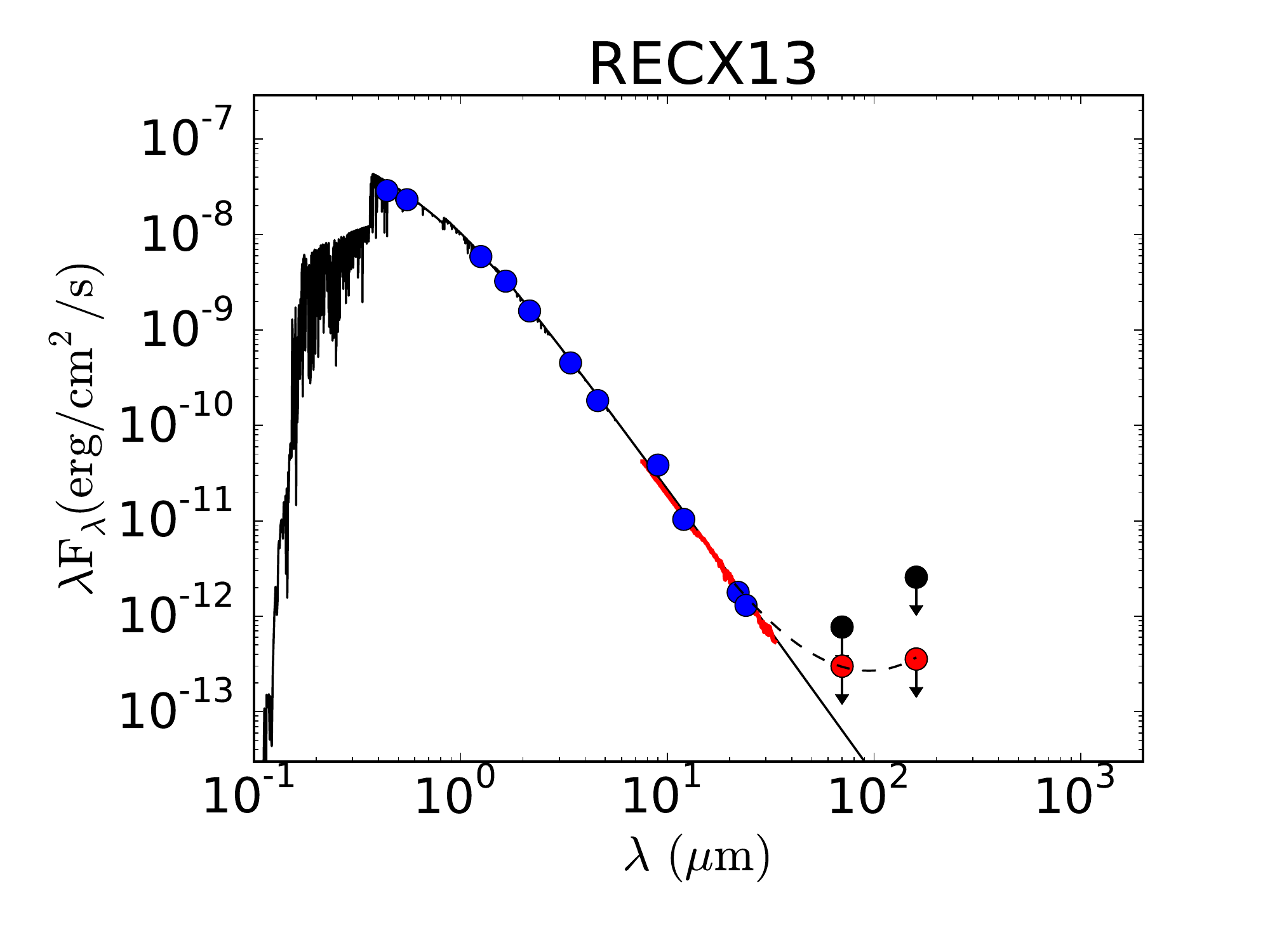}\\
\includegraphics[scale=0.24,trim=8mm 0mm 0mm 0mm,clip]{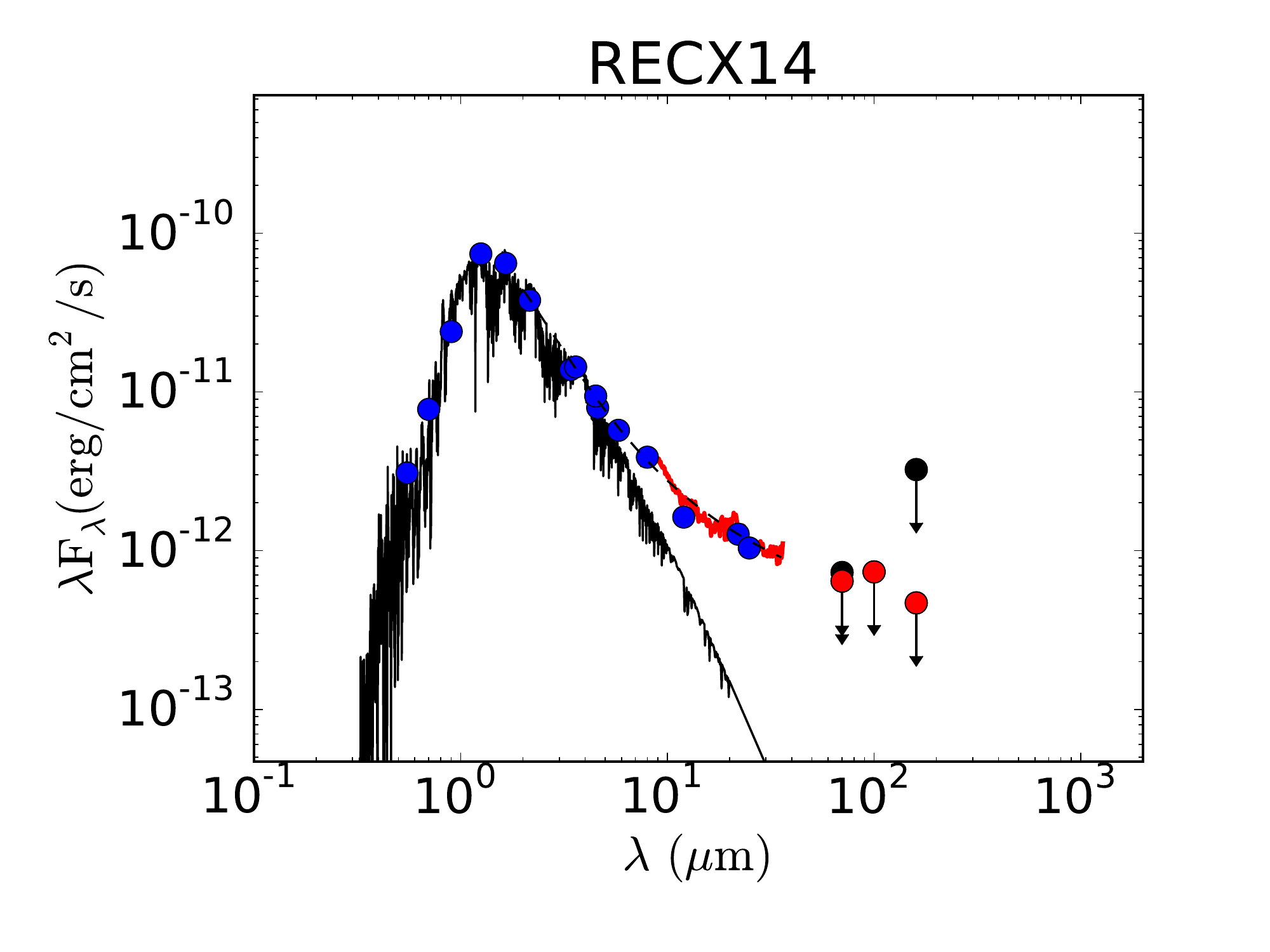}
\includegraphics[scale=0.24,trim=20mm 0mm 0mm 0mm,clip]{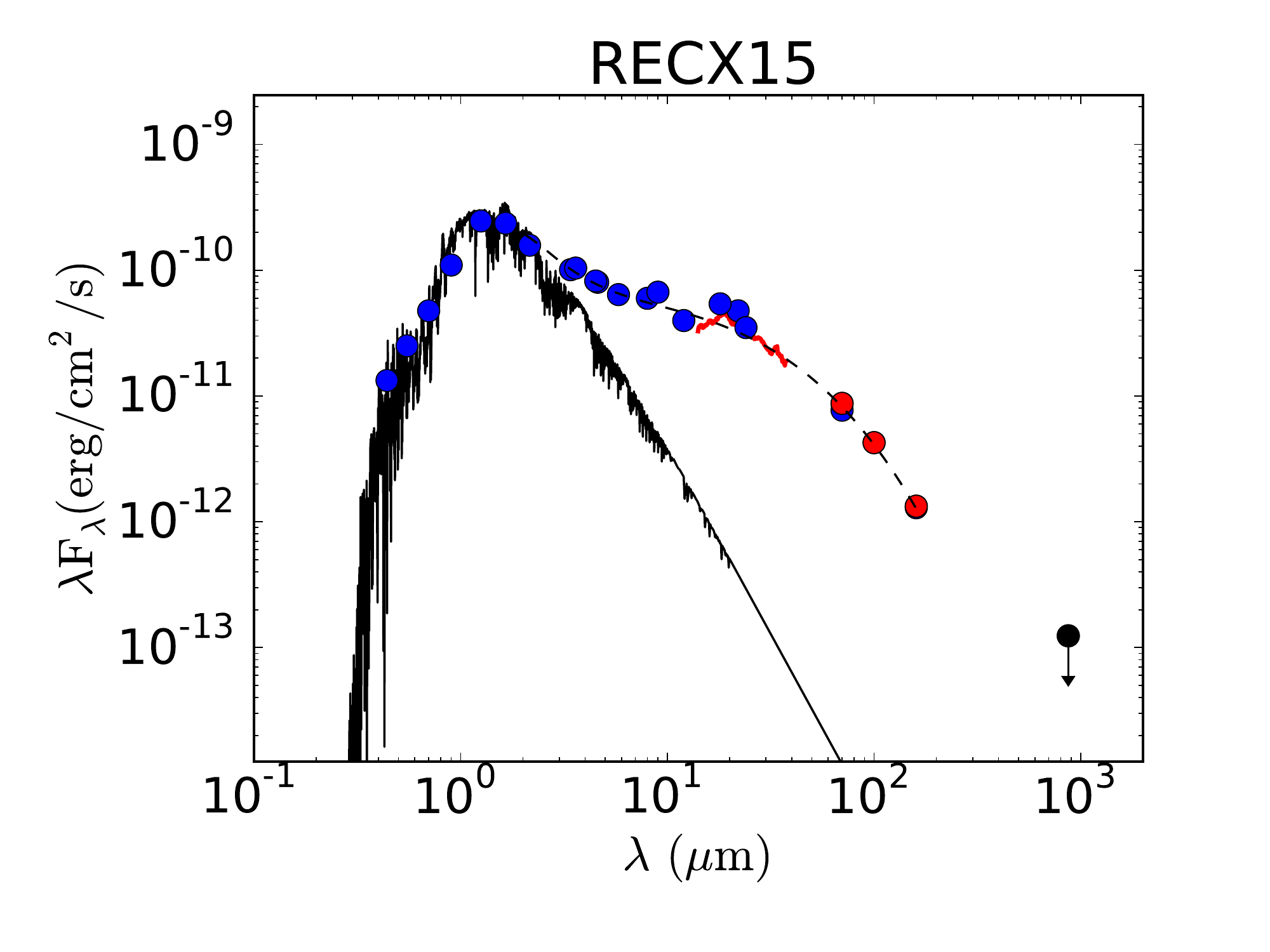}
\includegraphics[scale=0.24,trim=20mm 0mm 0mm 0mm,clip]{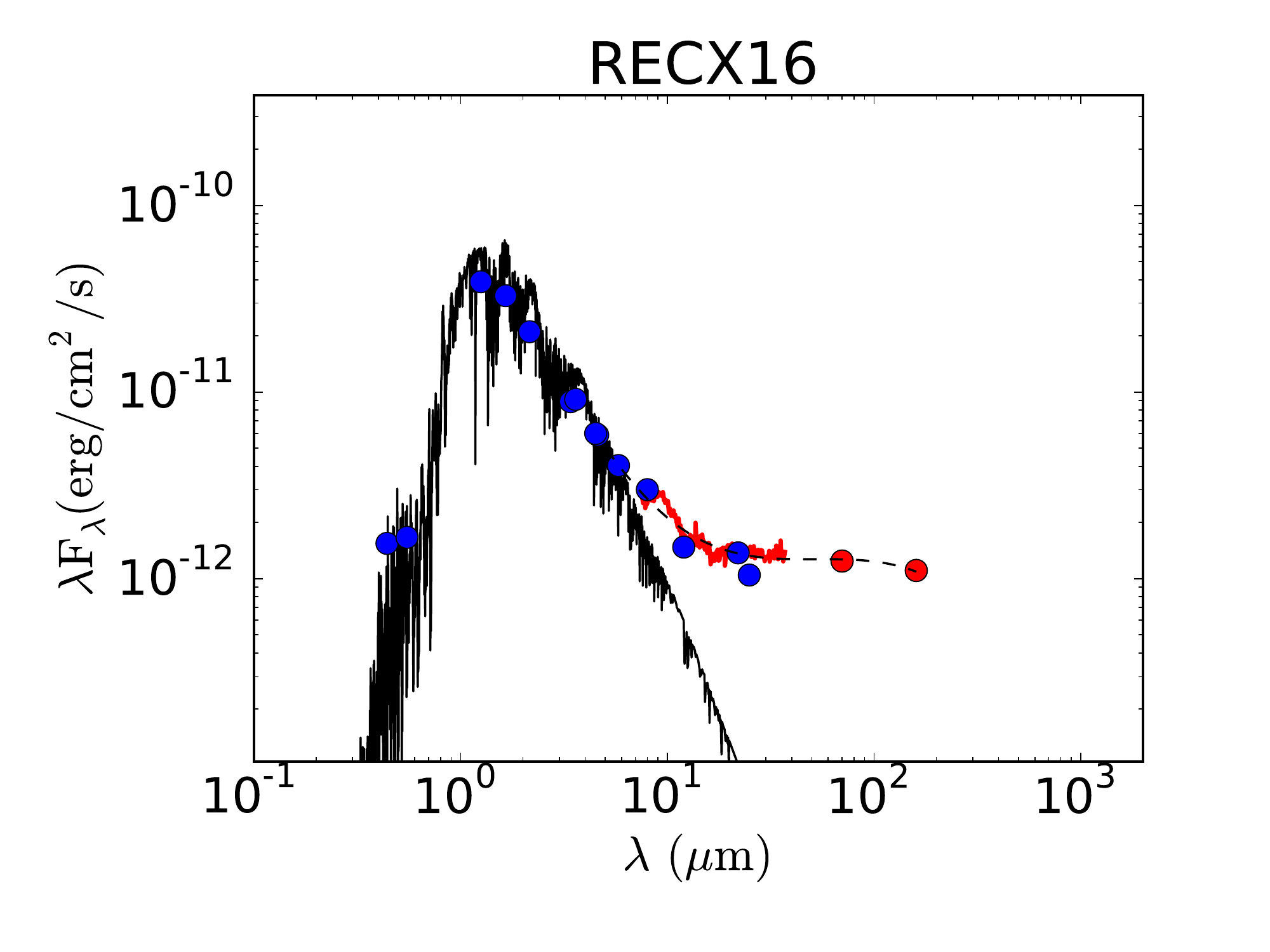}
\includegraphics[scale=0.24,trim=20mm 0mm 0mm 0mm,clip]{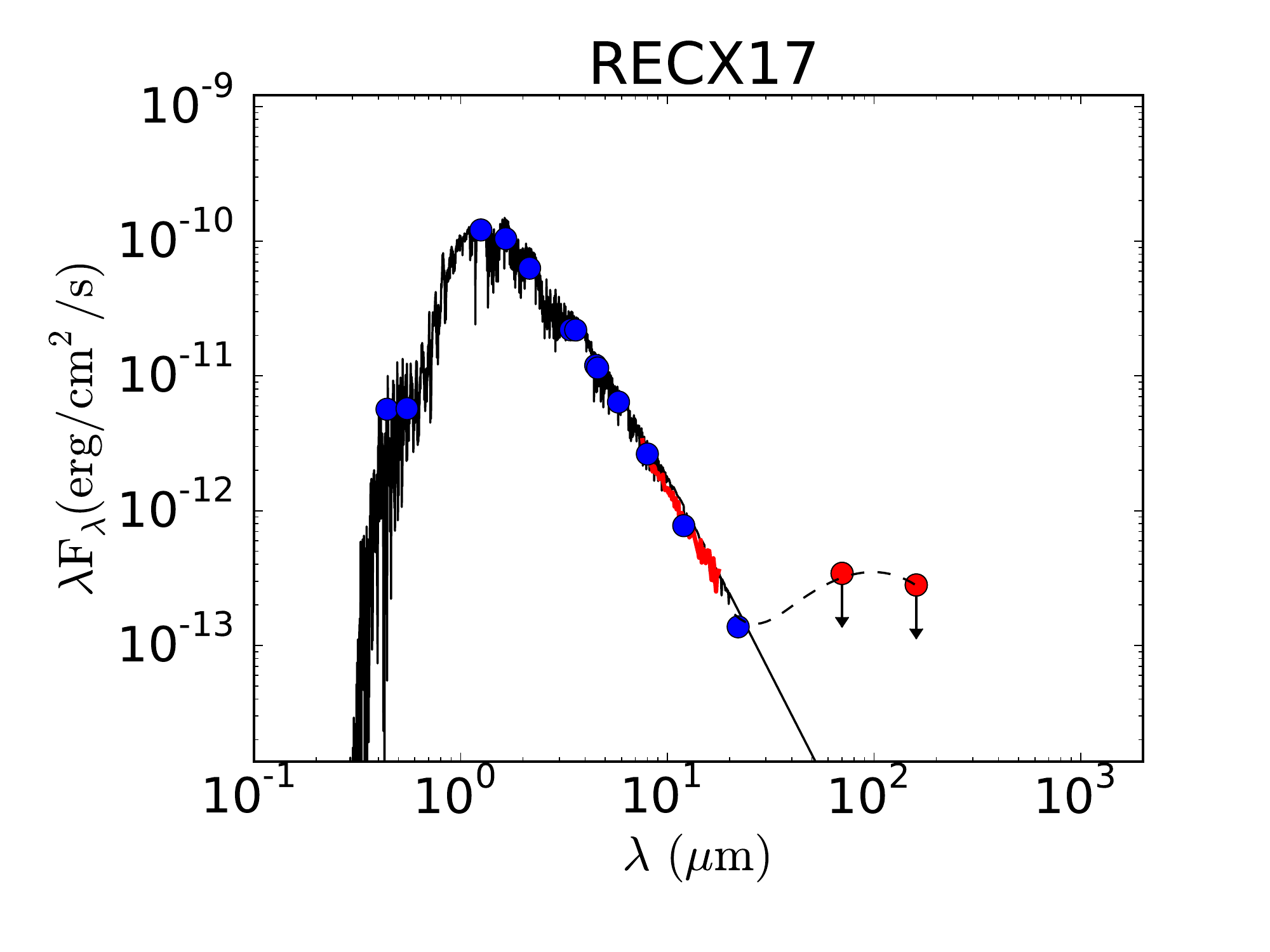}\\
\includegraphics[scale=0.24,trim=8mm 0mm 0mm 0mm,clip]{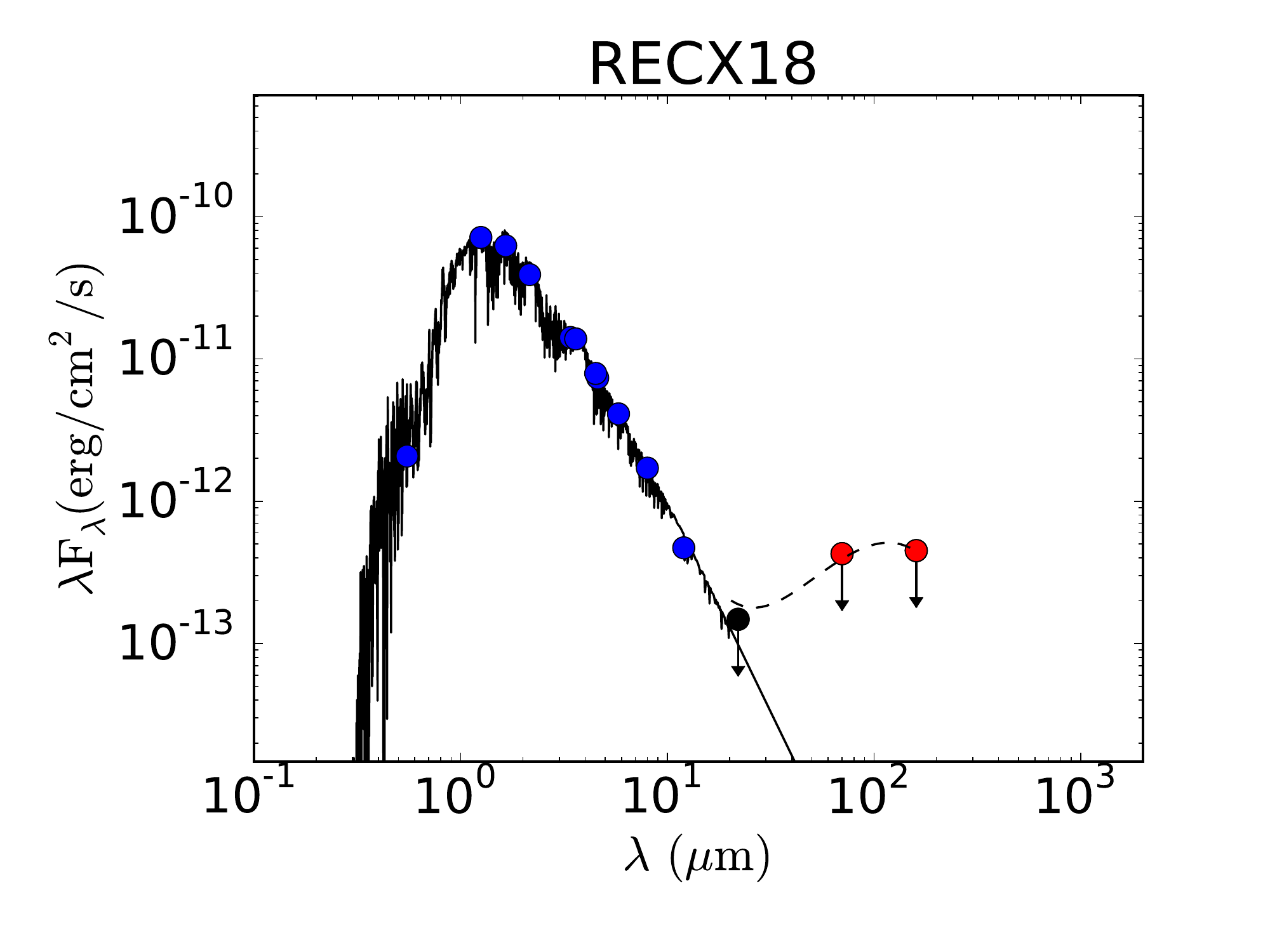}\\
\caption{SEDs of Eta Cha members observed with PACS photometry. PACS observations are shown as red dots. Blue dots depict photometry at different wavelengths from 2MASS, WISE, AKARI and the literature (see Sec. \ref{Sec:Sample}). The blue squares in RECX 3 and RECX 4 SEDs depict PACS photometry from \cite{Cieza2013}. 3$\sigma$ upper limits are shown as black arrows. The red curve depicts the IRS spectrum. The black solid curve is the photospheric model. The black dashed curve depicts the 3rd degree spline used to compute the infrared excess.  The source names are shown at the top of each panel. }
   \label{EtaCha_SED}
\end{center}
\end{figure*}

\begin{figure}[!t]
\begin{center}
\includegraphics[scale=0.4]{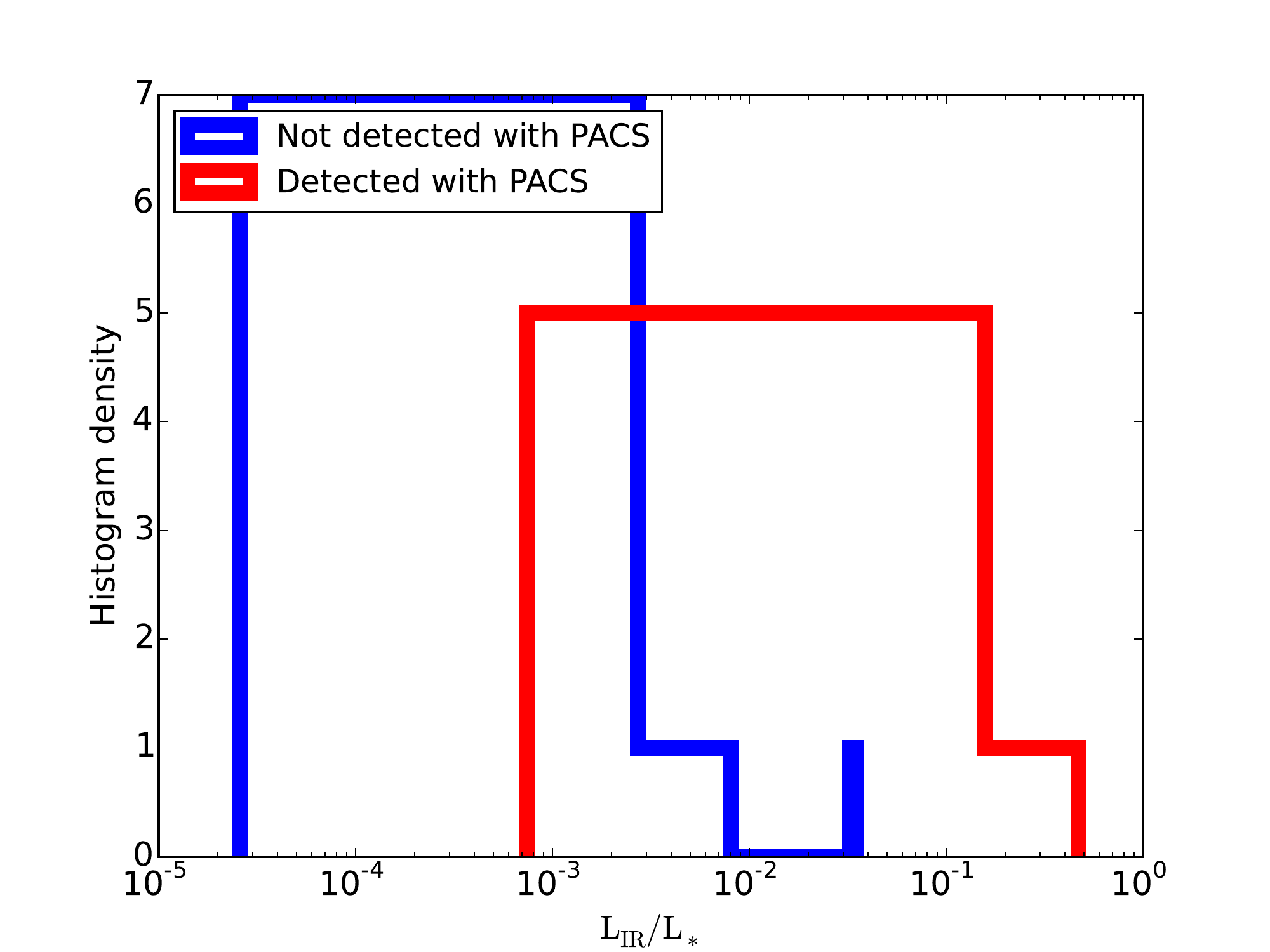}
\caption{Histogram showing the distribution of IR excesses for $\rm \eta$ Cha members observed with PACS.}
   \label{Fig:irhist}
\end{center}
\end{figure}

\subsection{Herschel-PACS photometry} 
We detected four out of sixteen sources observed at 70 $\rm \mu m$, leading to a detection fraction of $\rm 0.25^{+0.13}_{-0.08}$ \citep[with errors coming from binomial population distributions, see, e. g.][]{Burgasser2003}. Two sources in the sample, namely RECX 3 and RECX 4, were not detected at 70 $\rm \mu m$ in our survey, but \cite{Cieza2013} detected both at 70 $\rm \mu m$. Fluxes are shown in brackets in Table \ref{PACS_phot}. Including these 70 $\rm \mu m$ detections, the detection fraction grows to $\rm 0.47^{+0.12}_{-0.11}$, compatible with the fraction computed by \cite{Sicilia-Aguilar2009} using \textit{Spitzer} data. Since RECX 3 and RECX 4 resemble debris discs (see Sec. \label{Sec:SED}), the primordial disc fraction at 70 $\rm \mu m$ remains 0.25. Only eight objects were observed at 100 $\rm \mu m$, and we detected four of them ($\rm 0.50 \pm 0.16$ detection fraction). Finally, we detected 
five objects at 160 $\rm \mu m$ out of seventeen observed, leading to a detection fraction of $\rm 0.29^{+0.13}_{-0.08}$. 

All the detected fluxes are in excess above the photosphere and we observe no correlation of the excess with $\rm T_{eff}$ or spectral type, as already observed for TW Hydra Associations (TWA) and Beta Pictoris Moving Group (BPMG) stars \citep{Riviere2013, Riviere2014}. For RECX 16 this is the first detection of the source in the far-IR, both at 70 and 160 $\rm \mu m$. For RECX 9 and RECX 15, PACS fluxes are the first detections at 100 and 160 $\rm \mu m$, but 70 $\rm \mu m$ emission was previously detected with \textit{Spitzer} \citep{Gautier2008}. For these three sources, PACS detections add important data to the SED, and allow for a more accurate characterisation. Fluxes at 70 $\rm \mu m$ for three objects observed with PACS and MIPS \citep{Gautier2008} agree well within the errors, except for RECX 11, where the PACS flux is larger, even when errors are included, leading to a difference of 20\%. Interestingly, the change at 160 $\rm \mu m$ goes in the other direction, with the MIPS flux being a 20 \% larger than the PACS flux. Confirming if this is real variability requires more observations. If the variability is real it may reflect a fast change in disc properties.

To test whether the emission was extended at any wavelength, we performed azimuthally averaged radial profiles of the sources, and compared the resulting profiles with that of the standard star $\rm \alpha$ Boo. None of the sources  showed extended emission at any wavelength.

\begin{figure*}[!t]
\begin{center}
\includegraphics[scale=0.25,trim=0mm 30mm 10mm 0mm,clip]{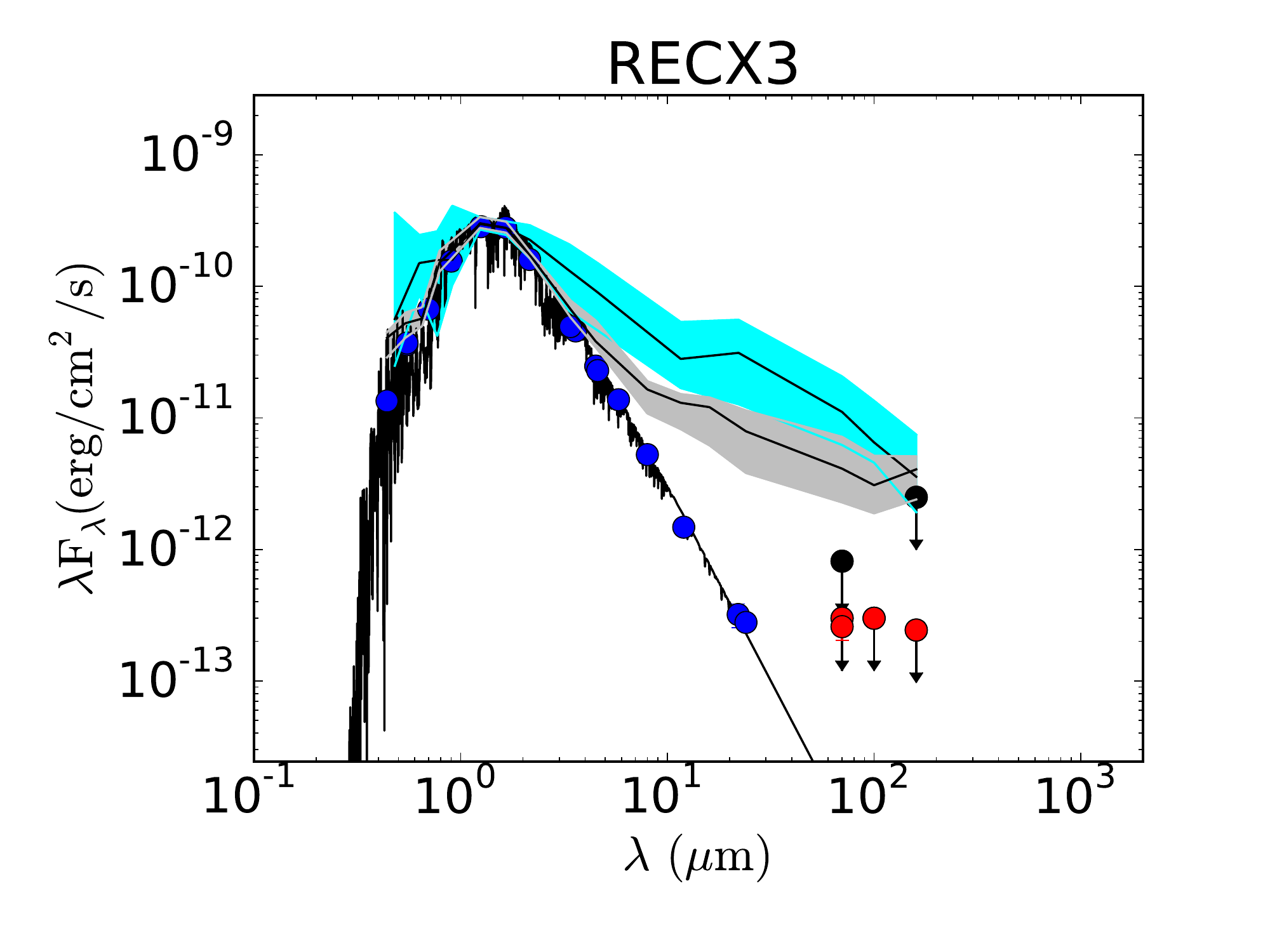}\includegraphics[scale=0.25,trim=20mm 30mm 10mm 0mm,clip]{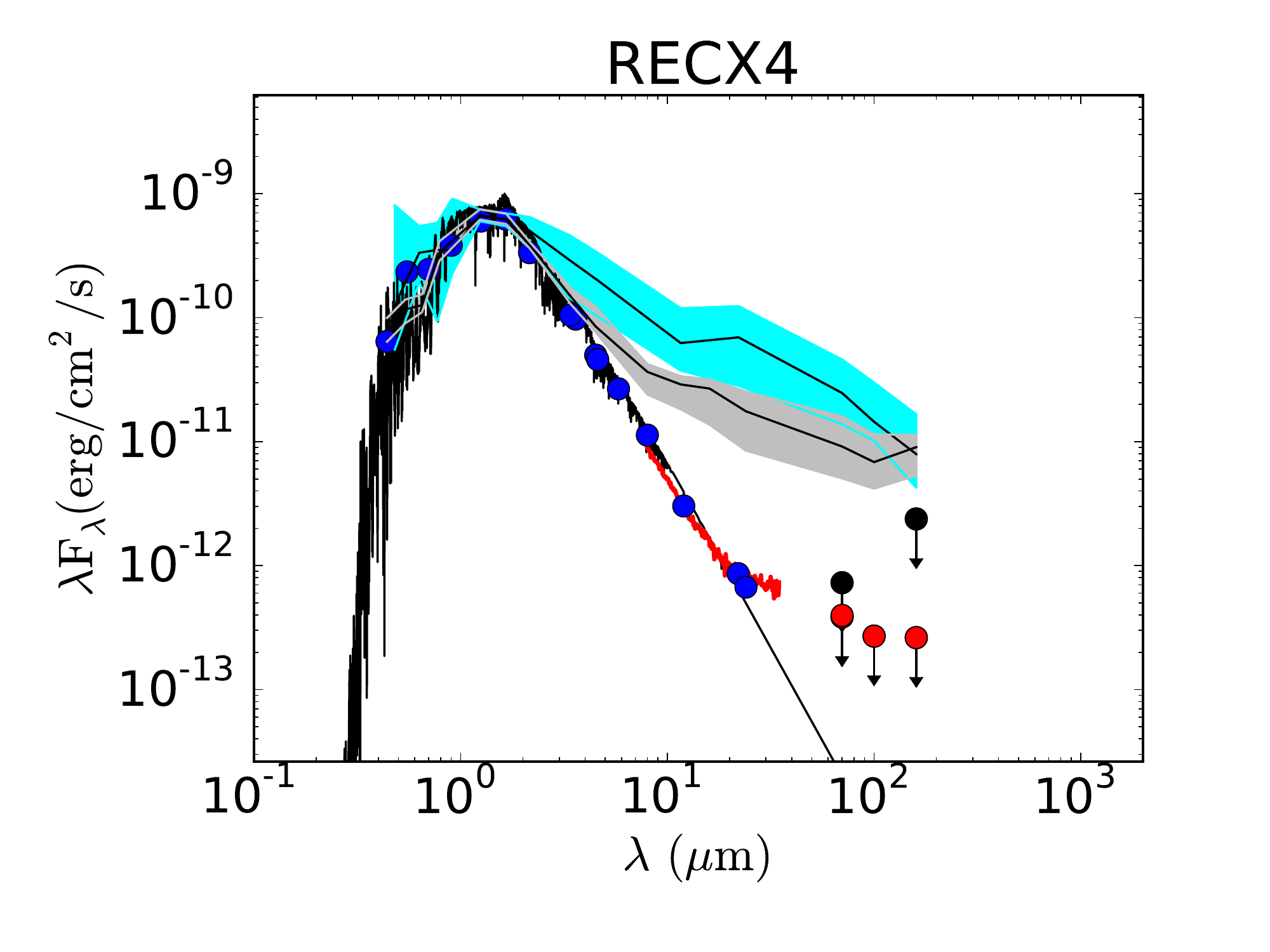}\includegraphics[scale=0.25,trim=20mm 30mm 10mm 0mm,clip]{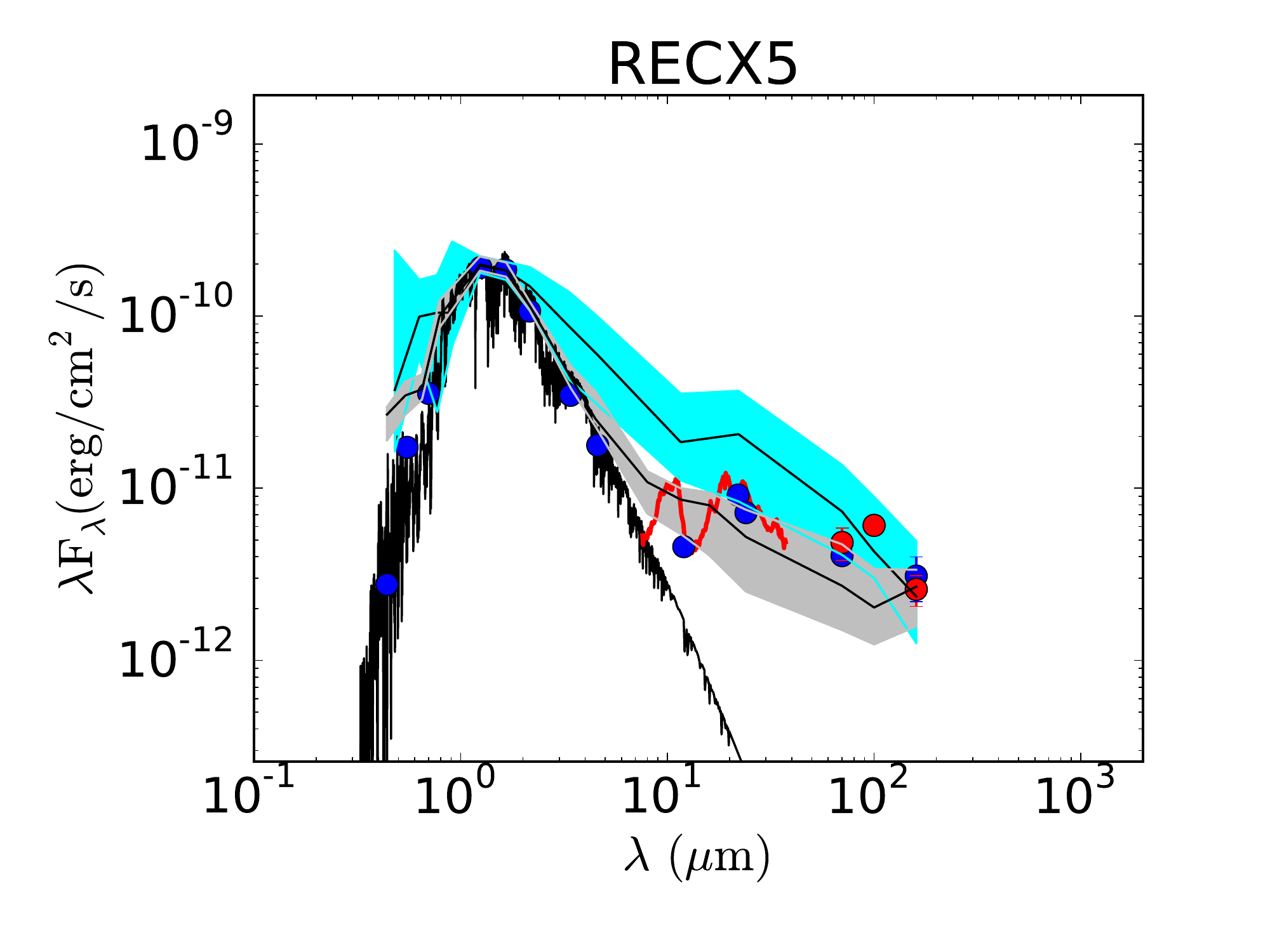}\includegraphics[scale=0.25,trim=20mm 30mm 10mm 0mm,clip]{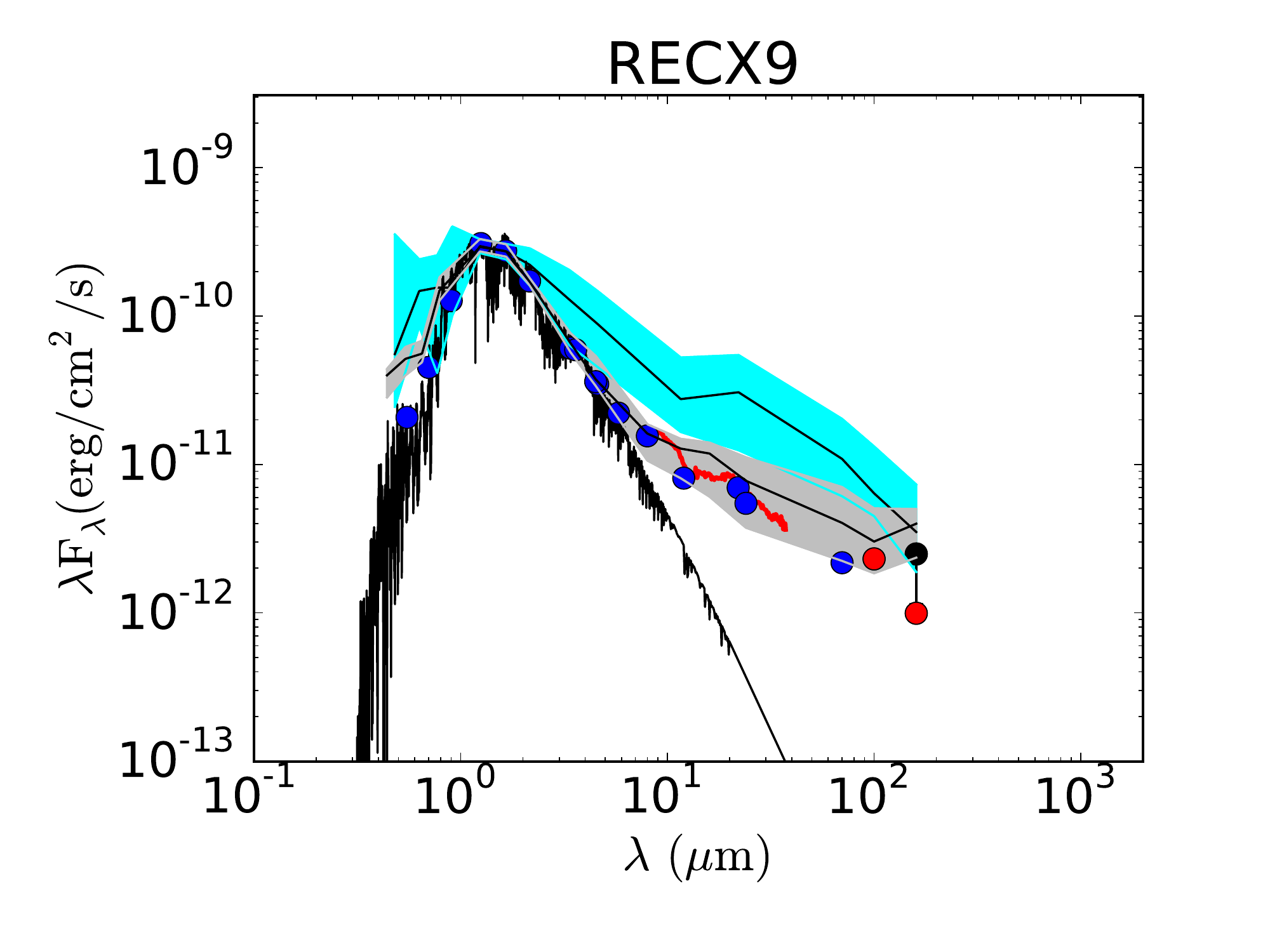}\\

\includegraphics[scale=0.25,trim=0mm 0mm 10mm 0mm,clip,clip]{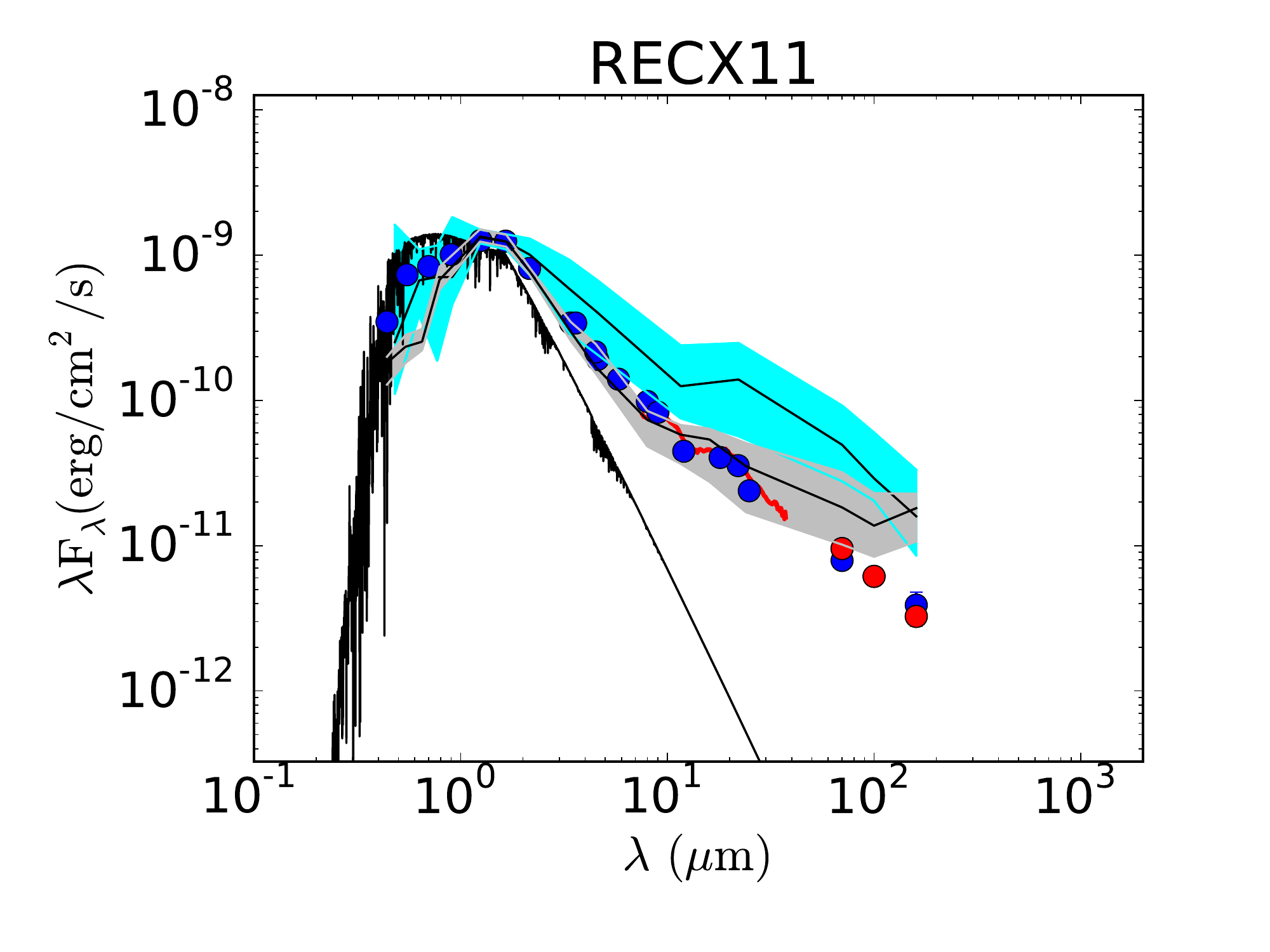}\includegraphics[scale=0.25,trim=20mm 0mm 10mm 0mm,clip]{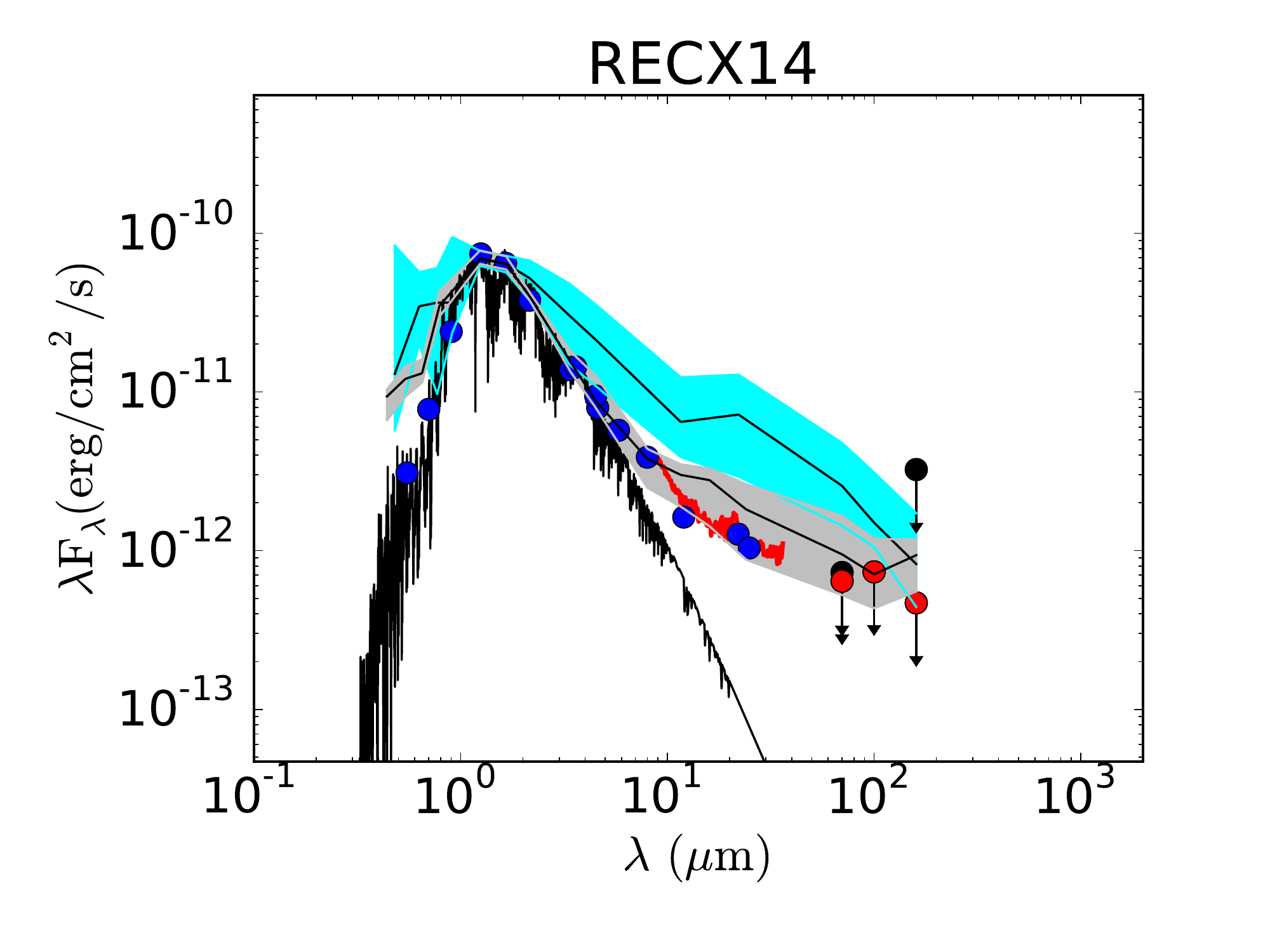}\includegraphics[scale=0.25,trim=20mm 0mm 10mm 0mm,clip]{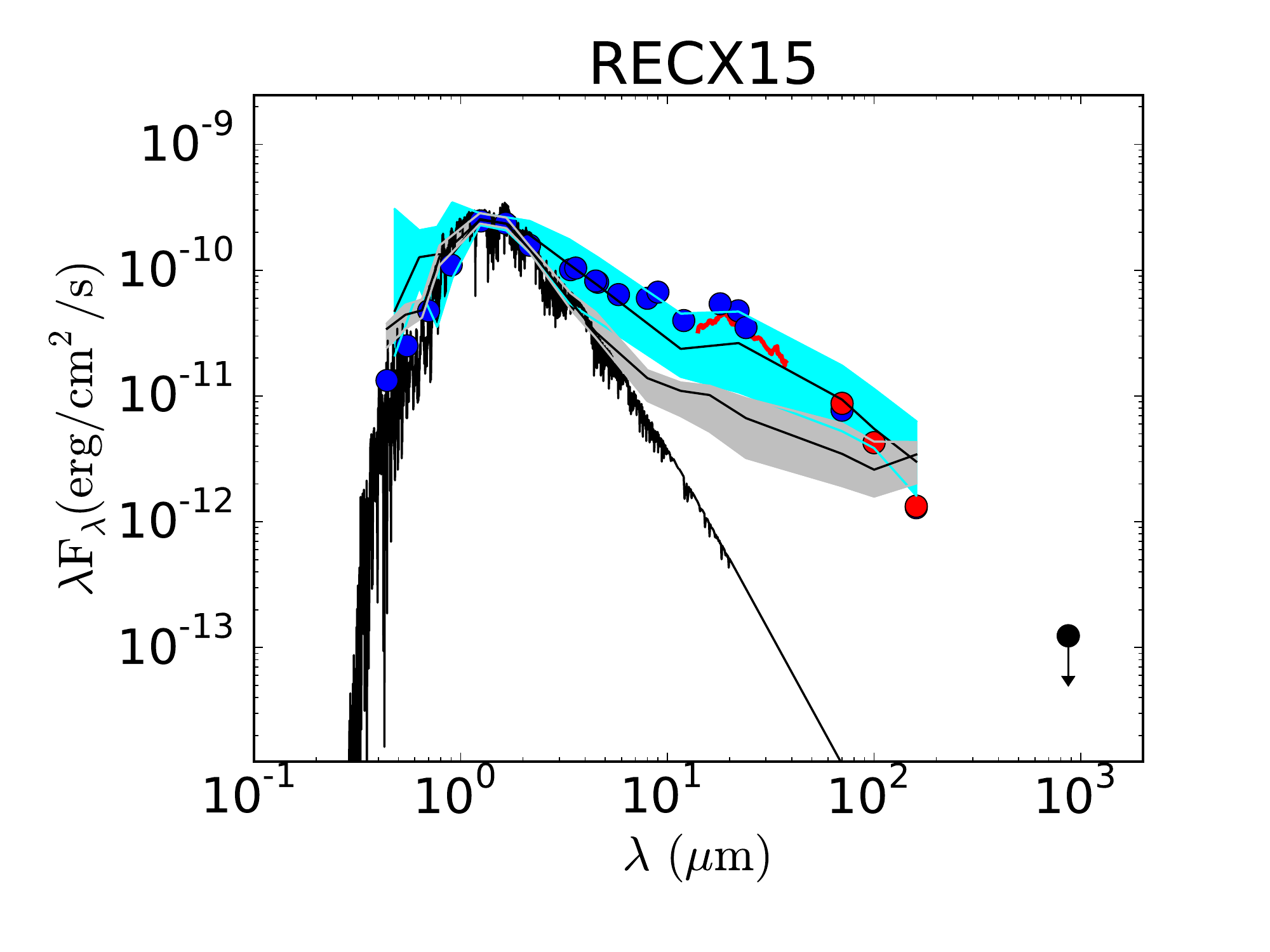}\includegraphics[scale=0.25,trim=20mm 0mm 10mm 0mm,clip]{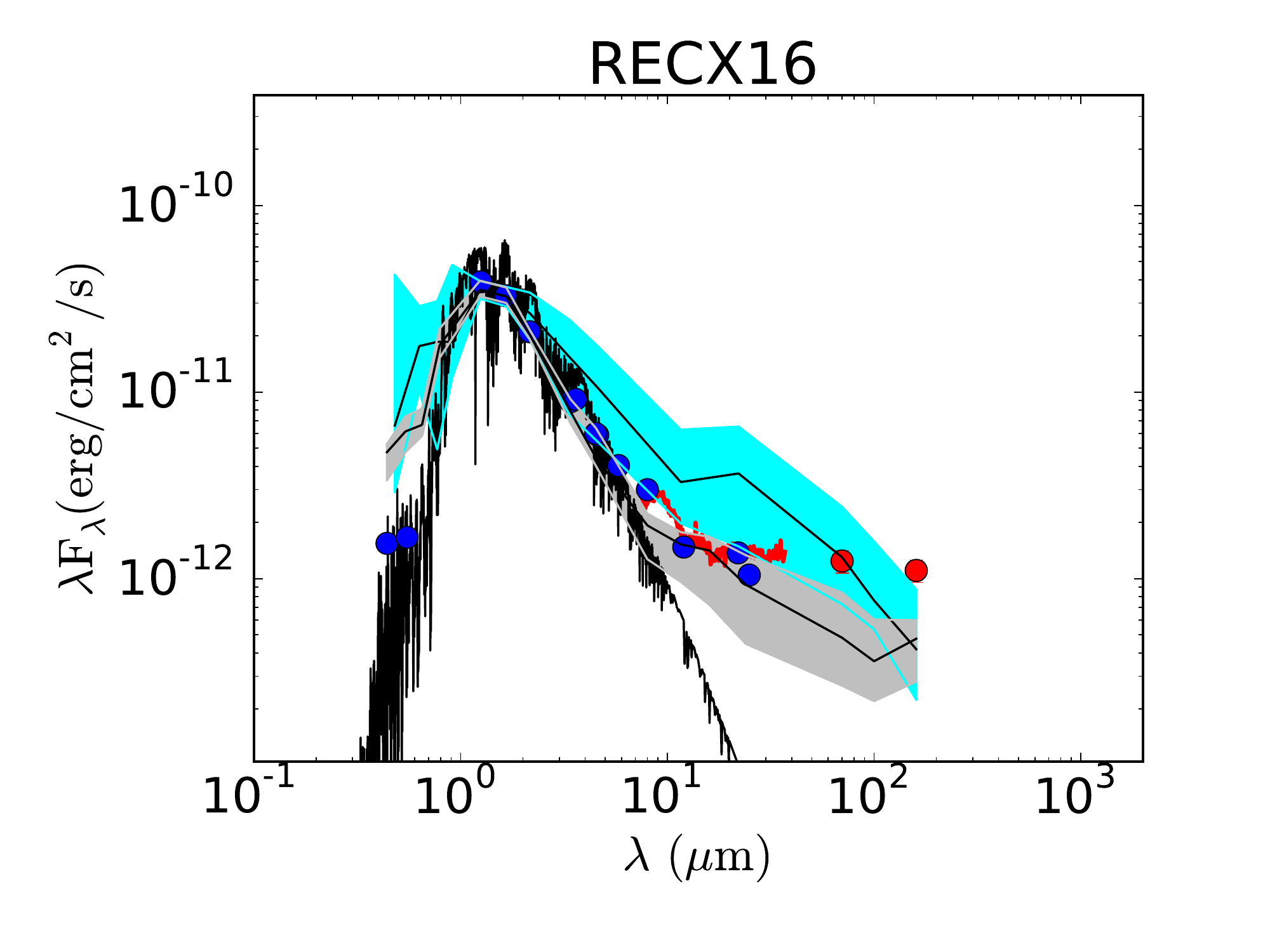}\\ 
\caption{SEDs of $\rm \eta$ Cha members detected with PACS photometry compared to the median SED in Taurus (cyan) and Upper Scorpius (grey) from \cite{Mathews2013}. Blue dots depict archival photometry, while the red ones are PACS photometry from the present study. Black arrows are upper limits. The red curve depicts the IRS spectrum, while the black one represents the photospheric model.}. 
   \label{EtaCha_SED_comp}
\end{center}
\end{figure*}

\subsection{Herschel-PACS spectroscopy} \label{Sec:specResults}
We have detected [OI] emission at 63.18 $\rm \mu m$ towards only one of the sources observed, leading to a total detection fraction of $\rm 0.08^{+0.14}_{-0.03}$. The detection fraction is low when compared to younger associations like Taurus \citep[$\rm 0.57 \pm 0.09$,][]{Howard2013} and Cha II \citep[$\rm 0.37^{+0.12}_{-0.09}$,][]{Riviere2015}. This decrease with age is indicative of an evolutionary trend. The fraction of sources with an IR-excess that show an [OI] detection is $\rm 0.17^{+0.23}_{-0.06}$. No $\rm H_{2}O$ emission was detected in the sample. The only source with an [OI] detection at 63.185 $\rm \mu m$ is RECX 15, which was discussed in detail in \cite{Woitke2011}. [OI] emission at 63 $\rm \mu m$ is observed only in the central spaxel. We show in the top panel of Fig. \ref{RECX15_contours} the spatial distribution of [OI] emission at 63.185 $\rm \mu m$ (colored contours), where we can see that the [OI] emission coincides with the continuum emission (line contours) at 63 $\rm \mu m$. To further test whether the emission was extended, we performed a test based on the method by \cite{Podio2012}, which makes use of the ratio of the line to continuum emission in each spaxel compared to that where the source is centreed. The result of this test is shown in the bottom panel of Fig. \ref{RECX15_contours}.  No significative residual emission is detected outside the central spaxel, therefore [OI] line emission is centreed at the position of the source (to the limit of PACS spatial resolution) and is not extended compared to the model PSF.

No species were detected at longer wavelengths for any of the sources (see Fig. \ref{EtaCha_range_spec}). Upper limits on line fluxes for wavelengths in the range 72 to 180 $\rm \mu m$ are provided in Table \ref{tabSpecRange}. \cite{Dent2013} showed that the detection fraction in the GASPS sample was $\rm \sim 50\%$ compared to $\rm \sim 10\%$ for the other lines. Therefore, the lack of detections for the other transitions fits well with our expectations. 

\section{Discussion}\label{Sec:Discussion}
\subsection{Spectral energy distributions and infrared excesses}\label{Sec:SED}
We complemented PACS observations with photometry at different wavelengths from public catalogues and the literature, including optical, near- and far-IR photometry (see Sec. \ref{Sec:Sample}). We also obtained \textit{Spitzer}-IRS data from the Cornell Atlas of Spitzer-IRS Sourcess \citep[CASSIS][]{Lebouteiller2011} and included them in the SED. The resulting SEDs are shown in Fig. \ref{EtaCha_SED}. 

\cite{Gautier2008} used \textit{Spitzer} observations to compute fractional disc luminosities for $\rm \eta$ Cha members up to 70 $\rm \mu m$. We extended the exercise up to 160 $\rm \mu m$ by including \textit{Herschel}-PACS observations presented in this paper. We computed the fractional disc luminosity by fitting a third degree spline to the observed data and the photospheric model. The photospheric contribution spline is then subtracted from the observed data spline to get the excess luminosity. For objects that were not-detected with PACS we used the lowest upper limits available as detections to derive upper limits on the IR-excess. IR-excesses are shown in Table \ref{PACS_phot}. Since some sources were only detected at 70 $\rm \mu m$ (namely RECX 3, RECX 4) we also computed the fractional disc luminosities using only photometry $\rm \leq 70 ~\mu m$. Luminosities computed in this way are shown in brackets in Table \ref{PACS_phot}.
 
In Fig. \ref{Fig:irhist} we show histograms of the fractional disc luminosities ($\rm L_{IR}/L_{*}$, computed using the whole spectral range available) for detected sources and of upper limits for non-detected sources, with bin sizes computed using the Freedman-Diaconis rule \citep{Freedman1981}. For RECX 3 and RECX 4 the fluxes by \cite{Cieza2013} were used. There is small overlap between both distributions due to the very small fractional disc luminosities shown by RECX 3 and RECX 4 ($\rm L_{IR}/L_{*} \sim 7 \times 10^{-4}$). $\rm L_{IR}/L_{*} < 10^{-3}$ for most non-detected sources, with only two non-detected sources with excesses in the range $\rm 10^{-3}-10^{-2}$, namely RECX 17 and RECX 18. 

Among the five objects detected at 160 $\rm \mu m$ with PACS, only RECX 5 and RECX11 were previously detected at 160 $\rm \mu m$ with \textit{Spitzer}-MIPS by \cite{Gautier2008}. Our infrared-excess for RECX 5 is 0.085, in good agreement with \cite{Gautier2008}. For RECX 11, however, our fractional disc luminosity is more than four times larger. RECX 9, 15 and 16 were not detected at 160 $\rm \mu m$ by \textit{Spitzer}-MIPS, so our detections at 160 $\rm \mu m$ are crucial for deriving more accurate fractional disc luminosities. For RECX 9 the fractional disc luminosity remains unchanged, while for RECX 15 and 16 our excesses are 2.0 and 1.7 times larger, respectively. RECX 3 and RECX 4 show fractional disc luminosities typical of optically thin debris discs. Our fractional disc luminosities for these sources are larger than the values computed by \cite{Gautier2008} using MIPS at 24 $\rm \mu m$.

\cite{Gautier2008} showed that the fractional disc luminosity in RECX 15 is typical of T Tauri stars. RECX 5, 9, 14 and 16 show fractional disc luminosities in the range 0.03 to 0.1. \cite{Gautier2008} points out that these excesses are intermediate between those of optically thick discs surrounding young stars (like in RECX 15 and 11) and debris discs, and propose that a likely explanation for this intermediate excesses is an optically thick disc that has experienced flattening. We will further see in Sec. \ref{Subsec:flat} how flat discs can explain the lack of [OI] detections for those sources. The low-IR excesses ($\rm L_{IR}/L_{*}<10^{-3}$) shown by RECX 3 and RECX 4 are typical of optically thin debris discs.

In Fig. \ref{EtaCha_SED_comp} we compare the SED's of $\rm \eta$ Cha stars with a detected excess in the far- or mid-IR with the median SED of Taurus and Upper Scorpius from \cite{Mathews2013}. Both RECX 3 and RECS 4 SED's fall well below the median SED for Upper Scorpius. The SED of RECX 5 is intermediate between the Taurus and Upper Scorpius median SED. The SED of RECX 9 is similar to the median SED in Upper Scorpius, but shows a comparably smaller flux at 160 $\rm \mu m$. RECX 11 and RECX 14 SEDs' are similar to the median SED in Upper Scorpius up to 70 $\rm \mu m$, but show smaller fluxes at longer wavelengths. RECX 16 show mid-IR fluxes that are intermediate between Taurus and Upper Scorpius, but far-IR fluxes similar to those in Taurus. Finally, RECX 15 shows near- and mid-IR fluxes typical of discs in Taurus, and far-IR fluxes similar to Upper Scorpius discs. Overall, $\rm \eta$ Cha SEDs' show a variety of shapes for coeval discs, from Class II SEDs to debris discs. \cite{Sicilia-Aguilar2009} compared the median SED up to 24 $\rm \mu m$ of $\rm \eta$ Cha members with that of young stellar associations in the range 1 to 12 Myr. The authors computed different median SEDs for Solar-type stars and M stars and concluded that the evolution of the SED depends on the spectral type. 

\subsection{Gas diagnostics in $\eta$ Cha discs}

\subsubsection{Accretion and $\rm H_{\alpha}$ profiles}\label{Sec:Acc}
The equivalent width (EW) of the $\rm H_{\alpha}$ line has been traditionally used as an indicator of accretion. \cite{Lawson2004} performed the first study of accretion in $\rm \eta$ Cha. Based on the method by \cite{WhiteBasri2003}, they concluded that RECX 5, 9, 11 and 15 are still accreting at the age of $\rm \eta$ Cha. In a later study, \cite{Jay2006} concluded that RECX 9,11 and 13 are accreting, but RECX 5 is not. Therefore, observations of the same accretion indicator at different epochs can change the classification of a system, indicating that accretion is highly episodic. It must be noted, however, that $\eta$ Cha members are active stars, and therefore flaring can also, and most likely does, contribute to EW variations in the $\rm H{\alpha}$  line.

We have gathered literature and archival data for low- and high-resolution spectroscopy (respectively) covering the $\rm H_{\alpha}$ line for a total of 11 sources. We will first discuss the estimations we can provide for the sample considering only the low-resolution spectroscopy, and secondly, the more detailed analysis conducted for targets with multiple epoch high-S/N, high-resolution spectroscopy.

\begin{figure}[!h]
\begin{center}
\includegraphics[scale=0.4]{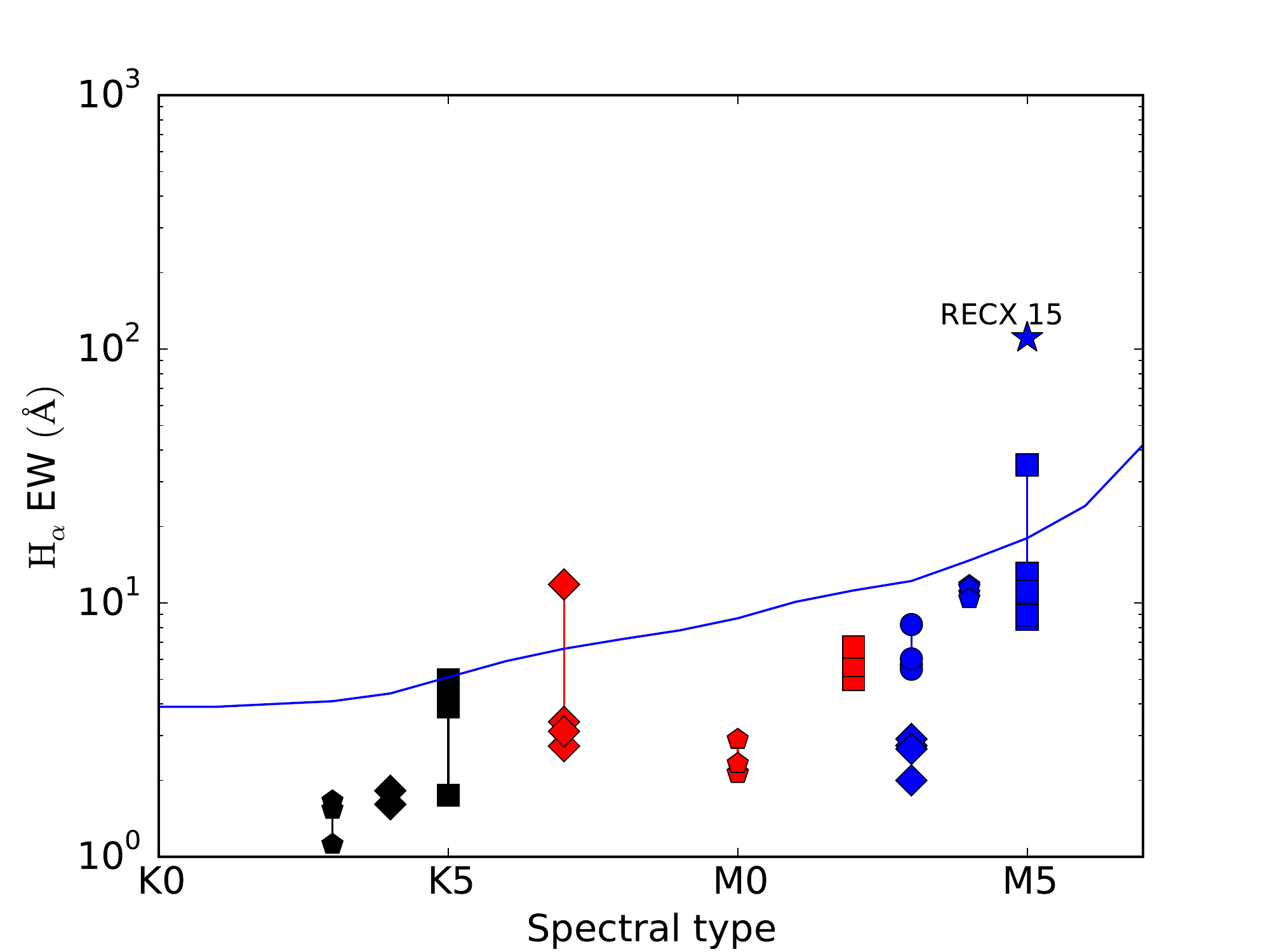}\\
\includegraphics[scale=0.4]{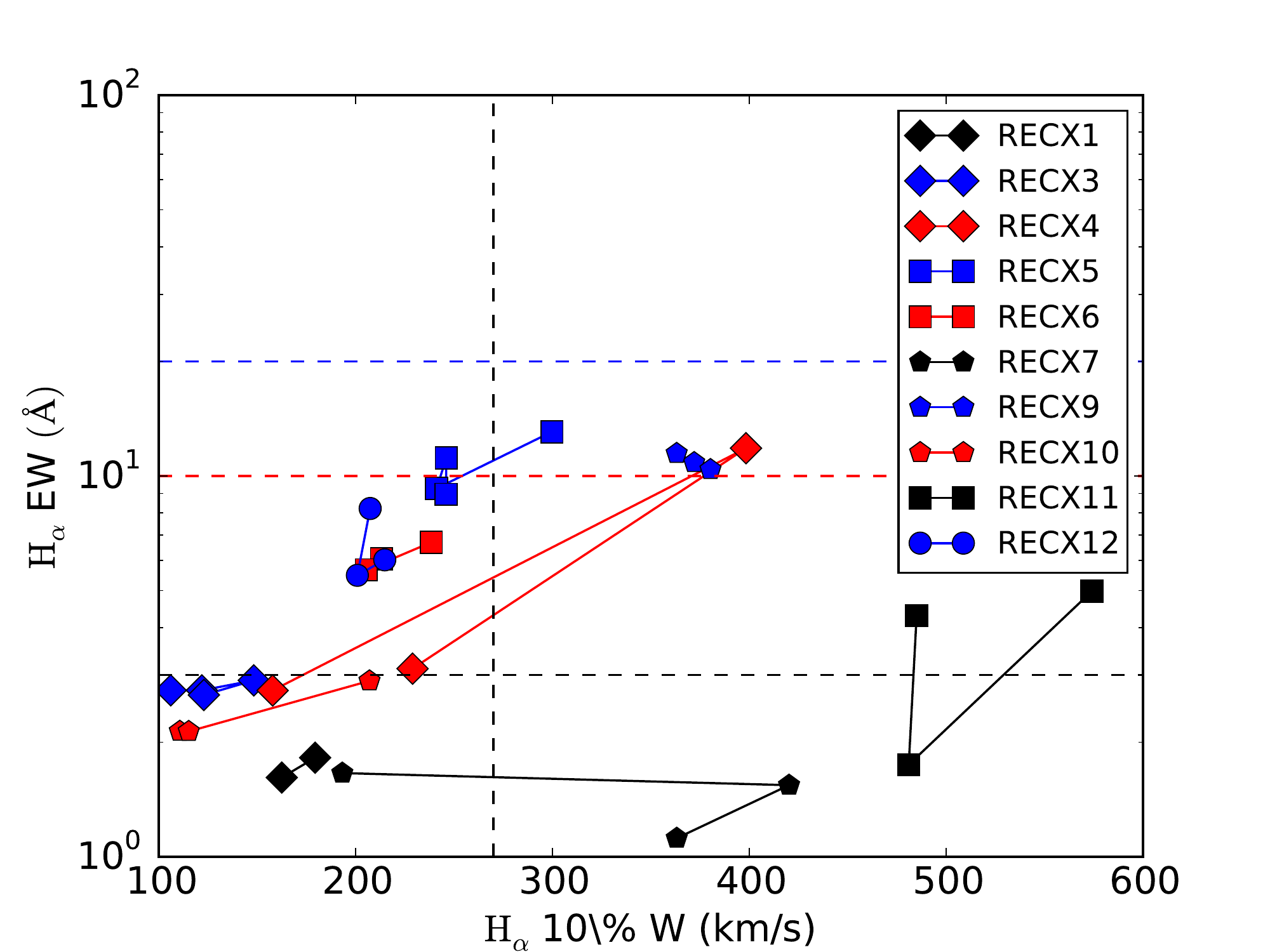}
\caption{Top: $\rm H_{\alpha}$  equivalent width  versus spectral type for $\rm \eta$ Cha members. The blue star depicts the position of RECX 15 in the diagram. The solid curve represents the saturation criterion. Bottom:  $\rm H_{\alpha}$  equivalent width  versus  $\rm H_{\alpha}$ 10\%-width. The vertical dashed line marks the position of the 10\% velocity width limit for accretion.}
   \label{Fig:accretion}
\end{center}
\end{figure}

We started our analysis by using the saturation criterion by \cite{Barrado2003}, where the EW boundary between accretors and non-accretors is a function of spectral type \citep[see][for a more detailed description of the method]{Barrado2003}. In Fig. \ref{Fig:accretion} we represent the $\rm H_{\alpha}$ equivalent widths of $\rm \eta$ Cha members versus the spectral type, with $\rm H_{\alpha}$ equivalent widths coming from   \cite{Lawson2004}  and \cite{Jay2006}, as well as from UVES spectra analyzed in this paper. We compile the equivalent widths from the literature and from the present work in Table \ref{Tab:Ha_EW}. We have multiple epochs for all the targets but one, namely RECX 15. Therefore, we can study variability in accretion indicators.  Among all the sources, RECX 4, RECX 5, RECX 11 and RECX 15 are actively accreting according to the saturation criterion in at least one epoch.  RECX 4, RECX 5 and RECX 11 show highly variable $\rm H_{\alpha}$ emission and no [OI] emission at 63 $\rm \mu m$. For RECX 4, the average EW from all epochs is (4.4$\rm \pm$3.7) $\rm \AA$, with a maximum value of 9.44 $\rm \AA$. In contrast, the average EW from non-accretion epochs is (2.8$\rm \pm$0.5) $\rm \AA$. In the case of RECX 5, the average EW from all epochs is (13$\rm \pm$10) $\rm \AA$, with a maximum EW of 35.0 $\rm \AA$, while the average EW from non-accretion epochs is $\rm 9.2\pm1.2~\AA$. RECX 11 is classified as an accreting system in one epoch, and falls near the boundary in the study by \cite{Lawson2004}. Therefore, we have indications of episodic accretion in at least three $\rm \eta$ Cha members. \cite{Murphy2011} concluded that the variability of the $\rm H_{\alpha}$ line in $\rm \eta$ Cha covers both short and long timescales, in agreement with what we see in the sample of UVES observations. RECX 15 is the only system detected in [OI] emission at 63 $\rm \mu m$. RECX 9 was classified as an accreting system both by  \cite{Lawson2004}  and \cite{Jay2006}, but it does not satisfies the saturation criterion. 

\begin{table}
\caption{Compilation of equivalent widths, velocity widths at 10\% of the $\rm H_{\alpha}$ line and mass accretion rates for $\rm \eta$ Cha members}             
\label{Tab:Ha_EW}      
\centering          
\begin{tabular}{llllll}     
\hline \hline  
Source  & EW & W-10\% & log$\dot M_{acc}$  & Epoch  & Ref. \\
\hline
RECX & ($\rm \AA$ )& (km/s) & (M$\rm_{\odot}$/yr) &(dd/mm/yy) & \\
\hline
1 & -1.2 & 103 & -11.9 & & 2 \\
1 &  -1.82 & 179 & -11.2 & 15/10/2011 & 1 \\
1 &  -1.61 & 162 & -11.3 & 22/10/2011 & 1 \\
3 & -2.0 & 116 & -11.8 &  & 2 \\
3 & -2.2 & 85 & -12.1 & & 3 \\
3 & -2.74 & 106 & -11.9 & 15/10/2011 & 1 \\
3 & -2.74 & 122 & -11.7 &  22/10/2011 & 1 \\
3 & -2.91 & 148 & -11.4 & 13/12/2011 & 1 \\
3 & -2.66 & 123 & -11.7 &  30/12/2011 & 1 \\
4 & -3.4 & 147 & -11.5 & & 2  \\
4 & -2.3 & 105 & -11.9 & & 3  \\
4 & -2.73 & 158 & -11.4 & 22/10/2011 & 1 \\
4 & -11.83&  398 & -9.02 &  30/12/2011 & 1 \\
4 & -3.12 & 229 & -10.7 &  06/01/2012 & 1 \\
5 & -8.6 &  194 & -11.0 &  & 2  \\
5 & -35.0 &  330 & -9.7 &  & 3  \\
5 & -13.06 & 300 & -10.0 & 13/12/2011 & 1 \\
5 & -9.28 & 241  & -10.6 & 30/12/2011 & 1 \\
5 & -11.15 & 246 & -10.5 & 06/01/2012 & 1 \\
5 & -8.95 & 246 & -10.5 & 07/01/2012 & 1\\
6 & -5.0 &  145 & -11.5 & &  2  \\
6 & -3.6 &  155 & -11.4 & &  3  \\ 
6 & -6.06 & 213 & -10.8 & 30/12/2011 & 1 \\
6 & -5.67 & 205 & -10.9 & 07/01/2012 & 1 \\
6 & -6.70 & 238 & -10.6 & 09/01/2012 & 1 \\
7 & -1.0 & 291 & -10.1 & &  2 \\
7 & -0.4 & 120 & -11.7 & &  3 \\
7 & -1.12 & 363 & -9.4 & 15/12/2011 & 1 \\
7 & -1.54 & 420 & -8.8 & 07/01/2012 & 1 \\
7 & -1.66 & 193 & -11.01 & 09/01/2012 & 1 \\
9 & -11.7 &  389 & -9.1 & &  2 \\
9 & -10.0 &  300 & -10.0 &  & 3 \\
9 & -10.85 & 372 & -9.3 & 07/01/2012 & 1 \\
9 & -11.46 & 363 & -9.4 & 09/01/2012 & 1 \\
9 & -10.40 & 380 & -9.2 &  10/01/2012 & 1 \\
10 & -1.2 &  102 & -11.9 & &  2 \\
10 & -1.0 &  110 & -11.8 & &  2 \\
10 & -2.14 & 112 & -11.8 & 15/12/2011 & 1 \\ 
10 & -2.13 & 115 & -11.7 & 06/01/2012 & 1  \\
10 & -2.90 & 207 & -10.9 & 07/01/2012 & 1  \\
11 & -3.9  &  345 & -9.5 & &  2 \\
11 & -3.0  &  330 & -9.7 & &  3 \\
11 & -4.98 & 574 & -7.3 & 06/01/2012 & 1 \\ 
11 & -1.75 & 481 & -8.2 &07/01/2012 & 1\\ 
11 & -4.30 & 485 & -8.2  & 09/01/2012 & 1\\
12 & -5.7 & 154 & -11.4 & &  2 \\
12 & -4.2 & 160 & -11.3 & &  3 \\
12 & -8.22 & 207 & -10.9 & 15/12/2011 & 1 \\
12 & -5.48 & 201 & -10.9 & 06/01/2012 & 1 \\
12 & -6.02 & 215 & -10.8 & 07/01/2012 & 1 \\
15 & -90 & 530 & -7.7 &  & 3 \\
 \hline                  
\end{tabular}
\tablefoot{References are (1): this work; (2): \cite{Jay2006}; (3): \cite{Lawson2004} }
\end{table}

\begin{figure*}[!t]
\begin{center}
\includegraphics[scale=0.2,trim=0mm 0mm 0mm 0mm,clip]{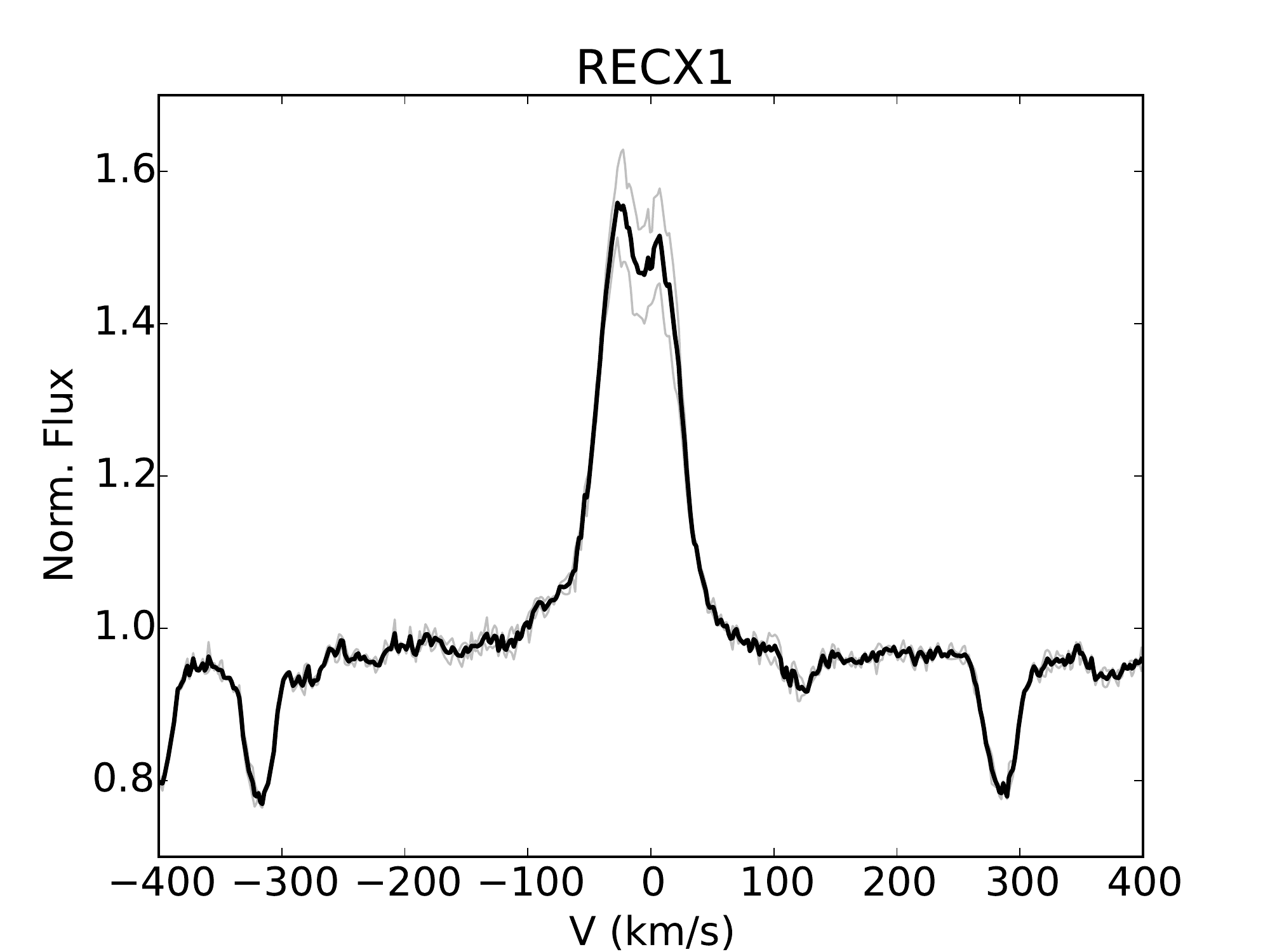}\includegraphics[scale=0.2,trim=0mm 0mm 0mm 0mm,clip]{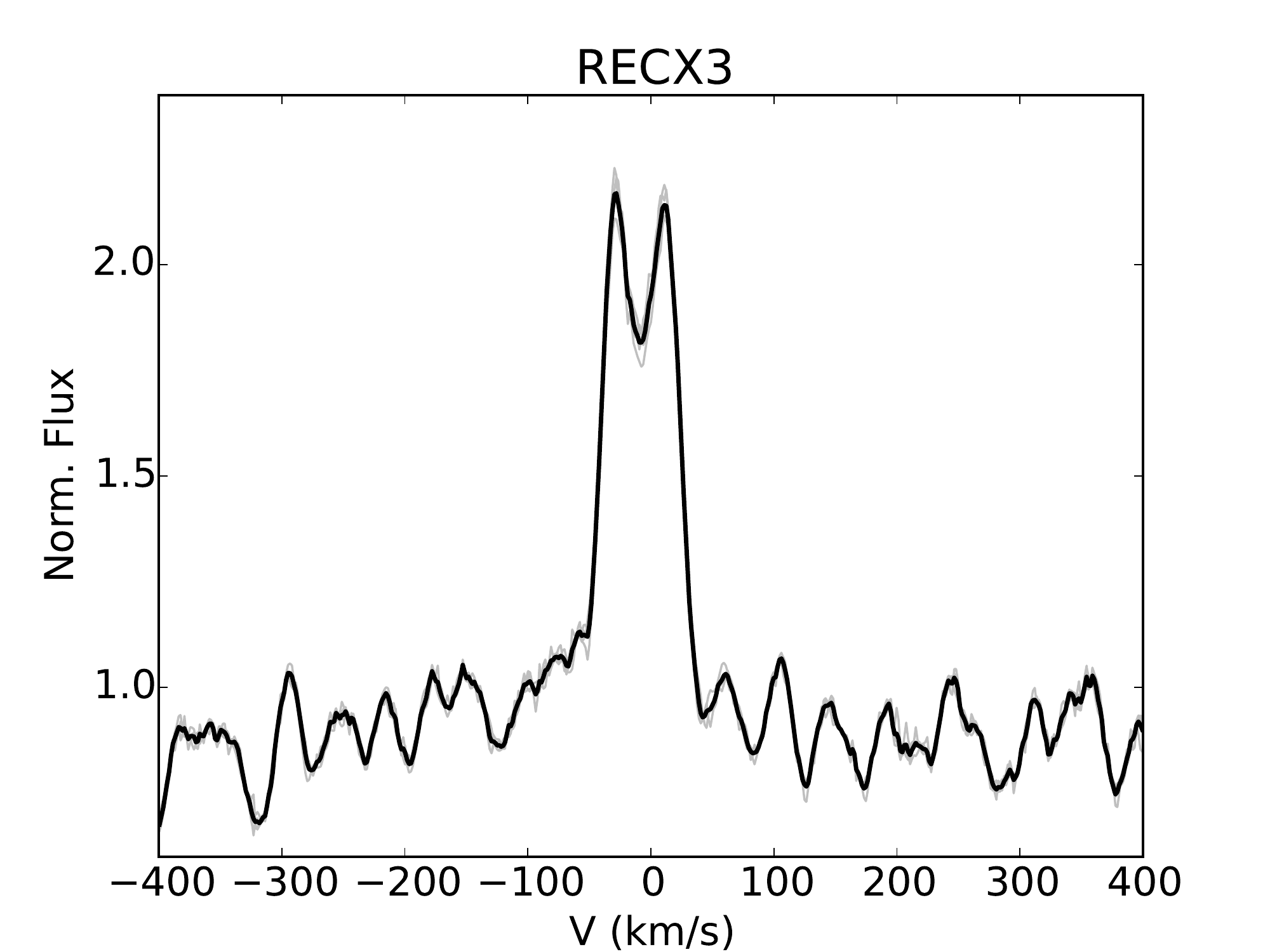}\includegraphics[scale=0.2,trim=0mm 00mm 0mm 0mm,clip]{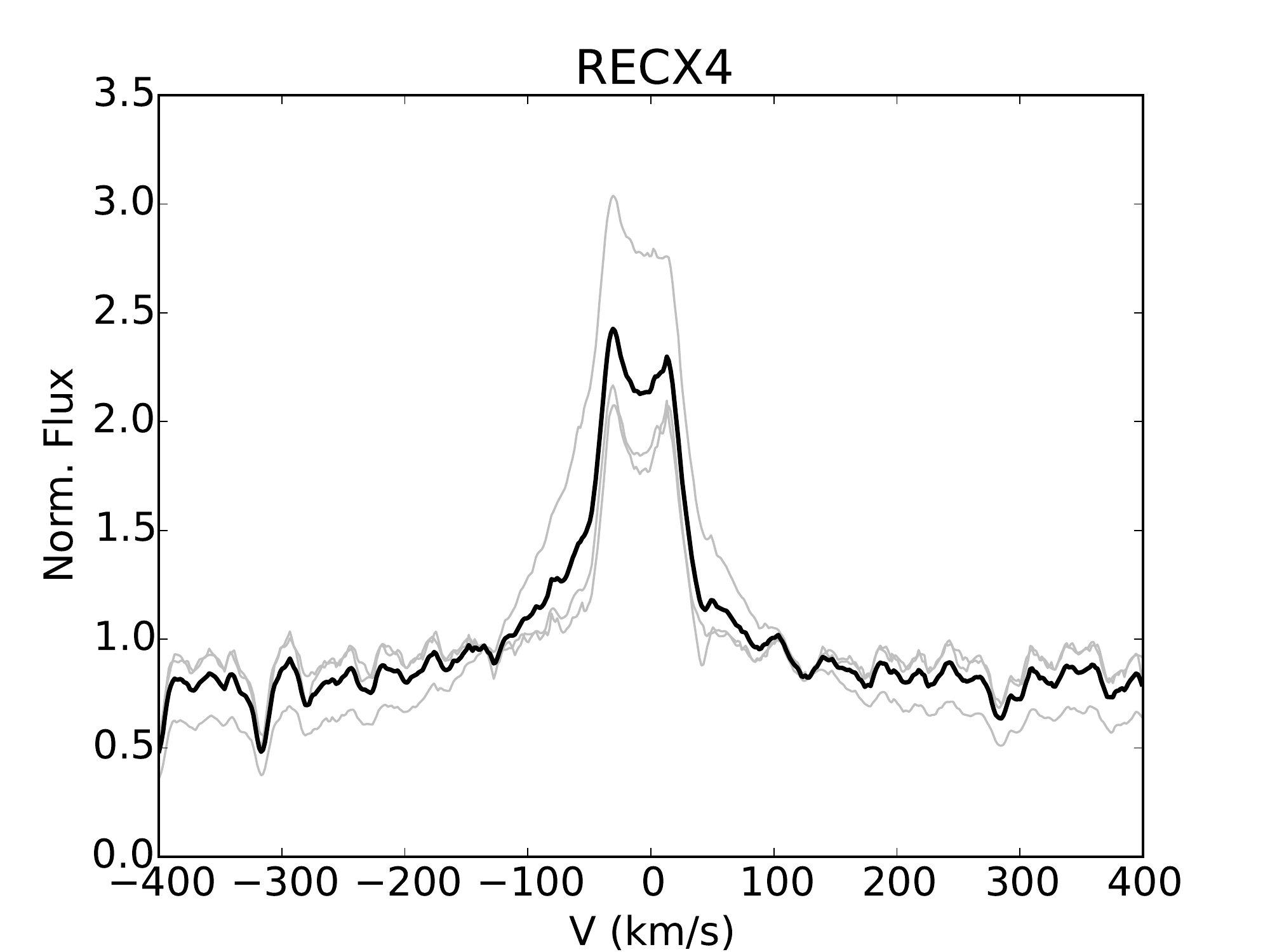}\includegraphics[scale=0.2,trim=0mm 00mm 0mm 0mm,clip]{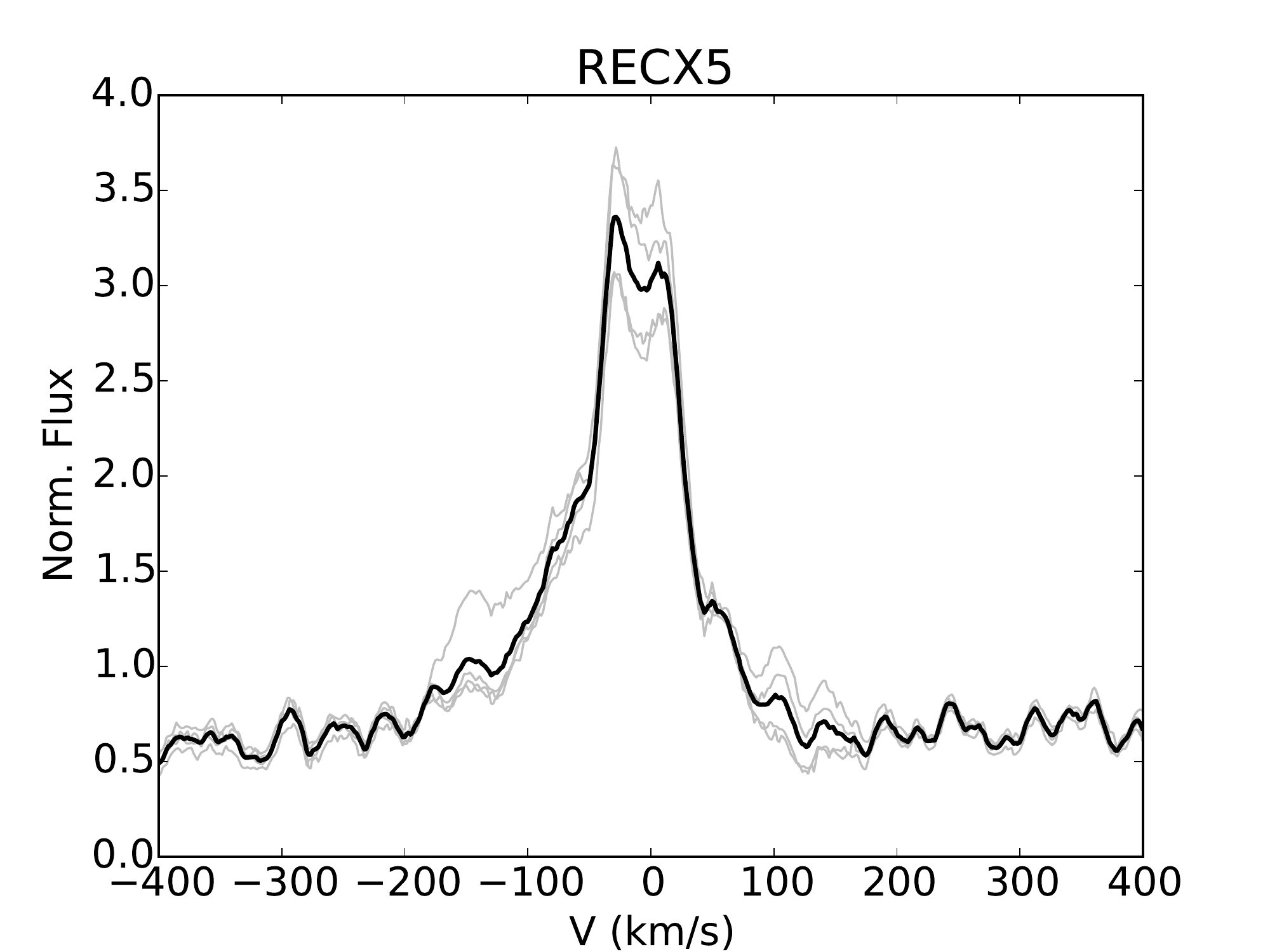}\\
\includegraphics[scale=0.2,trim=0mm 00mm 0mm 0mm,clip]{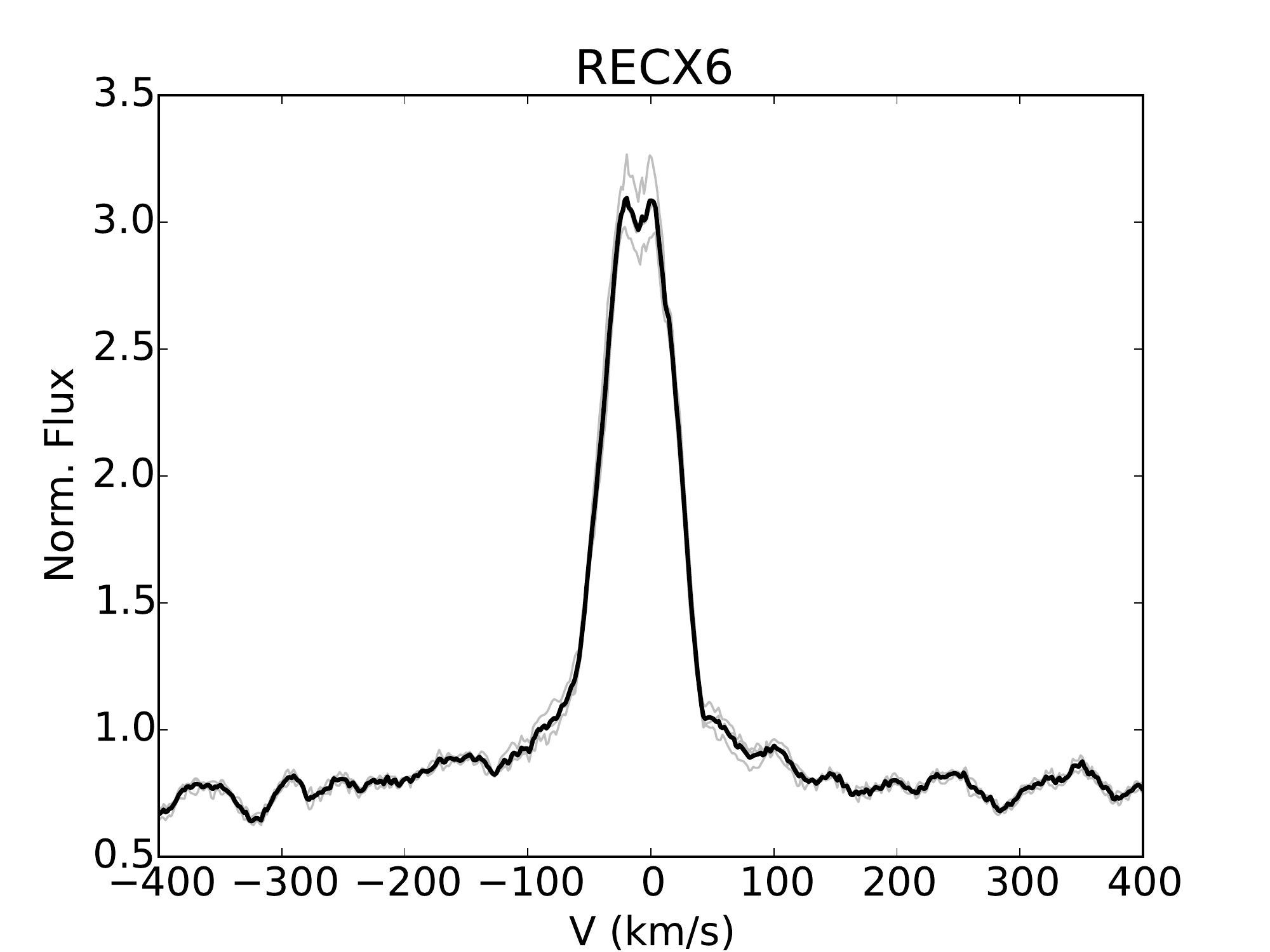}\includegraphics[scale=0.2,trim=0mm 00mm 0mm 0mm,clip]{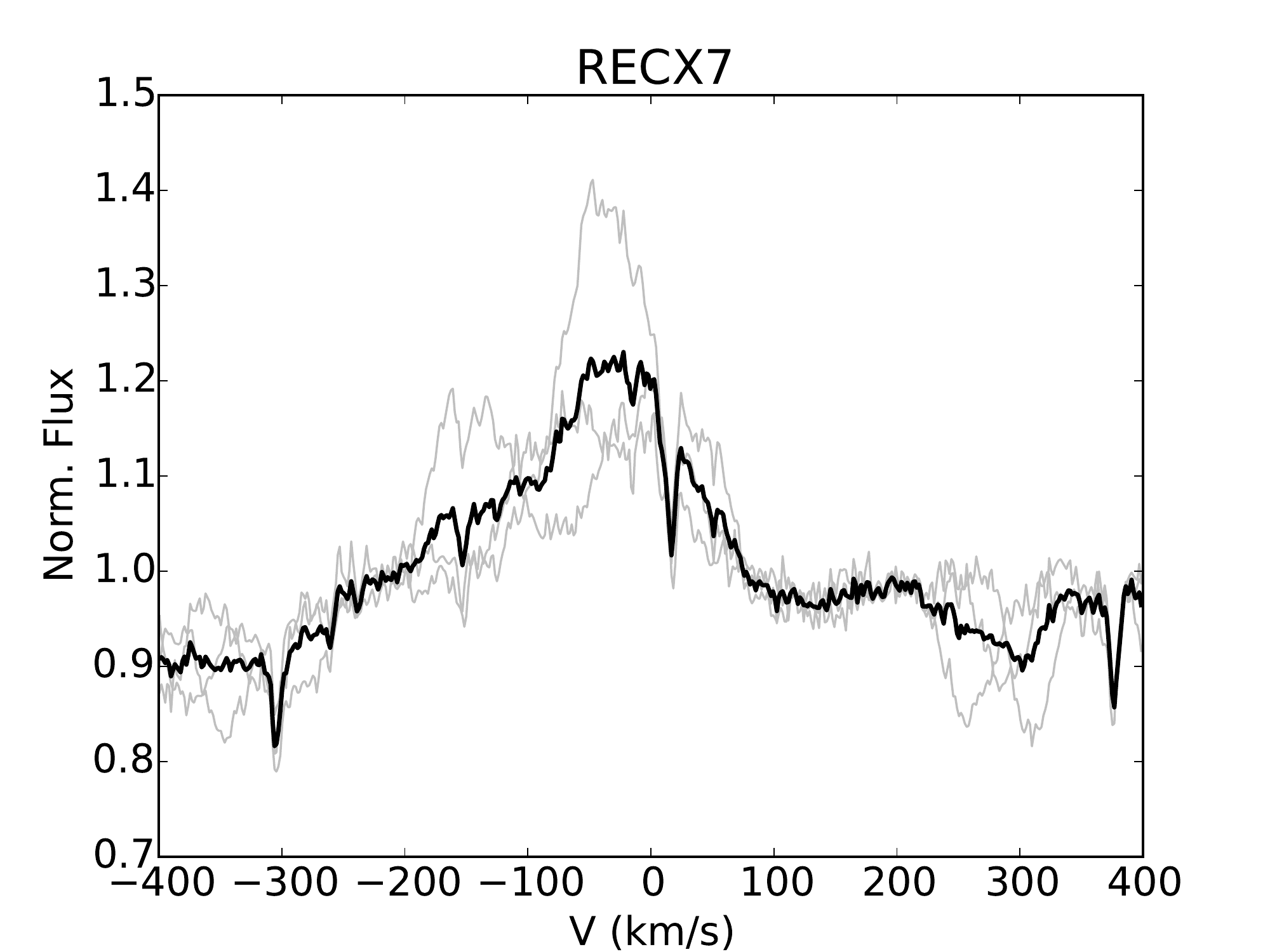}\includegraphics[scale=0.2,trim=0mm 00mm 0mm 0mm,clip]{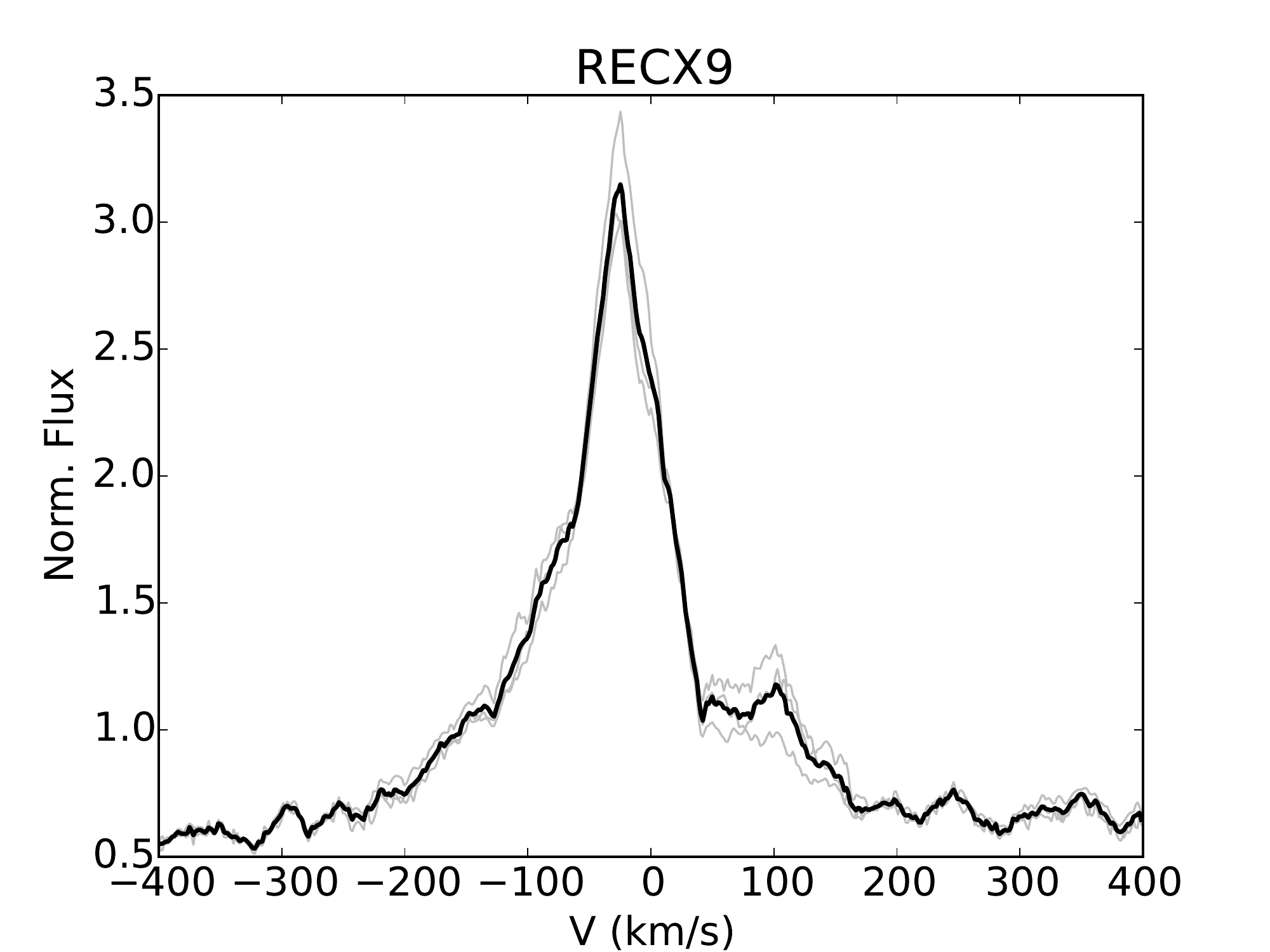}\includegraphics[scale=0.2,trim=0mm 00mm 0mm 0mm,clip]{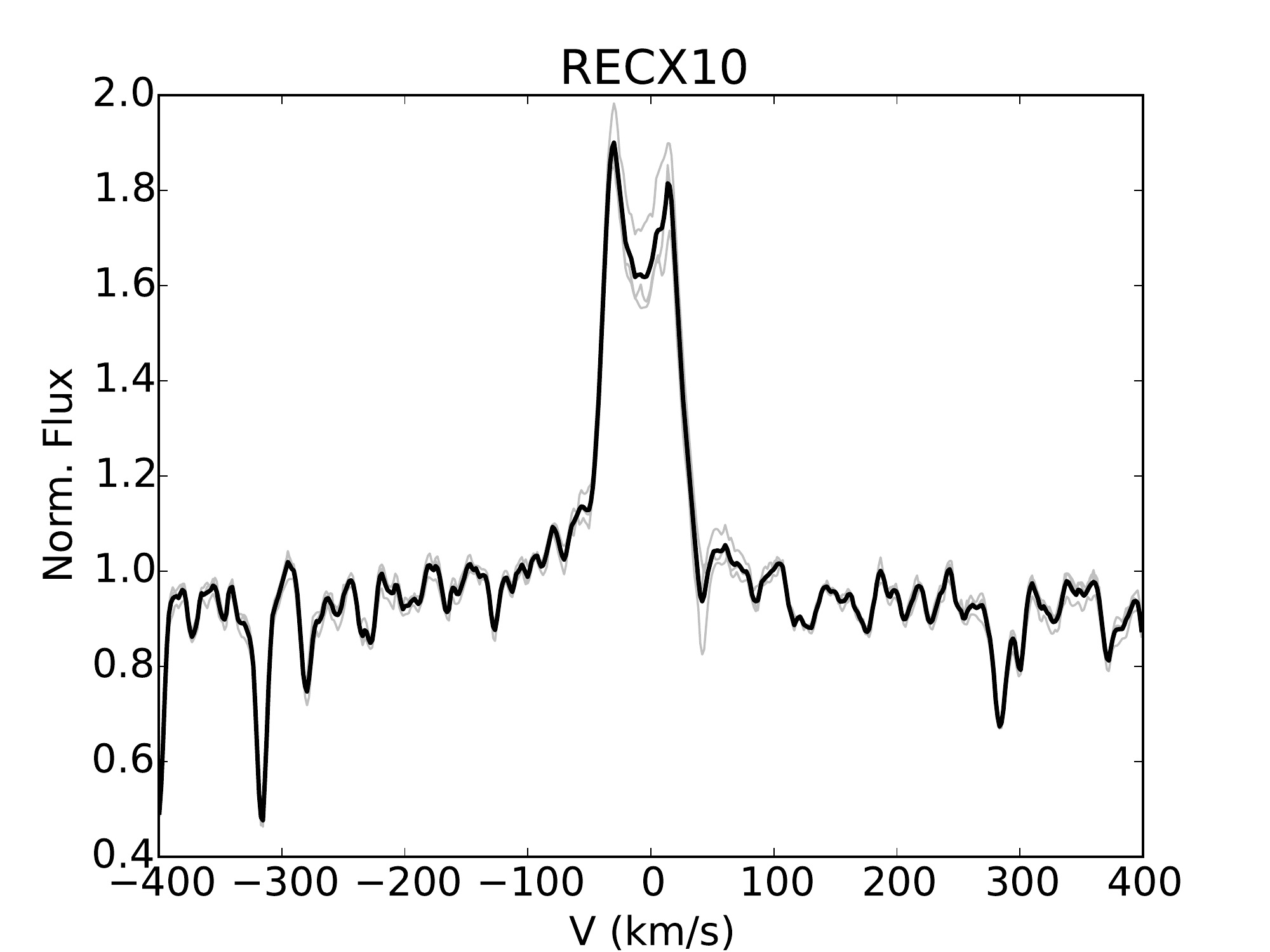}\\
\includegraphics[scale=0.2,trim=0mm 00mm 0mm 0mm,clip]{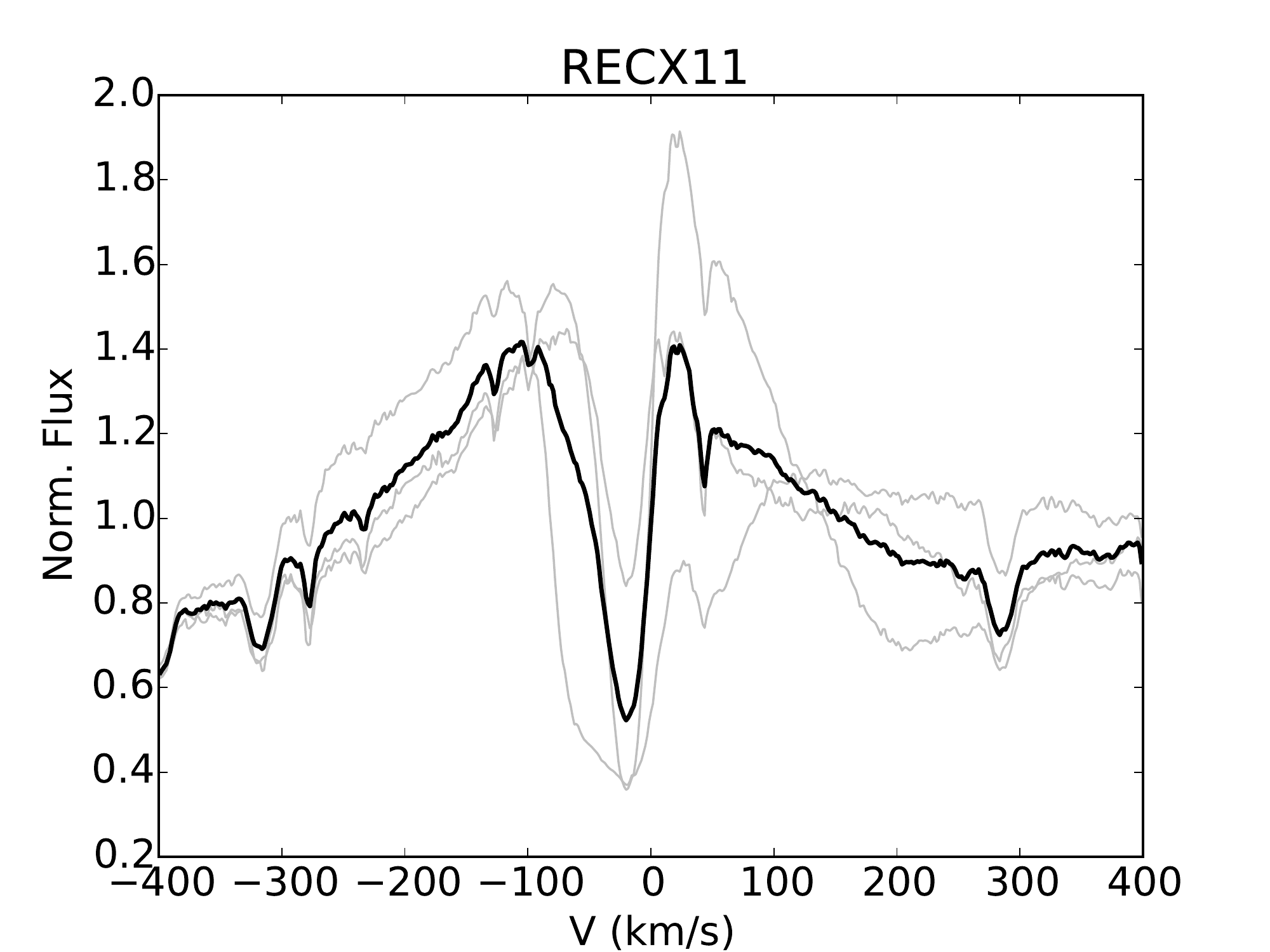}\includegraphics[scale=0.2,trim=0mm 00mm 0mm 0mm,clip]{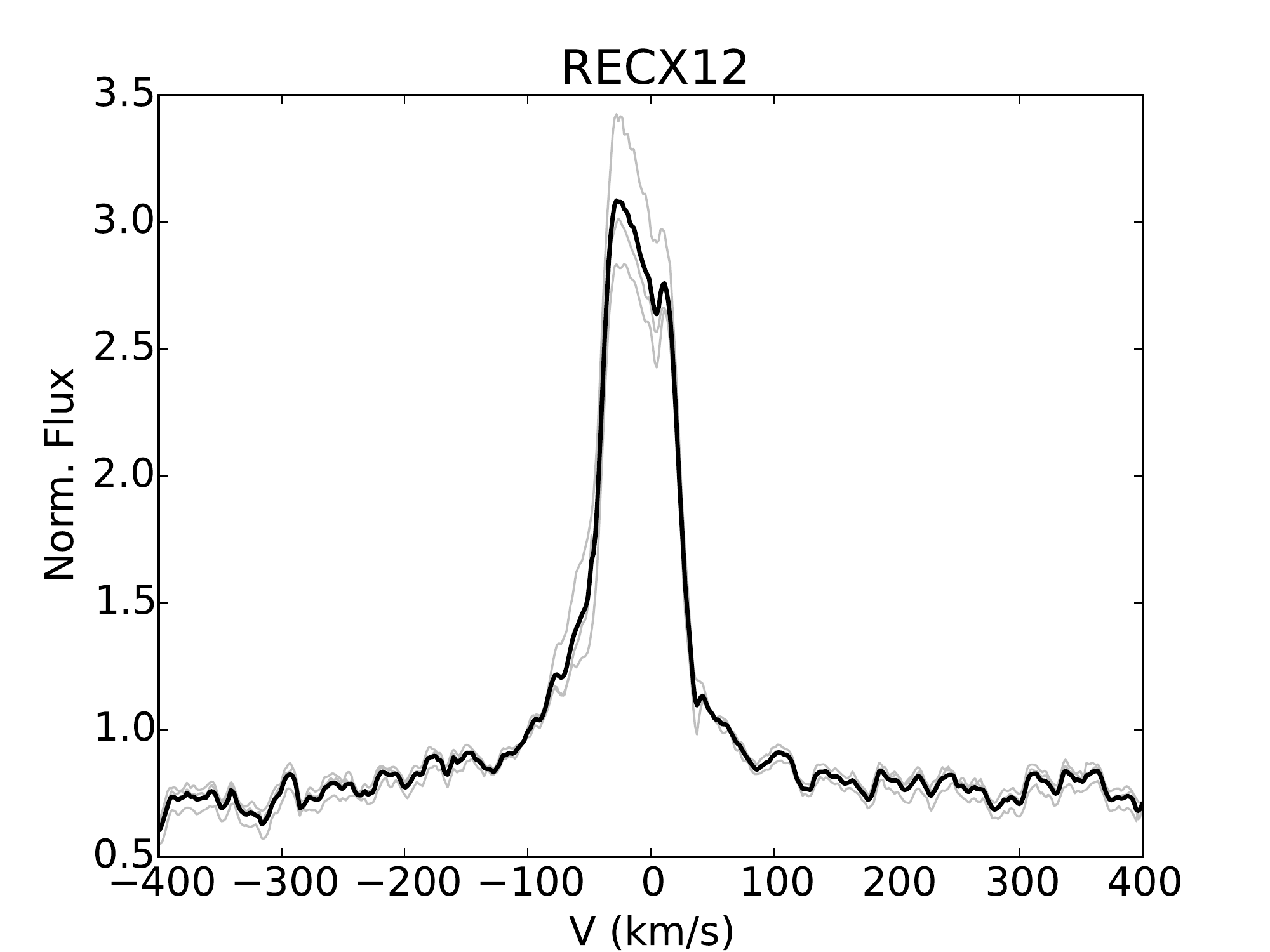}
\caption{UVES spectra of $\eta$ Cha members at 6560 $\rm \mu m$ centreed on the $\rm H_{\alpha}$ line. Gray curves show individual epochs, while the black curve shows the average spectrum from all epochs.}
   \label{EtaCha_UVES_Ha}
\end{center}
\end{figure*}

For 10 sources in the sample we have multi-epoch UVES archival observations of $\rm H_{\alpha}$ at high resolution. The spectra of these sources around the $\rm H_{\alpha}$ line are shown in Fig. \ref{EtaCha_UVES_Ha}. We used  these observations to analyze accretion signatures with more details. The most prominent facts are the variety of profiles present  and the high variability of the $\rm H_{\alpha}$ line, regardless of the shape of the profile. We distinguish three groups of spectra according to the shape of the $\rm H_{\alpha}$ emission. First, there is a group of simple double-peaked features, with peaks that are mostly symmetric, typical of chromospheric activity \citep[see, e. g.][]{Frasca2008}.  This group includes RECX 1, RECX 3, RECX 6, RECX 10 and RECX 12. None of the sources in the first group have a detected IR excess  neither is classified as actively accreting. Furthermore, their SEDs are photospheric at least until 30 $\rm \mu m$, an indication that if any disc exists, it must be cold, with a large inner gap devoid of dust.

 The second group consist of features with a P Cygni profile, indicative of the presence of an outflow, and RECX 11 is the only member. We recall that RECX 11 shows a prominent excess starting at the near-IR. 
 
 The third group is made of single- or double-peaked profiles with high velocity components and high variability, and includes RECX 4, RECX 5 and RECX 9. In the third group we find the most interesting profiles. First, RECX 4 shows high variability between different epochs, including one with high velocity wings (EW = 9.4 compared to the average value of $\rm \langle EW \rangle = 2.5 \pm 0.1$ from the other two UVES epochs), with velocities as high as 100 km/s, both in the blue- and red-shifted part of the spectrum. This velocity is not high enough to conclude whether it is of stellar origin, or is it due to winds and accretion. RECX 5 shows a blue-shifted component at $\rm \sim -150 km/s$ in one epoch, indicative of accretion. Finally, RECX 9 shows a single peaked profile with a high-velocity blue-shifted wing at $\rm \sim -80 km/s$ and an isolated red-shifted feature around 100 km/s. The shape of the feature can be due to a very deep self-absorption in the red side of the spectra. All the stars in this group show mid- IR excess, while RECX 5 and RECX 9 also show far-IR excess. The variability of the line and the presence of high-velocity wings is likely due to the presence of infalling gas.
 
The spectra for RECX 7 are very hard to interpret, and so we treat them separately. The different epochs show prominent variability. RECX 7 has no detected excess, therefore the variability is purely chromospheric. Furthermore, RECX 7 is a double-lined spectroscopic binary  \citep[SB2, ][]{Lyo2003}, a fact that explains the broad shape of the line compared to its low equivalent width, as well as variability along the orbital phase.
 
 To complement the qualitative comments in the previous paragraphs, we show in the bottom panel of Fig. \ref{Fig:accretion} equivalent widths versus widths at 10 \% for the $\rm H_{\alpha}$ line, together with the criterion by \cite{WhiteBasri2003}. Five $\rm \eta$ Cha members are accreting according to this criterion, namely RECX 4, RECX 5, RECX 7, RECX 9 and RECX 11. RECX 4 and RECX 5 show 10\% widths in agreement with ongoing accretion in just on epoch, RECX 7 in two epochs, and RECX 9 and  RECX 11 in all the available epochs. Interestingly, RECX 9 is not classified as an accretor when equivalent widths are used, but falls near the saturation criterion. The final list of sources with indications of accretion is then RECX4, RECX 5, RECX 7, RECX 9, RECX 11 and RECX 15, and in at least three sources we have evidence of variable accretion, namely RECX 4, RECX 5, and RECX 11. Using the relation by \cite{Natta2004}, we computed accretion rates for $\rm \eta$ Cha members. The resulting rates are shown in Table \ref{Tab:Ha_EW}, and range from $\rm \sim  10^{-12}~M_{\odot}/yr$ for non-accretors to $\rm \sim  10^{-8}~M_{\odot}/yr$ for sources that are actively accreting.
 
\begin{figure}[!h]
\begin{center}
\includegraphics[scale=0.3,trim=0mm 0mm 0mm 0mm,clip]{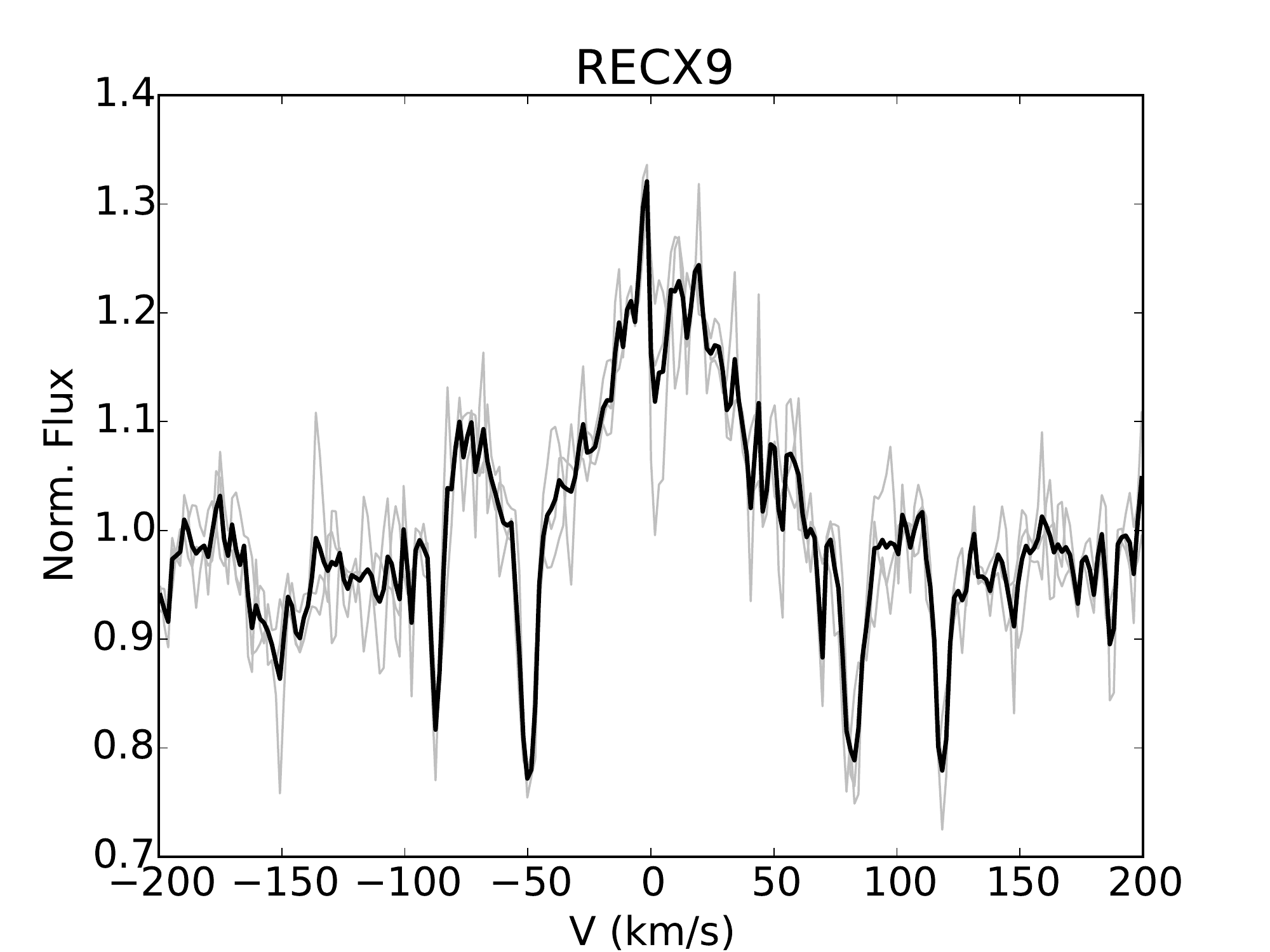}\\
\includegraphics[scale=0.3,trim=0mm 0mm 0mm 0mm,clip]{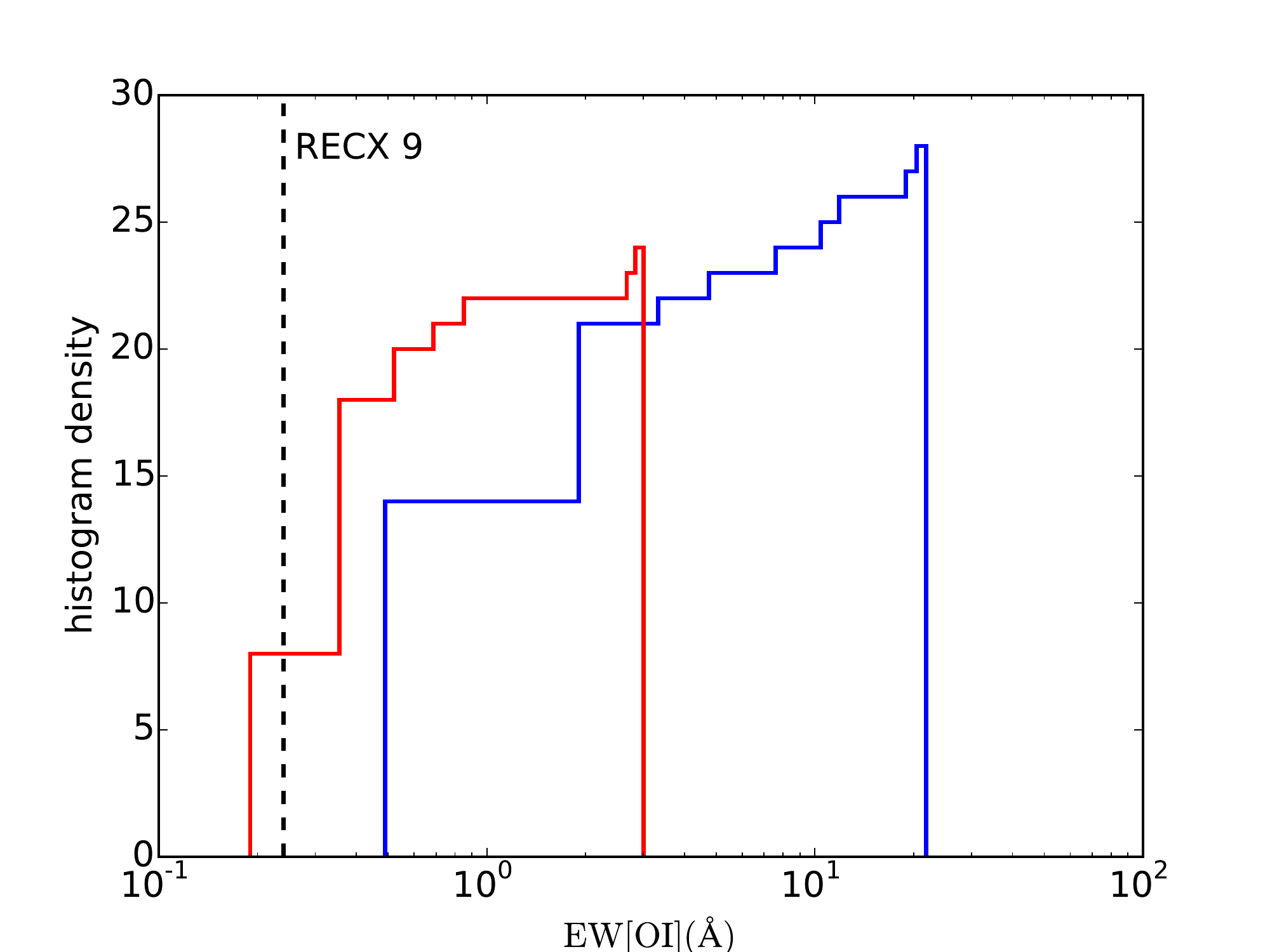}
\caption{Top: UVES spectra of RECX 9 members at 6300 $\rm \AA$ centreed on the [OI] line. Gray curves show individual epochs, while the black curve shows the average spectrum from all epochs. Bottom: cumulative histogram of [OI] EW at 6300 $\rm \AA$ from \cite{Hartigan1995}. Sources with an identified jet are shown in blue, while sources without a jet signature are shown in red. The vertical dashed line marks the position of RECX 9 [OI] EW at 6300 $\rm \AA$.}
   \label{RECX9_OI_6300}
\end{center}
\end{figure}

To further investigate gas properties and dynamics, we looked for [OI] emission at 6300 $\rm \AA$. However, only RECX 9 showed hints of emission (see Fig. \ref{RECX9_OI_6300}). The emission is detected in all three epochs, with $\rm EW=0.24,0.23~and~0.25~ \AA$, with an average error of 0.05 $\rm \AA$. \cite{Hartigan1995} studied [OI] emission at 6300 $\rm \AA$ in a sample of Classical and Weak T Tauri stars. In the bottom panel of Fig. \ref{EtaCha_UVES_Ha} we show cumulative histograms of the [OI] EW at 6300 $\rm \AA$. The sample is divided onto jet and non-jet sources following \cite{Howard2013}. EWs for jet sources are in the range 0.5 to 22 $\rm \AA$, with a typical value around 4 $\rm \AA$, while non-jet sources show values in the range 0.2 to 3.0 $\rm \AA$. We therefore conclude that [OI] emission in RECX 9 is not associated to a jet. 

\subsubsection{Flattened discs and [OI] emission at 63 $\rm \mu m$}\label{Subsec:flat}
Three parameters govern the [OI] flux at 63 $\rm \mu m$ in a protoplanetary disc: the total gas mass, the UV radiation field and the flaring geometry of the disc \citep{Woitke2010}. Therefore, the lack of [OI] at 63 $\rm \mu m$ does not necessarily imply that they are gas empty. Rather, three alternative explanations are possible. First, these discs can be particularly flat, in agreement with the argument used by \cite{Gautier2008} to explain intermediate IR excesses in $\eta$ Cha sources.  \cite{Woitke2010} showed that for the same disc model, changing the flaring index from 1.0 to 1.2 results in an increase of almost two orders of magnitude in [OI] flux for the high mass discs ($\rm M_{gas} > 10^{-4}~M_{\odot}$). \cite{Keane2014} studied a sample of transitional discs observed with PACS and concluded that the lower [OI] luminosity in transitional discs can result either from a decrease in flaring or from a decrease in the gas-to-dust ratio, which strengthens the option of flat discs to explain the lack of [OI] emission. An alternative explanation is a very weak UV field. A change in [OI] line flux similar to the one produced by the change in flaring index is produced when moving from low UV models ($\rm f_{UV}=0.001$, where $\rm f_{UV}$ is the UV excess, defined as $\rm f_{UV}=L_{UV}/L_{*}$) to  high UV models \citep[$\rm f_{UV}=0.1$, see Fig. 2 in ][]{Woitke2010}. The third option is that these discs are devoid of gas. A combination of the three scenarios can also explain the lack of [OI] emission.

A specially interesting case is that of $\rm \eta ~Cha$ members where we detected signs of accretion yet no [OI] emission at 63 $\rm \mu m$ is observed (RECX 4, RECX 5, RECX 9 and RECX 11). A similar case is \object{TWA~03A}, also known as Hen 3-600, where no [OI] emission is detected while $\rm H_{\alpha}$ emission agrees with an actively accreting system \citep{Riviere2013}. The detection of active accretion favours the possibility of a high $\rm f_{UV}$ disc, since the accretion luminosity correlates with the $\rm f_{UV}$ \citep{Yang2012}. However, the correlation is subject to large scatter, and the previous conclusion is only valid from a statistical point of view, leaving open the possibility of active accretion with low $\rm f_{UV}$ in some systems. However, it is very unlikely that all the active accretors lacking [OI] emission show very low $\rm f_{UV}$ ($\rm \sim 0.001$). The most likely explanation for the lack of [OI] emission in this sources is then a flat disc. Alternative explanations have to face the problem of active accretion still ongoing, even if it is episodic.

Confirmation of the conclusions regarding transitional and/or flattened discs might come from high spatial resolution ALMA observations in the continuum, to get insight onto the dust geometry, and  certain species like CO, which help to understand the gas distribution. The power of ALMA to solve the geometry of the disc by means of thermal emission and line emission maps has been demonstrated in the last years \citep[see e. g., ][]{deGregorio2013, vanderMarel2013}.

\subsubsection{Gas content in RECX 15}
RECX 15 was already known to show [OI] emission at 63 $\rm \mu m$ \citep{Woitke2011}. However, no other far-IR line emission was detected in the source. According to our previous section, the system is actively accreting. Furthermore, the work of \cite{Lawson2004} indicates the presence of a blue-shifted component, likely attributable to a wind, and derives a mass accretion rate of $\rm 10^{-9}~M_{\odot}/year$. \cite{Ramsay2007} detected $\rm H_{2}$ emission at 2.1218 $\rm \mu m$ and derived a mass of hot gas of $\rm 5 \times 10^{-9}~M_{\odot}$, almost one order of magnitude larger than the mass of hot gas that they derive for TW Hya. The authors concluded that the line was emitted by gas in Keplerian rotation at 2 AU from the central star. Observations of other six $\rm \eta~Cha $ members were reported in the same study, with negative results, including RECX 5, 9 and 11. 

Observations of the source at multiple wavelengths were used to  model in detail the disc around RECX 15 by \cite{Woitke2011}, including PACS line observations at 63 $\rm \mu m$ and range spectroscopic observations, PACS photometric observations and observations of CO with APEX at 867 $\rm \mu m$, amongst others. However, due to the high number of non-detections, the authors could only derive a range of gas masses compatible with observations, $\rm 5 \times10^{-5} < M_{gas}/M_{\odot} <3 \times 10^{-3}$, and a gas-to-dust ratio of more than 2000.

The analysis of [OI] emission at 63 $\rm \mu m$ in Taurus sources by \cite{Howard2013} showed that it is correlated with the continuum emission at 63 $\rm \mu m$, and that for the same continuum level, outflow sources show a higher [OI] flux. With a flux density of 0.204 Jy at 70 $\rm \mu m$ and line flux of $\rm 2.4 \times 10^{-17}~W/m^{2}$, RECX 15 falls in the outflow region of the plot. The presence of blue-shifted components in the [OI] 6300 $\rm \AA$ further suggest the presence of a jet or outflow \citep{Woitke2011}. Jet sources can show [OI] at 63 $\rm \mu m$ extended emission along the jet direction \citep{Podio2012}. However, we only detect [OI] emission in the central spaxel (see Fig. \ref{RECX15_contours}).

\subsection{Multiplicity in $\rm \eta$ Cha} 
Multiplicity can have a huge impact on disc evolution \citep{Kraus2012}. In order to better place a context for the peculiar characteristics of the $\eta$ Cha discs we have performed a detailed study of the multiplicity of our sample. Previous works looked for the presence of binaries in $\rm \eta$ Cha. As pointed out by \cite{Bouwman2006}, the presence of a disc seems to be linked to single stars. In the following we overview the results from previous studies:
\begin{itemize}
\item  \cite{Kohler2002} looked for visual binaries among $\rm \eta$ Cha members, detecting companions for RECX 1 and RECX 9 at physical separations of $\rm \sim$17 and $\rm \sim$22 AU, respectively.
\item  \cite{Lyo2003} showed that RECX 7 is a spectroscopic binary with a projected physical separation of $\rm \sim$0.1 AU.
\item \cite{Andersen1975} concluded that RECX 8 is a spectroscopic binary, with a projected separation of $\rm \sim$0.04 AU.  
\item \cite{Brandeker2006} shows that RECX 12 is a visual binary. From the asymmetric profile of the NACO image the authors derive a separation of $\rm \sim$0$\arcsec$.04. 
\item \cite{Brandeker2006} presented observations of all $\rm \eta ~Cha$ members except RECX 4, looking for visual companions using adaptive optics (AO) imaging with the nasmyth adaptive optics system (NACO) instrument.  These observations were sensitive down to substellar and planetary-mass companions at angular separations approximately 0.3$\arcsec$ and 0.5$\arcsec$. The authors concluded that there is a deficit of wide binaries at separations $\rm >$20 AU in the cluster.
\end{itemize}

\begin{figure}[!h]
\begin{center}
\includegraphics[scale=0.3,trim=0mm 0mm 0mm 0mm,clip]{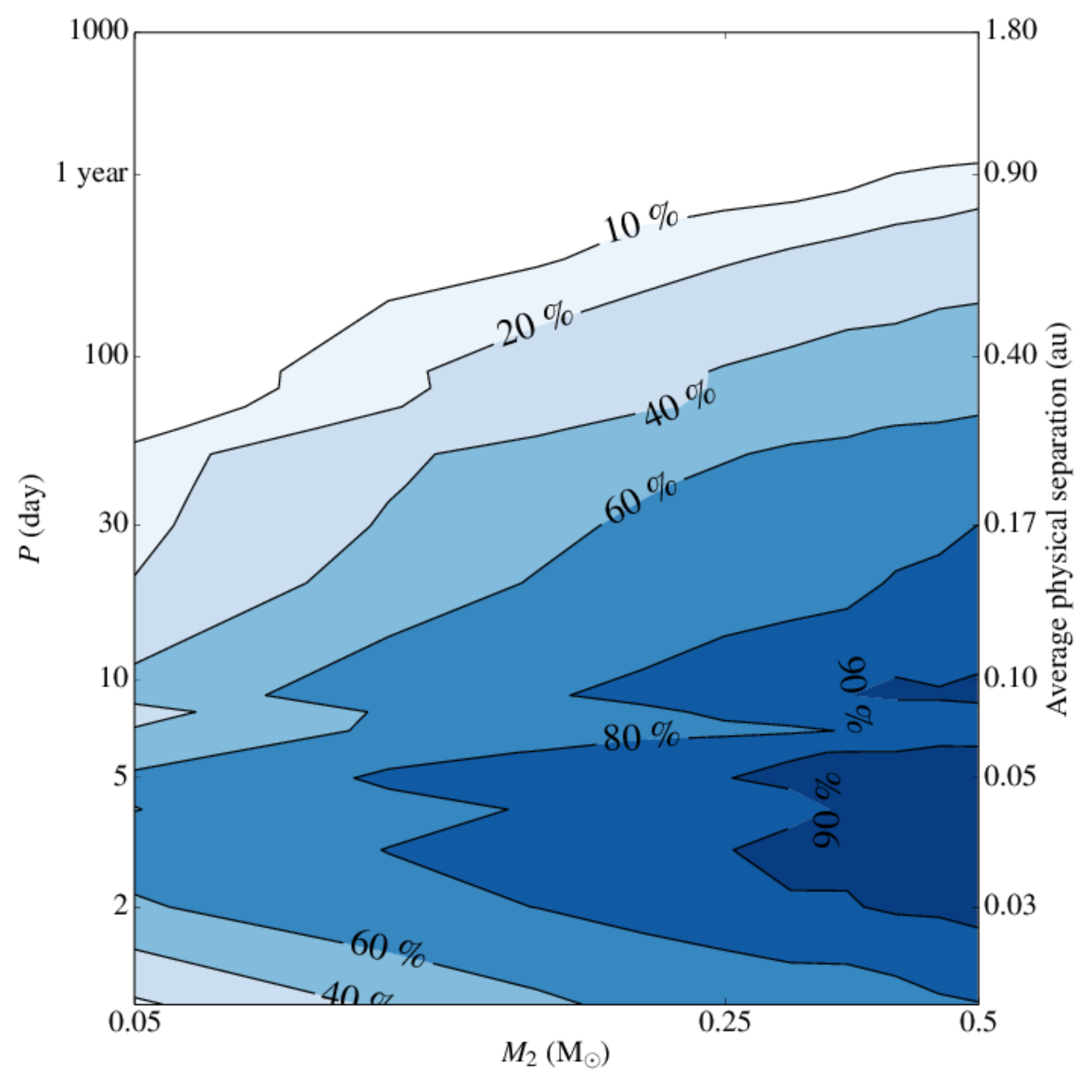}
\caption{Probability density maps for $\rm M_{2}$.}
   \label{Fig:multiplicity}
\end{center}
\end{figure}

We now use the UVES radial velocity measurements from Sec. \ref{Subsec:UVES_reduction}  to tackle the presence of close-in binaries among $\rm \eta ~Cha$ members. Multiple epochs were available for each of the 10 objects (2-4 separate observations).  This provides sensitivity to RV variations resulting from single-lined spectroscopic binaries (SB1s) as well as double- and potentially triple- lined spectroscopic binaries (SB2s and SB3s).  The separation in time between these observations ranges from 1 day to 1 month.  This uneven time sampling affects our sensitivity to systems of different configurations (inclination, mass, period). To account for the sensitivity in detecting binary systems as a result of our irregularly-spaced spectroscopic data we created a set of synthetic binary systems. We then "observe" these systems using the dates of UVES observations and determine whether or not the RV variation would have been significant enough to be detected given the difference in time.  Details of the methodology are described in \cite{Duquennoy1991} and more recently in \cite{Elliott2014}. We used an average mass of the primary  0.5 $\rm M_{\odot}$, the average sample mass using the evolutionary tracks of \cite{Baraffe1998}, assuming an age of 7 Myr. The resulting analysis is shown in Fig. \ref{Fig:multiplicity}, where we represent the probability distribution for different combinations of period (P) and mass of the secondary  ($\rm M_{2}$). 
According to our calculations, we have a $\rm >$60\% probability of detecting companions with $\rm M_{2} \gtrsim 0.1 M_{\odot}$ in orbits with $\rm P < 30$ d. Our results for $\eta$ Cha members are:

\begin{itemize}
\item RECX 1: we find no-evidence for new companions.
\item RECX 3, 4, 5, 6, 10 and 11: no signs of multiplicity from imaging studies nor from the present spectroscopic study.
\item RECX 7: we confirm the presence of a spectroscopic binary. Deriving the orbital parameters is out of the scope of this paper.
\item RECX 9, 12: no signs of spectroscopic binaries
\end{itemize}

Four $\eta$ Cha members with discs were covered by our multiplicity study: RECX4, RECX 5, RECX 9, and RECX 11. To our sensitivity limits ($\rm M_{2} \gtrsim 0.25 M_{\odot}$, $\rm a \gtrsim 0.8 ~AU$), the claim by \cite{Bouwman2006} that circumstellar discs are linked to single stars remains true, exception made of RECX 9. In total, we analyzed UVES data for 10 $\eta$ Cha members and found only one spectroscopic binary, leading to a SB detection fraction of 0.10$\rm^{+0.17}_{-0.03}$  \citep[0.18$\rm^{+0.16}_{-0.06}$, if we include RECX 8 from ][]{Andersen1975}, compatible with the fractions found for the SACY \citep[Search for Associations Containing Young stars,][]{Torres2006} associations in \cite{Elliott2014}, which are in the range 0.03-0.14. However, the lack of binaries wider that 20 AU is puzzling: none of the 17 members of the association studied by \cite{Brandeker2006} showed signs of wider binaries ($\rm a>20AU$), leading to a detection fraction $\rm 0^{+0.1}$. In contrast, the detection fraction for SACY   members in the same parameter space is $\rm 0.23^{+0.04}_{-0.03}$ \citep{Elliott2015}. The $\eta$ Cha association is then a strange association in terms of wide binaries distribution. However, the difference is only 2$\sigma$, therefore it has to be considered with caution.

\subsection{Disc evolution in $\rm \eta$ Cha}
Discs in $\eta$ Cha show a wide range of fractional disc luminosities, from  $\rm < 10^{-3}$ to $\rm > 0.1$, which implies that both debris and primordial discs are found around coeval stars. Therefore, $\eta$ Cha could represent an intermediate evolutionary stage for clusters where debris have started to form while primordial discs still survive, similar to TWA. This leaves open the possibility that the second generation dust from collisional grinding that constitutes debris discs forms while the primordial disc is still present. 

{One disc in the association, namely RECX 4, shows a fractional disc luminosity that is typical of debris discs while showing one epoch of active accretion (see Sec. \ref{Sec:Acc}), which makes it particularly interesting, since debris discs are thought to be gas empty or very poor. This could represent an intermediate stage between the exhaustion of the primordial disc and the formation of the debris disc, with second generation dust and primordial gas present at the same time. RECX 4 then joins the short list of old, second generation debris discs with accretion signatures detected so far \citep{Looper2010a,Looper2010b,Zuckerman2014,Reiners2009}.

The high number of infrared excess detections in $\rm \eta$ Cha \citep[$\rm \sim 50\%$][]{Sicilia-Aguilar2009}, as well as the shape of the SED for detected sources challenges our knowledge of disc evolution, as most of the circumstellar material is supposed to dissipate in timescales shorter than the age of the association. According to \cite{Sicilia2006b}, 50 \% of low-mass stars lose their near- and mid-IR excesses by 3 Myr. Most sources in $\rm \eta$ Cha do not show near-IR excess, RECX 11 and 15 being the exceptions. However, 50 \% of the cluster members show mid-IR excess, indicative of the presence of a warm disc. At 70 $\rm \mu m$ the detection fraction is $\rm 0.33^{+0.13}_{-0.10}$, compatible with the mid-IR fractions when uncertainties are included. \cite{Fang2013} shows that disc evolution and dispersal proceeds slowly in loose stellar associations compared to denser ones. At an age of 7 Myr, $\rm \eta$ Cha circumstellar show a variety of properties, from primordial discs with an IR-excess more prominent than the median SED in Taurus to debris discs. The presence of primordial discs agrees well with $\rm \eta$ Cha being also a loose stellar association where the evolution proceeds slowly. According to the study by  \cite{Ribas2014}, however, the detection fraction in $\eta$ Cha seems to fit within the general behaviour \citep[see Fig. 2 in ][]{Ribas2014}. This study was based in mid-IR detection fractions, which tend to be lower than the far-IR detection fractions due to the detection of a population of debris discs that are unseen in the mid-IR. However, this does not affect the detection fraction of primordial discs ($\rm 0.35^{+0.12}_{-0.09}$), since debris discs are second generation. Furthermore, the disc fraction in $\eta$ Cha could change in the future if undiscovered members are detected (see Sec. \ref{Sec:Introduction}). A complete census of $\eta$ Cha members is needed to tackle the problem of the different timescale in loose environments.

\section{Summary}\label{Sec:Summary}
We have observed 17 members of the $\rm \eta ~Cha$ cluster with PACS in photometric mode at 70, 100 and 160 $\rm \mu m$. A subsample of 13 members was also observed in line spectroscopic mode, aiming to detect [OI] and o-$\rm H_{2}O$ emission at 63 $\rm \mu m$. Three sources in this subsample were also observed with PACS in range spectroscopic mode. The main results and conclusions of the present study are:

\begin{itemize}
	\item[1)] We detected four out of sixteen discs observed at 70 $\rm \mu m$, four out of eight at 100 $\rm \mu m$ and five out of seventeen at 160 $\rm \mu m$, leading to a far-IR excess detection fraction of $\rm 0.29^{+0.13}_{-0.08}$.
	\item[2)] We detected [OI] emission towards RECX 15. The emission is not extended, but the position of the source in a $\rm F_{[OI],63 \mu m}$ versus $\rm F_{70 \mu m}$ diagram is consistent with outflow sources.
	\item[4)] $\eta$ Cha members show a variety of SEDs, including Class II discs, transitional discs and debris discs. The survival of transitional and primordial discs could be linked to the loose nature of the $\rm \eta$ Cha, as proposed by previous authors. 
	\item[5)] We studied $\rm H_{\alpha}$ profiles and computed equivalent widths, finding accretion signatures in RECX 4, 5, 9, 11 and 15. 
	\item[6)] We have detected signatures of accretion in one debris disc in the association, namely RECX 4. This is one of the few cases where active accretion is found around a debris disc.
	\item[7)] The intermediate IR excess shown by RECX 5, RECX 9 and RECX 16 can be explained as the result of a flattened geometry, which can also explain the lack of [OI] detections for these sources. The cases of RECX 5 and RECX 9 are especially challenging, as they show signs of episodic accretion, indicating for the presence of gas. A flat disc can easily explain the detection of episodic accretion together with the lack of [OI] emission. However, it is also possible that the gas clearing timescale is shorter than the dust one. Furthermore, low UV irradiation can also be responsible for the lack of [OI] detections in some sources. 
	\item[8)] We looked for close-in companions of $\eta$ Cha members in UVES data and found no new spectroscopic binaries among ten $\eta$ Cha members studied. We confirmed that RECX 7 is a spectroscopic binary. The fraction of spectroscopic binaries is then 0.18$\rm^{+0.16}_{-0.06}$, compatible with other nearby associations. No disc is found around spectroscopic binaries in $\eta$ Cha.
	
\end{itemize}

\acknowledgements
 PRM acknowledges funding from the ESA Research Fellowship Programme.  AB acknowledges  financial support from the Proyecto Fondecyt de Iniciación 11140572 and support from the Millenium Science Initiative, Chilean Ministry of Economy, Nucleus RC130007. IK acknowledges funding from the European Union Seventh Framework Programme FP7-2011 under grant agreement no 284405. This publication makes use of VOSA, developed under the Spanish Virtual Observatory project supported from the Spanish MICINN through grant AyA2011-24052. AR acknowledges funding from the ESAC Science Operations Division research funds with code SC 1300016149 and support from the ESAC Space Science Faculty. 
\bibliographystyle{aa} 
\bibliography{biblio.bib}

\end{document}